\documentclass[useAMS,usenatbib]{mn2e}

\usepackage{verbatim,graphicx,epsfig,dcolumn}
\newcolumntype{.}{D{.}{.}{4}}
\newcolumntype{,}{D{.}{.}{2}}
\newcolumntype{;}{D{.}{.}{1}}

\newcommand{\lesssim}{{\lower-1.2pt\vbox{\hbox{\rlap{$<$}\lower5pt\vbox{\hbox{$\sim$}}}}}}
\newcommand{\gtrsim}{{\lower-1.2pt\vbox{\hbox{\rlap{$>$}\lower5pt\vbox{\hbox{$\sim$}}}}}}

\title[A calibration of RGB mass-loss efficiency]{Mass loss on the red giant branch: the value and metallicity dependence of Reimers' $\eta$ in globular clusters}
\author[I. McDonald, A. A. Zijlstra]{I.~McDonald$^{1}$\thanks{E-mail: iain.mcdonald-2@manchester.ac.uk}, A.~A.~Zijlstra$^{1}$\\
$^{1}$Jodrell Bank Centre for Astrophysics, Alan Turing Building, Manchester, M13 9PL, UK}

\begin{document}

\date{Accepted 9999 December 32. Received 9999 December 32; in original form 9999 December 32}

\pagerange{\pageref{firstpage}--\pageref{lastpage}} \pubyear{9999}

\maketitle

\label{firstpage}

\begin{abstract}
The impact of metallicity on the mass-loss rate from red giant branch (RGB) stars is studied through its effect on the parameters of horizontal branch (HB) stars. The scaling factors from Reimers (1975) and Schr\"oder \& Cuntz (2005) are used to measure the efficiency of RGB mass loss for typical stars in 56 well-studied Galactic globular clusters (GCs). The median values among clusters are, respectively, $\eta_{\rm R} = 0.477 \pm 0.070 ^{+0.050}_{-0.062}$ and $\eta_{\rm SC} = 0.172 \pm 0.024 ^{+0.018}_{-0.023}$ (standard deviation and systematic uncertainties, respectively). Over a factor of 200 in iron abundance, $\eta$ varies by $\lesssim$30 per cent, thus mass-loss mechanisms on the RGB have very little metallicity dependence. Any remaining dependence is within the current systematic uncertainties on cluster ages and evolution models. The low standard deviation of $\eta$ among clusters ($\approx$14 per cent) contrasts with the variety of HB morphologies. Since $\eta$ incorporates cluster age, this suggests that age accounts for the majority of the ``second parameter problem'', and that a Reimers-like law provides a good mass-loss model. The remaining spread in $\eta$ correlates with cluster mass and density, suggesting helium enrichment provides the third parameter explaining HB morphology of GCs. We close by discussing asymptotic giant branch (AGB) mass loss, finding the AGB tip luminosity is better reproduced and $\eta$ has less metallicity dependence if globular clusters are more co-eval than generally thought.
\end{abstract}

\begin{keywords}
stars: mass-loss --- stars: horizontal branch --- stars: winds, outflows --- globular clusters: general --- stars: evolution
\end{keywords}


\section{Introduction}
\label{IntroSect}

Mass loss is known to dominate the evolution on the asymptotic giant branch (AGB). However, the slower evolution of stars with $M \lesssim 1$ M$_\odot$ means more mass is lost during the red giant branch (RGB) phase than on the AGB. RGB mass loss has been much less studied, partly because it has only minor effects on RGB evolution. However, it is crucial for later evolution, setting the stellar temperature on the horizontal branch (HB) and dictating the length and path of its AGB evolution. 

RGB mass loss is a component of most stellar evolution codes, and is generally parameterised by simple relations of the stellar parameters (e.g.\ luminosity ($L$), radius ($R$), mass ($M$), surface gravity ($\log g$), effective temperature ($T_{\rm eff}$), etc.). The two parameterisations most commonly used are those of \citet{Reimers75} and \citet[][hereafter SC05]{SC05}, which we examine in this work. However, use of these laws requires a fiducial reference point, parameterised by a scaling factor, $\eta$, which determines the constant of proportionality in these scaling laws.

Calibrating $\eta$ is important in understanding the post-RGB stages of low-mass stellar evolution, particularly if $\eta$ varies with other stellar parameters, such as stellar metallicity. No direct variation of metallicity is included in the Reimers and SC05 relations, however metallicity indirectly affects the parameterised mass-loss rate through the consequent variation of stellar radius and the speed at which the star evolves. Mass loss from RGB stars is thought to occur via (magneto-)acoustic processes, hence it should be largely independent of metallicity. If a parameterisation like those of \citet{Reimers75} and SC05 provides is an accurate representation of the mass-loss rates of RGB stars, we can expect $\eta$ to be independent of metallicity too.

RGB mass loss is notoriously difficult to measure, relying mainly on three tracers. Blueshifted components in optical and near-IR lines trace mass motions close to the stellar surface, but this does not necessarily translate into the rate at which mass is ejected from the star (e.g.\ \citealt{MvL07,MAD09}). Sub-mm CO line strengths trace material far from the star, but for RGB stars is only sensitive to the handful nearest Earth (e.g.\ \citealt{Groenewegen14}). Infrared excesses trace dust in the wind, but are hard to measure conclusively and difficult to translate from a dust mass-loss rate to a total mass-loss rate, besides which little (if any) dust production is expected on the RGB (\citealt{MBvLZ11}; \citealt{MZB12}, and references therein). A better way to measure mass loss is to obtain a difference in stellar mass between two points in its evolution, and use this to determine a scaling parameter, such as $\eta$. This avoids many of the systematic uncertainties which limit the above methods and averages out any short-term variations in the mass-loss rate.

Few places offer such unique laboratories to study RGB mass loss as globular clusters. As comparatively simple, resolved stellar systems, we can probe the properties of their individual stars with accurately known distances, metallicities, abundances and ages. The stellar mass and evolutionary history can easily be determined for individual stars. Measuring the mass of horizontal branch stars in globular clusters allows us to determine the $\eta$ appropriate for $\sim$0.8-M$_\odot$ stars over their previous evolution.

Determining $\eta$ is also important in solving the so-called second parameter problem \citep{FPB97}. HB morphology of Galactic globular clusters is largely set by a single parameter (stellar metallicity). However, a second parameter is needed to explain the variations in morphological structure among clusters of a similar metallicity. Cluster age and helium abundance can both affect cluster morphology. \citet[][hereafter GCB+10]{GCB+10} find cluster age can account for most of the remaining differences. If metallicity and age account for the entirety of HB morphology, we should expect little variation in $\eta$ among different clusters, as both age and metallicity are taken into account in the stellar evolution models we will use.

This manuscript is structured as follows: in Section \ref{DataSect} we collate the existing HB star masses, stellar compositions and cluster ages required to compute $\eta$. We also discuss the models we use to describe mass loss from RGB stars. In Section \ref{EvolSect} we describe existing and new stellar evolution models used to derive the mass-loss histories of each cluster's stars. In Section \ref{AnalSect} we calculate our final values of $\eta$ and their uncertainties. In Section \ref{DiscSect}, we compare these results to previous empirical and semi-empirical determinations, discuss their implications for the evolution of AGB stars in globular clusters, and the veracity of the variations with metallicity we find.

During the course of writing this manuscript, it was found that publsihed stellar evolution codes were insufficient for our purposes. This led us to produce the new evolutionary tracks described in Section \ref{EvolSect}. The application of all stellar evolution models impart poorly quantified errors to the results. We have quoted estimates for these errors in the text, but provide appendices which contain more detailed discussion. In Appendix \ref{AgesSect} we explore the various determinations of cluster ages. In Appendix \ref{ModelsSect} we compare our new evolutionary tracks to other models. In Appendix \ref{TrackCompareSect} we discuss various errors associated with our derivation of $\eta$.

\section{Data collection and homogenisation}
\label{DataSect}

\subsection{Data}
\label{DataDataSect}

\begin{center}
\begin{table*}
\caption{Data for each cluster, showing metallicities, derived ages, initial masses for the TP-AGB stars, observed current HB masses (from space- (S) and ground- (G) based observations), and mass-loss efficiencies $\eta_{\rm R}$ and $\eta_{\rm SC}$. Additional systematic errors of $\Delta \eta_{\rm R} = ^{+0.050}_{-0.062}$ and $\Delta \eta_{\rm SC} = ^{+0.018}_{-0.023}$ apply.}
\label{DataTable}
\begin{tabular}{cccccccccccc}
    \hline \hline
Cluster & [Fe/H] & Age  & M$_{\rm init}$ & \multicolumn{1}{c}{Median HB mass}  & \multicolumn{2}{c}{$\eta$ for median stars (internal errors)} \\
\ 	& (dex)  &(Gyr) & (M$_\odot$)    & (M$_\odot$)          		& $\eta_{\rm R}$ & $\eta_{\rm SC}$ \\
    \hline

 NGC 104 & --0.72 & 11.95 & 0.69 & 0.886 $^{+0.031}_{-0.032}$ & 0.648 $\pm$ 0.001 & 0.452 $^{+0.026}_{-0.031}$ & 0.175 $^{+0.008}_{-0.005}$ \\
NGC 1261 & --1.27 &  9.99 & 0.50 & 0.876 $^{+0.024}_{-0.023}$ & 0.683 $\pm$ 0.007 & 0.473 $^{+0.058}_{-0.066}$ & 0.183 $^{+0.020}_{-0.020}$ \\
NGC 1851 & --1.18 & 10.07 & 0.47 & 0.880 $^{+0.024}_{-0.022}$ & 0.664 $\pm$ 0.005 & 0.505 $^{+0.052}_{-0.056}$ & 0.192 $^{+0.019}_{-0.017}$ \\
NGC 1904 & --1.60 & 12.14 & 1.04 & 0.814 $^{+0.030}_{-0.026}$ & 0.607 $\pm$ 0.003 & 0.520 $^{+0.085}_{-0.093}$ & 0.189 $^{+0.024}_{-0.030}$ \\
NGC 2298 & --1.92 & 12.50 & 0.77 & 0.801 $^{+0.029}_{-0.028}$ & 0.663 $\pm$ 0.007 & 0.393 $^{+0.074}_{-0.093}$ & 0.143 $^{+0.022}_{-0.029}$ \\
NGC 2419 & --2.15 & 12.62 & 0.93 & 0.796 $^{+0.031}_{-0.030}$ & 0.676 $\pm$ 0.004 & 0.445 $^{+.....}_{-.....}$ & 0.158 $^{+.....}_{-.....}$ \\
NGC 2808 & --1.14 & 10.36 & 0.70 & 0.876 $^{+0.029}_{-0.028}$ & 0.608 $\pm$ 0.000 & 0.587 $^{+0.072}_{-0.072}$ & 0.208 $^{+0.030}_{-0.023}$ \\
 NGC 288 & --1.32 & 11.30 & 0.41 & 0.843 $^{+0.021}_{-0.020}$ & 0.603 $\pm$ 0.004 & 0.541 $^{+0.049}_{-0.053}$ & 0.193 $^{+0.016}_{-0.013}$ \\
NGC 3201 & --1.59 & 10.86 & 0.49 & 0.840 $^{+0.020}_{-0.018}$ & 0.654 $\pm$ 0.006 & 0.492 $^{+0.065}_{-0.071}$ & 0.181 $^{+0.021}_{-0.019}$ \\
 NGC 362 & --1.26 & 10.05 & 0.56 & 0.875 $^{+0.026}_{-0.024}$ & 0.680 $\pm$ 0.011 & 0.475 $^{+0.072}_{-0.077}$ & 0.183 $^{+0.024}_{-0.023}$ \\
NGC 4147 & --1.80 & 11.98 & 0.72 & 0.812 $^{+0.025}_{-0.023}$ & 0.648 $\pm$ 0.006 & 0.454 $^{+0.072}_{-0.087}$ & 0.167 $^{+0.021}_{-0.030}$ \\
NGC 4372 & --2.17 & 12.73 & 0.65 & 0.794 $^{+0.026}_{-0.025}$ & 0.664 $\pm$ 0.000 & 0.394 $^{+0.047}_{-0.072}$ & 0.141 $^{+0.015}_{-0.028}$ \\
NGC 4590 & --2.23 & 11.70 & 0.64 & 0.813 $^{+0.024}_{-0.023}$ & 0.704 $\pm$ 0.001 & 0.351 $^{+0.055}_{-0.086}$ & 0.127 $^{+0.016}_{-0.029}$ \\
NGC 4833 & --1.85 & 12.33 & 0.73 & 0.805 $^{+0.027}_{-0.025}$ & 0.650 $\pm$ 0.007 & 0.431 $^{+0.072}_{-0.088}$ & 0.158 $^{+0.020}_{-0.029}$ \\
NGC 5024 & --2.10 & 12.42 & 0.62 & 0.800 $^{+0.026}_{-0.025}$ & 0.657 $\pm$ 0.001 & 0.426 $^{+0.053}_{-0.071}$ & 0.153 $^{+0.019}_{-0.026}$ \\
NGC 5053 & --2.27 & 11.88 & 0.53 & 0.809 $^{+0.022}_{-0.022}$ & 0.709 $\pm$ 0.004 & 0.325 $^{+0.055}_{-0.090}$ & 0.113 $^{+0.015}_{-0.027}$ \\
NGC 5272 & --1.50 & 11.59 & 0.39 & 0.829 $^{+0.018}_{-0.017}$ & 0.670 $\pm$ 0.001 & 0.405 $^{+0.033}_{-0.055}$ & 0.157 $^{+0.009}_{-0.015}$ \\
NGC 5466 & --1.98 & 12.53 & 0.51 & 0.799 $^{+0.026}_{-0.025}$ & 0.705 $\pm$ 0.009 & 0.281 $^{+0.064}_{-0.102}$ & 0.104 $^{+0.022}_{-0.035}$ \\
NGC 5694 & --1.98 & 13.64 & 0.91 & 0.780 $^{+0.030}_{-0.028}$ & 0.644 $\pm$ 0.006 & 0.382 $^{+0.073}_{-0.092}$ & 0.136 $^{+0.025}_{-0.026}$ \\
NGC 5824 & --1.91 & 13.06 & 0.52 & 0.791 $^{+0.024}_{-0.024}$ & 0.651 $\pm$ 0.004 & 0.392 $^{+0.045}_{-0.068}$ & 0.143 $^{+0.012}_{-0.020}$ \\
NGC 5897 & --1.90 & 13.00 & 0.00 & 0.792 $^{+0.015}_{-0.015}$ & 0.648 $\pm$ 0.001 & 0.402 $^{+0.008}_{-0.032}$ & 0.146 $^{+.....}_{-.....}$ \\
NGC 5904 & --1.29 & 11.06 & 0.62 & 0.850 $^{+0.025}_{-0.024}$ & 0.627 $\pm$ 0.001 & 0.513 $^{+0.054}_{-0.057}$ & 0.189 $^{+0.020}_{-0.015}$ \\
NGC 5927 & --0.49 & 11.45 & 0.71 & 0.910 $^{+0.051}_{-.....}$ & 0.642 $\pm$ 0.000 & 0.548 $^{+.....}_{-.....}$ & 0.196 $^{+.....}_{-.....}$ \\
NGC 5946 & --1.29 & 11.63 & 1.88 & 0.839 $^{+0.053}_{-0.046}$ & 0.612 $\pm$ 0.000 & 0.507 $^{+0.148}_{-0.107}$ & 0.185 $^{+0.052}_{-0.029}$ \\
NGC 5986 & --1.59 & 11.87 & 0.72 & 0.819 $^{+0.024}_{-0.021}$ & 0.611 $\pm$ 0.001 & 0.525 $^{+0.061}_{-0.074}$ & 0.190 $^{+0.017}_{-0.023}$ \\
NGC 6093 & --1.75 & 12.59 & 0.74 & 0.802 $^{+0.024}_{-0.022}$ & 0.641 $\pm$ 0.010 & 0.431 $^{+0.077}_{-0.092}$ & 0.159 $^{+0.023}_{-0.032}$ \\
NGC 6101 & --1.98 & 11.85 & 0.63 & 0.812 $^{+0.027}_{-0.026}$ & 0.665 $\pm$ 0.002 & 0.433 $^{+0.058}_{-0.076}$ & 0.155 $^{+0.020}_{-0.022}$ \\
NGC 6121 & --1.16 & 11.81 & 0.50 & 0.843 $^{+0.024}_{-0.022}$ & 0.664 $\pm$ 0.011 & 0.402 $^{+0.046}_{-0.058}$ & 0.160 $^{+0.013}_{-0.015}$ \\
NGC 6171 & --1.02 & 12.39 & 0.83 & 0.843 $^{+0.031}_{-0.028}$ & 0.650 $\pm$ 0.001 & 0.404 $^{+0.038}_{-0.047}$ & 0.160 $^{+0.010}_{-0.011}$ \\
NGC 6205 & --1.53 & 12.00 & 0.42 & 0.820 $^{+0.017}_{-0.017}$ & 0.601 $\pm$ 0.004 & 0.531 $^{+0.050}_{-0.063}$ & 0.191 $^{+0.015}_{-0.017}$ \\
NGC 6218 & --1.37 & 12.51 & 0.74 & 0.817 $^{+0.027}_{-0.025}$ & 0.594 $\pm$ 0.008 & 0.501 $^{+0.069}_{-0.071}$ & 0.182 $^{+0.020}_{-0.017}$ \\
NGC 6235 & --1.28 & 11.70 & 1.88 & 0.838 $^{+0.053}_{-0.046}$ & 0.617 $\pm$ 0.051 & 0.749 $^{+.....}_{-.....}$ & 0.260 $^{+.....}_{-.....}$ \\
NGC 6254 & --1.56 & 11.78 & 0.66 & 0.822 $^{+0.024}_{-0.021}$ & 0.596 $\pm$ 0.008 & 0.557 $^{+0.076}_{-0.084}$ & 0.198 $^{+0.020}_{-0.024}$ \\
NGC 6266 & --1.18 & 11.96 & 0.78 & 0.839 $^{+0.029}_{-0.027}$ & 0.609 $\pm$ 0.002 & 0.491 $^{+0.055}_{-0.053}$ & 0.181 $^{+0.017}_{-0.011}$ \\
NGC 6273 & --1.74 & 12.28 & 0.45 & 0.808 $^{+0.019}_{-0.018}$ & 0.613 $\pm$ 0.000 & 0.512 $^{+0.041}_{-0.056}$ & 0.185 $^{+0.011}_{-0.018}$ \\
NGC 6284 & --1.26 & 11.24 & 0.58 & 0.848 $^{+0.025}_{-0.024}$ & 0.583 $\pm$ 0.005 & 0.575 $^{+0.071}_{-0.068}$ & 0.199 $^{+0.027}_{-0.018}$ \\
NGC 6341 & --2.31 & 12.62 & 0.57 & 0.795 $^{+0.023}_{-0.022}$ & 0.686 $\pm$ 0.005 & 0.346 $^{+0.057}_{-0.089}$ & 0.116 $^{+0.016}_{-0.029}$ \\
NGC 6352 & --0.64 & 11.21 & 0.59 & 0.912 $^{+0.029}_{-0.049}$ & 0.667 $\pm$ 0.007 & 0.468 $^{+0.033}_{-.....}$ & 0.180 $^{+0.012}_{-.....}$ \\
NGC 6362 & --0.99 & 12.04 & 0.47 & 0.853 $^{+0.023}_{-0.021}$ & 0.652 $\pm$ 0.011 & 0.419 $^{+0.038}_{-0.046}$ & 0.164 $^{+0.010}_{-0.009}$ \\
NGC 6397 & --2.02 & 12.56 & 0.40 & 0.798 $^{+0.024}_{-0.023}$ & 0.635 $\pm$ 0.004 & 0.468 $^{+0.046}_{-0.063}$ & 0.165 $^{+0.016}_{-0.018}$ \\
NGC 6535 & --1.79 & 12.11 & 0.79 & 0.810 $^{+0.026}_{-0.024}$ & 0.642 $\pm$ 0.023 & 0.460 $^{+0.120}_{-0.132}$ & 0.169 $^{+0.035}_{-0.046}$ \\
NGC 6584 & --1.50 & 11.37 & 0.73 & 0.833 $^{+0.025}_{-0.024}$ & 0.683 $\pm$ 0.003 & 0.390 $^{+0.060}_{-0.082}$ & 0.152 $^{+0.017}_{-0.026}$ \\
NGC 6637 & --0.64 & 11.52 & 0.92 & 0.905 $^{+0.037}_{-0.905}$ & 0.671 $\pm$ 0.006 & 0.447 $^{+0.047}_{-.....}$ & 0.175 $^{+0.015}_{-.....}$ \\
NGC 6652 & --0.81 & 11.86 & 0.67 & 0.877 $^{+0.030}_{-0.029}$ & 0.650 $\pm$ 0.001 & 0.444 $^{+0.028}_{-0.032}$ & 0.173 $^{+0.008}_{-0.006}$ \\
NGC 6681 & --1.62 & 12.23 & 0.80 & 0.812 $^{+0.025}_{-0.022}$ & 0.603 $\pm$ 0.000 & 0.525 $^{+0.063}_{-0.075}$ & 0.190 $^{+0.018}_{-0.024}$ \\
NGC 6712 & --1.02 & 11.00 & 0.00 & 0.871 $^{+0.014}_{-0.012}$ & 0.668 $\pm$ 0.004 & 0.447 $^{+.....}_{-.....}$ & 0.173 $^{+.....}_{-.....}$ \\
NGC 6723 & --1.10 & 12.26 & 0.86 & 0.839 $^{+0.030}_{-0.030}$ & 0.644 $\pm$ 0.000 & 0.419 $^{+0.041}_{-0.053}$ & 0.164 $^{+0.012}_{-0.014}$ \\
NGC 6752 & --1.54 & 12.04 & 0.44 & 0.818 $^{+0.018}_{-0.017}$ & 0.607 $\pm$ 0.005 & 0.518 $^{+0.052}_{-0.066}$ & 0.188 $^{+0.016}_{-0.018}$ \\
NGC 6779 & --1.98 & 12.79 & 0.67 & 0.795 $^{+0.029}_{-0.028}$ & 0.652 $\pm$ 0.001 & 0.409 $^{+0.051}_{-0.070}$ & 0.146 $^{+0.017}_{-0.021}$ \\
NGC 6809 & --1.94 & 12.49 & 0.57 & 0.800 $^{+0.027}_{-0.025}$ & 0.658 $\pm$ 0.005 & 0.408 $^{+0.056}_{-0.075}$ & 0.148 $^{+0.017}_{-0.023}$ \\
NGC 6838 & --0.78 & 11.57 & 0.74 & 0.887 $^{+0.030}_{-0.030}$ & 0.659 $\pm$ 0.000 & 0.447 $^{+0.031}_{-0.034}$ & 0.174 $^{+0.009}_{-0.007}$ \\
NGC 6934 & --1.47 & 10.90 & 0.59 & 0.845 $^{+0.023}_{-0.022}$ & 0.673 $\pm$ 0.008 & 0.440 $^{+0.069}_{-0.078}$ & 0.169 $^{+0.020}_{-0.022}$ \\
NGC 6981 & --1.42 & 11.02 & 0.63 & 0.844 $^{+0.025}_{-0.023}$ & 0.680 $\pm$ 0.000 & 0.414 $^{+0.048}_{-0.059}$ & 0.162 $^{+0.014}_{-0.017}$ \\
NGC 7078 & --2.37 & 12.33 & 0.75 & 0.800 $^{+0.027}_{-0.025}$ & 0.688 $\pm$ 0.004 & 0.361 $^{+0.070}_{-0.097}$ & 0.115 $^{+0.034}_{-0.027}$ \\
NGC 7089 & --1.65 & 11.59 & 0.83 & 0.823 $^{+0.026}_{-0.023}$ & 0.628 $\pm$ 0.010 & 0.511 $^{+0.094}_{-0.101}$ & 0.186 $^{+0.028}_{-0.032}$ \\
NGC 7099 & --2.27 & 12.83 & 0.65 & 0.792 $^{+0.025}_{-0.025}$ & 0.665 $\pm$ 0.001 & 0.391 $^{+0.049}_{-0.076}$ & 0.135 $^{+0.013}_{-0.029}$ \\
    \hline
\end{tabular}
\end{table*}
\end{center}

\subsubsection{Horizontal branch masses}
\label{HBMassSect}

GCB+10 provide the minimum, median and maximum masses of horizontal branch stars in 78 globular clusters, comprising of 74 clusters with \emph{Hubble Space Telescope} imagery, and 49 clusters for which they have ground-based observations, with some overlap between the two. These 78 clusters represent our base data set, from which we use those with accurately determiend ages and metallicities in published literature (59 clusters).

Strictly, the `minimum' and `maximum' values of GCB+10 are the 5th and 95th centile masses. This equates to 1.65$\sigma$ if one assumes a Gaussian mass distribution, though this is not implied by their observations. These masses assume that the stars have primordial helium abundance (GCB+10 also give the required helium enrichment to explain the horizontal branch morphology at fixed stellar mass). They are reproduced in Table \ref{DataTable}.

GCB+10, in their Section 2.3, note that there is a large uncertainty in the maximum mass of HB stars in each cluster. For each star, they derive a mass based on that star's colour, then apply a mass-based correction to revert that star to its position on the zero-age HB (ZAHB). The uncertainty in that correction, and in the amount of time the star has been on the HB, leads to an uncertainty in the ZAHB mass, which is most severe for the maximum-mass HB stars (their equation (8)). This large uncertainty precludes us from calculating accurate values for $\eta$ for the high end of the mass distribution within each cluster. 

Conversely, the potential for helium enrichment among the stars with the minimum HB mass means that estimates of the evolutionary history of these stars are highly uncertain. This precludes us from calculating average values of $\eta$ for this opposite end of the HB either. For the majority of the following discussion, we are therefore limited to the median HB masses of GCB+10.

\subsubsection{Stellar compositions}
\label{CompSect}

In order to obtain a reliable initial stellar mass from evolutionary models, we need to know the star's composition. Typically, this is reduced to three parameters: the global metallicity, the helium abundance, and the $\alpha$-element enhancement, as these have the largest effect on the stellar isochrones.

[Fe/H] is usually taken as the tracer for global metallicity. We adopt this from the 2010 edition of the Harris globular cluster catalogue \citep{Harris96}\footnote{The unpublished documentation for this edition can be found at \tt{http://arxiv.org/abs/1012.3224}.}. Harris uses the metallicity scale of \citet{CBG+09b}, which is based on more-recent, higher-resolution spectroscopy than the traditional \citet{ZW84} or \citet{CG97} scales (hereafter ZW84 and CG97). In practice, we can expect the uncertainty of the adopted metallicities is of order $\pm$0.1 dex or less. We adopt this value as a global uncertainty on the absolute [Fe/H].

We make the assumption that median HB stars have little helium enhancement due to internal pollution, which implies a star-formation rate that declines relatively quickly from the cluster's birth. The near-constant offset between the median and maximum masses in most clusters reported by GCB+10 suggests this is typically true. However, this may not be a valid assumption for a small number of clusters: GCB+10 single out NGC 2808 as a case in point. We return to this point in Section \ref{EtaVarySect}. The uncertainty that helium abundance gives to the mass of HB stars is complex and poorly determined. We enter a detailed discussion in Appendix \ref{EtaHeSect}, where we attribute an additional $\sim$0.01 M$_\odot$ uncertainty to the mass of HB stars.

The helium fraction is usually taken to be a similar formulism to that adopted by \citet{SW02} ($Y = 0.23 + 3Z$), which increases the global helium abundance only marginally from cosmological abundances, based on the cluster metallicity.
The picture in globular clusters is partly clouded by the slight spread in age within each cluster. The latest-forming stars in a globular cluster are likely to be enriched with helium, nitrogen and sodium and depleted in oxygen (among other elements; \citealt{FCK+04,KFS+06,GDOB+10}). These stars evolve faster and will not be well-represented by traditional stellar isochrones, nor will helium-enriched stars have the same zero-age HB temperature as their primordial-composition counterparts, creating some uncertainty in the HB mass (e.g.\ GCB+10). Potentially large variations in helium abundance that may exist ($\approx$24--40\%) but it is difficult to measure the abundance directly (e.g.\ \citealt{DA13,SDS14}). For our calculations, we adopt the default helium enrichment of each stellar evolution model, neglecting internal pollution. This is discussed in more detail in Appendix \ref{EtaHeSect}.

The $\alpha$-element (O, Si, Ca, Ti) enhancement is usually parameterised by the ratio [$\alpha$/Fe], which follows a known anti-correlation with [Fe/H] (e.g.\ \citealt{MBB+05}). However, \citet{Carney96}, and later \citet{Habgood01}, find a constant value of [$\alpha$/Fe] = +0.3 dex for both old and young, and halo and disc clusters. This is mirrored in \citet{RCGS13}, who find the same mean [$\alpha$/Fe] = +0.30 dex, with a standard deviation of 0.11 dex. Values in their compilation range from 0.02 dex (NGC 6626 = M28) to 0.53 dex (NGC 6218 = M12). Most isochrones either adopt [$\alpha$/Fe] = +0.2 or +0.4 dex (the typical range of [$\alpha$/Fe]).

The factors described in this section do not include some unusual abundance details (e.g.\ factors of up to 10 difference in the Cu/Fe ratio; \citealt{Sneden04}) but include the factors likely to significantly alter the evolution and appearance of the star.

\subsubsection{Cluster ages}
\label{AgeSect}

Most published ages of globular clusters are on a relative scale, however our approach requires an absolute age. Since that age directly sets the initial mass, hence $\eta$ that is derived, the choice of absolute age scale is an important systematic in the final result.

Ultimately, the absolute age is significantly affected by the exact metallicity scale used, typically either that of CG97 or ZW84. By comparing and combining different studies, we can provide an estimate for how accurate the absolute calibration of age is. We use ages of clusters derived from scaled averages of the following publications: \citet{SW02}; \citet{DAPC+05}; \citet{MFAP+09}; \citet{DSA+10}; and \citet{VBLC13} (hereafter SW02, DA+05, MF+09, D+10 and VBLC13). Together, they provide estimates via 11 subtly different methods for 89 globular clusters. In 55 clusters, the majority of these data are derived from the \emph{HST} ACS survey: the most-extensive homogeneous dataset of photometry of faint stars in globular clusters. Minor differences between the studies arise from different adopted metallicities, different methods of reddening and different stellar evolution codes. Details of each study can be found in Appendix \ref{AgesSect}.

\begin{figure*}
\centerline{\includegraphics[height=0.95\textwidth,angle=-90]{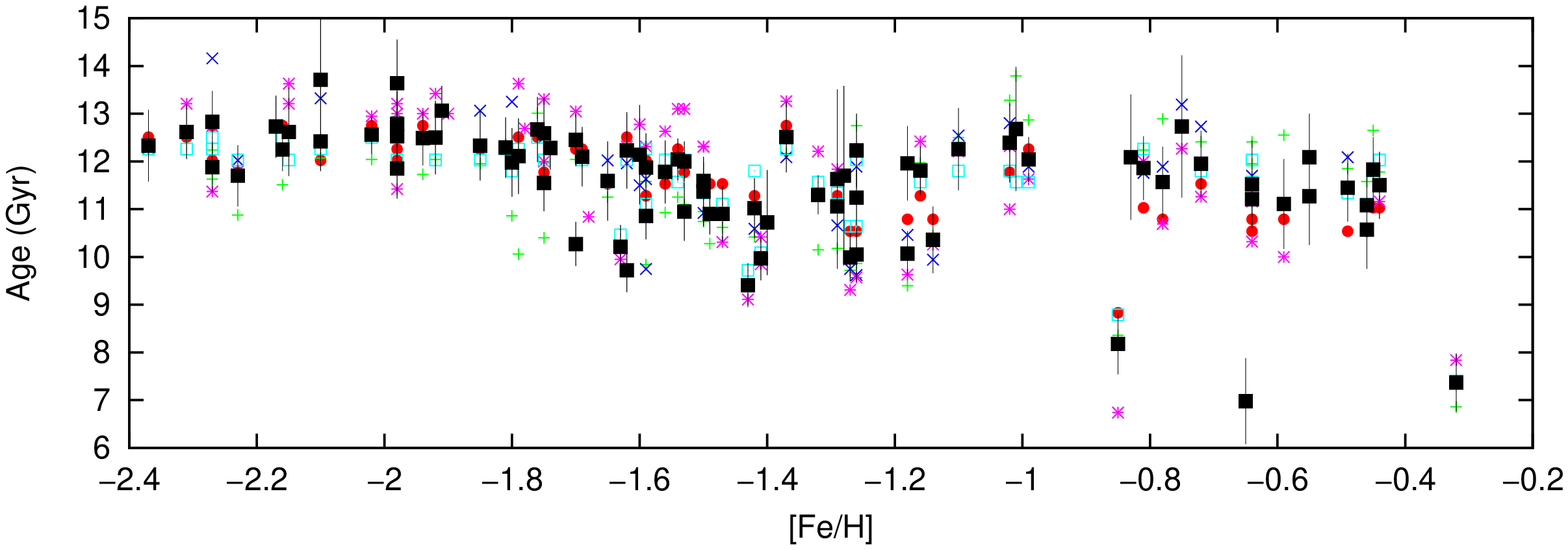}}
\centerline{\includegraphics[height=0.95\textwidth,angle=-90]{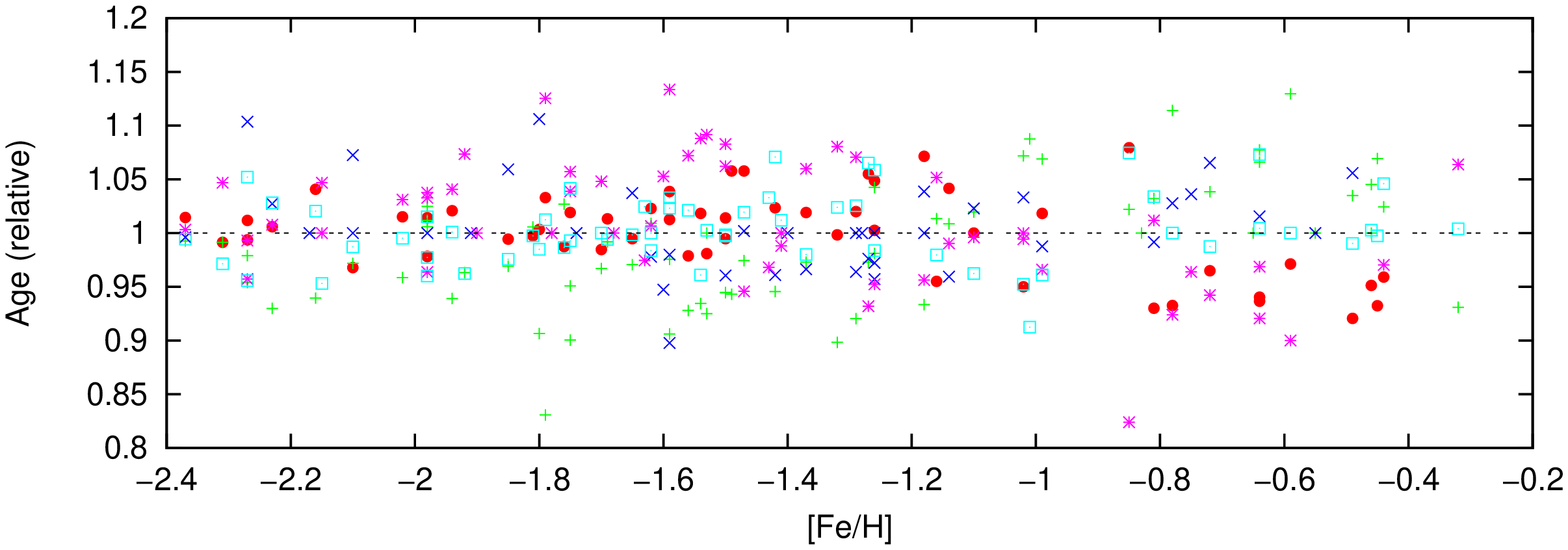}}
\caption{Adopted ages of globular clusters on an absolute scale (top panel) with the relative deviations in age of each included study (bottom panel). The large, black squares with error bars denote values and uncertainties adopted; smaller, coloured points denote the individual studies: SW02 -- magenta asterisks; DA+05 -- blue crosses; MF+09 -- green plus signs; D+10 -- cyan open squares; and VBLC13 -- red dots.}
\label{AgesFig}
\end{figure*}

Of the 89 clusters with ages published in the above sample, 18 have the full set of 11 age measurements. They are typically the largest, closest, best-observed clusters, with the best-determined parameters: NGC 104 (47 Tuc), 362, 1261, 3201, 4590 (M68), 5904 (M5), 6171 (M107), 6218 (M12), 6362, 6584, 6637 (M69), 6652, 6681 (M70), 6723, 6838 (M71), 6934, 7078 (M15) and 7099 (M30). These cover almost the full gamut of observed metallicities and derived ages. We use these clusters to calibrate the ages against each other.

We scale each set of cluster ages so that the average age of these 18 clusters is identical in each method of each study. We then average the values together for each study, and average the studies together to create a single age. We arrive at a normalisation constant of an age of 11.61 Gyr for the average cluster. The standard deviation among clusters is 1.19 Gyr. We take the standard deviation among studies of the same cluster (typically 0.66 Gyr) as the error in age of each cluster, except when only one study has measured the age, where we take that study's age uncertainty. Figure \ref{AgesFig} shows the adopted ages for each cluster and the relative differences between the ages found by each study. The adopted ages are listed along with other parameters of each cluster in Table \ref{DataTable}.

Among the different studies, the typical scatter in the ages is of order 4 per cent. The exceptions are for the metal-rich clusters, where the ages of MF+09 are noticeably (1--2 Gyr) older than those of SW02 and VBLC13, while the DA+05 and D+10 ages lie in between. VBLC13 address this difference in the age--metallicity gradient in detail in their Introduction and their section 6.1.1. They attribute it the sensitivity of their method to the adopted metallicity of the cluster which, being from \citet{RHS+97}, involves some extrapolation at the metal-rich end. We have maintained a straight average of the five studies for our results, accepting the increased uncertainty for the clusters with metallicities of [Fe/H] $\gtrsim$ --1.1 dex.

A compilation of published ages for 38 clusters, \citet{RCGS13}, was published during the preparation of this paper. They give ages which are relatively similar to ours: they are on average 0.66 Gyr older than ours, and the differences have a standard deviation of 0.88 Gyr. The most deviant ages are those with the highest uncertainties, as noted by \citet{RCGS13}. Excluding the outliers NGC 5946, 6235, 6342 and 6544, their values are 0.95 Gyr older with a standard deviation of 0.50 Gyr. We retain this set of ages as a separate comparison during our analysis.

\subsection{Mass-loss models}
\label{EtaModelSect}

While a variety of formulisms have been employed to generate mass loss from stars, by far the most enduring is that of \citet{Reimers75}:
\begin{equation}
\dot{M} = 4 \times 10^{-13} \, \eta_{\rm R} \, \frac{L R}{M} \ \ [{\rm M}_\odot\ {\rm yr}^{-1}],
\label{EtaEq}
\end{equation}
where $\dot{M}$ is the mass-loss rate, in solar units, and $L$ is the luminosity, $R$ the radius and $M$ the mass of the star in solar units. The factor $\eta$ is a scaling parameter that describes the \emph{efficiency} with which mass is lost. 

Many attempts have been made to improve on Reimers' formulism. Most have some restrictions which make them inapplicable, either: (a) they do not cover the low-mass regime we are interested in; (b) they are tuned to main sequence stars, and do not reproduce the dominant RGB mass loss well; (c) they do not separate AGB stars undergoing dust-driven winds from dustless RGB stars; or (d) they focus solely upon dust-driven AGB winds (examples can be found in the comparative work by \citealt{SC07}). These later stages are not relevant to deriving $\eta_{\rm R}$. However, SC05 covers the regime of globular cluster RGB stars, and is hence relevant to our current study.


SC05 take what they describe as a ``physical approach'', adapting Reimers' model based on scaling laws of chromospherically driven winds. They derive the following formula:
\begin{eqnarray}
\dot{M} &=& 4 \times 10^{-13} \, \eta_{\rm SC} \, \frac{L R}{M} \nonumber \\
&& \left( \frac{T_{\rm eff}}{4000 {\rm K}} \right)^{3.5} \left(1+ \frac{g_\odot}{4300 g} \right) [{\rm M}_\odot\ {\rm yr}^{-1}].
\label{SCEtaEq}
\end{eqnarray}
where $g$ is the stellar surface gravity. In SC05, these authors also derive $\eta_{\rm SC}$ for the two globular clusters NGC 5904 and 5927 (see Section \ref{EtaDiscSect}), fitting $\eta_{\rm SC} \approx 0.2$, qualitatively suggesting that this provides a better fit than Reimers' law. Further work in \citet{SC07} suggests that this law also provides a better fit of observed mass-loss rates over a wide variety of masses and metallicities. By extending this to a wider sample of globular clusters, we can determine whether this remains true over considerably lower masses and lower metallicities.

Though there is no physical reason why $\eta$ cannot be time-variant, the laws are defined such that $\eta$ should be constant over time. Making this assumption, one can calculate $\eta$ from $\Delta M$ by solving the time integral of the above equations, namely:
\begin{equation}
\Delta M_{\rm R} = 4 \times 10^{-13} \eta_{\rm R} \int_0^t \frac{L^{1.5}}{M (T / T_\odot {\rm K})^2} {\rm d} t \ \ [{\rm M}_\odot].
\label{Eta3Eq}
\end{equation}
and:
\begin{eqnarray}
\Delta M_{\rm SC} \!\!&=&\!\! 4 \times 10^{-13} \, \eta_{\rm SC} \!\! \int_0^t \!\!\! \frac{L^{1.5}}{M (T_{\rm eff}/T_\odot)^2} \nonumber \\
&&\!\! \left( \!\frac{T_{\rm eff}}{4000 {\rm K}}\! \right)^{\!\!\!3.5} \!\! \left(\!1\!+\!\frac{1}{4300} \frac{L}{M(T_{\rm eff}/T_\odot)^4} \! \right) \! {\rm d} t \ [{\rm M}_\odot].
\label{SCEta3Eq}
\end{eqnarray}

Here, $R$ has been substituted for $\sqrt{L} / T_{\rm eff}^2$, where $T_{\rm eff}$ is the stellar effective temperature. The difference in mass, $\Delta M$, can be numerically integrated for a specific $\eta$ using stellar evolutionary tracks, allowing comparison to the observed change in mass, $\Delta M_{\rm obs}$, derived from the above observations.

\section{Stellar evolution models}
\label{EvolSect}

With the cluster parameters known, we now need to use stellar evolutionary models to calculate the total mass loss experienced to date by stars currently on the HB. 

\subsection{Published models}
\label{EvolReqSect}

Stellar evolutionary models can provide initial masses, but each comes with its own restrictions. It is important for our analysis that the evolutionary tracks correctly reproduce the stellar parameters ($L$, $R$, $T_{\rm eff}$, $g$ and $M$) and their time evolution. Most evolutionary tracks incorporate varying initial stellar mass and metallicity. However, not all allow enhancement of $\alpha$ elements which, in cool stars, primarily affects the atmospheric opacity, hence $R$, $g$ and $T_{\rm eff}$. A proper prescription of stellar mass loss on the RGB is also required, as this affects gravity, causing repercussive effects in $T_{\rm eff}$ and $R$. The internal pressure is also altered such that $L$ changes and, along with it, the timing of the helium flash that terminates the RGB.

No publicly available set of stellar models incorporates both a sufficiently variable [$\alpha$/Fe] ratio and a proper RGB mass-loss treatment. Of those that come close, the 2012 version of the Dartmouth models \citep{DCJ+08} incorporates a variable [$\alpha$/Fe], but does not apply mass loss while the star is on the RGB (A.\ Dotter, priv.\ comm.). Version 1.1 of the {\sc parsec} models \citep{BMG+12} includes mass loss in the form of a variable Reimers' $\eta$, but is restricted to stars of solar-scaled composition and publicly presents isochrones rather than evolutionary tracks. We compare to these and other publicly available model sets, the details of which are given in Appendix \ref{ModelsSect}. However, for a proper derivation of $\eta$, we must create our own set of models which incorporate both these effects.

\subsection{{\sc Mesa} evolutionary tracks}
\label{EvolReqSect}

\begin{center}
\begin{table}
\caption{Adopted elemental abundances for the {\sc mesa} models. All other elements set to [X/Fe] = 0.}
\label{MesaAbundTable}
\begin{tabular}{cc}
    \hline \hline
Abundance ratio & Adopted value \\
    \hline
$Y$	& 0.2485 + 1.78 $Z$ \\
\ [$Z$/Fe]	& +0.226 dex \\
\ [C/Fe]	& --0.4 dex \\
\ [N/Fe]	& +0.9 dex \\
\ [O/Fe]	& +0.2 dex \\
\ [Ne/Fe]	& +0.3 dex \\
\ [Na/Fe] & +0.3 dex \\
\ [Mg/Fe] & +0.4 dex \\
\ [Al/Fe] & +0.4 dex \\
\ [Si/Fe] & +0.4 dex \\
\ [S/Fe]  & +0.3 dex \\
\ [Ar/Fe] & +0.3 dex \\
\ [Ca/Fe] & +0.3 dex \\
\ [Ti/Fe] & +0.2 dex \\
\ [Cr/Fe] & +0.1 dex \\
\ [Ni/Fe] & 0 dex \\
    \hline
\end{tabular}
\end{table}
\end{center}

We have created a grid of evolutionary tracks using the {\sc mesa} (Modules for Experiments in Stellar Astrophysics) stellar evoulution code \citep{PBD+11,PCA+13}. This grid spans [Fe/H] = --2.6 to --0.2 dex in 0.2 dex intervals, and initial stellar masses of 0.76 to 0.94 M$_\odot$ at 0.02 M$_\odot$ intervals. The grid is not complete, but models are run to cover the entire range of globular cluster star initial masses and metallicities, including the range of likely uncertainties. Models were run for 15 Gyr of evolution, but generally only maintain convergence as far as the last or second-last thermal pulse. All models are complete at least to the start of the thermally pulsating AGB. 

Default {\sc mesa} parameters were assumed in most cases, with the following exceptions. Helium abundance was set at $Y = 0.2485 + 1.78 Z$ and $Z_\odot$ was taken as 0.0152. Metals were fractioned to impart [$\alpha$/Fe] $\sim$ +0.3 dex, but with detailed abundances based on \citet{RCGS13} and references therein, as listed in Table \ref{MesaAbundTable}. The initial pre-main-sequence models began with an initial core temperature of 5 $\times$ 10$^5$ K and were relaxed for 200 steps before further processing. `Type 2' opacities were used, and the {\tt kappa\_file\_prefix} option was set to {\tt gs98\_aF\_p3}, representing an [$\alpha$/Fe] = +0.3 dex \citep{GS98}. Henyey mixing-length theory was adopted \citep{HVB65} with $\alpha_{\rm MLT} = 1.92$ to match the RGB of 47 Tucanae (see Appendix \ref{MESACompareSect}). The {\tt which\_atm\_option} atmospheric boundary condition was set to {\tt photosphere\_tables}.

Mass loss can be included into {\sc mesa} in a variety of built-in parameterisations. We applied Reimers' mass-loss law to the entire stellar evolution. We compute two sets of models, for $\eta_{\rm R}$ = 0.4 and $\eta_{\rm R}$ = 0.5. The law of SC05 is not a presently available mass-loss prescription in {\sc mesa}. The inherent differences in stellar evolution between a star undergoing Reimers-like mass loss and a SC05-like mass loss impart an additional uncertainty to our determination of $\eta_{\rm SC}$.

A detailed discussion of the systematic uncertainties imparted by the choice of stellar evolution model, stellar composition and stellar physics can be found in Appendix \ref{TrackCompareSect}. A comparison of the {\sc mesa} tracks to the Dartmouth and {\sc parsec} models can be found in Appendix \ref{MESACompareSect}.

\subsection{Initial stellar masses}
\label{MassSect}

\begin{figure}
\centerline{\includegraphics[height=0.47\textwidth,angle=-90]{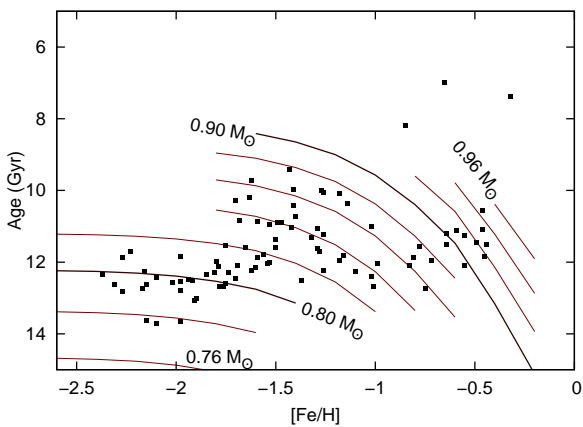}}
\caption{Lines of constant mass in metallicity--age space from the {\sc mesa} models. The literature-derived locations of globular clusters are shown as black points. }
\label{GridFig}
\end{figure}

Figure \ref{GridFig} shows the initial masses of RGB tip stars as taken from our {\sc mesa} models and the Dartmouth isochrones at [$\alpha$/Fe] = +0.3 dex (linearly interpolated between their +0.2 and +0.4 dex results). Overlain are the globular clusters. Most clusters have RGB tip stars with initial masses between 0.76 and 0.90 M$_\odot$. The most-metal-rich have higher masses, but still below 1 M$_\odot$. Only the three youngest clusters (in order of increasing age, Palomar 1, Terzan 7 and Palomar 12) have masses that approach or exceed 1 M$_\odot$. We do not model these younger clusters in this work.

Uncertainties in the derivation of initial mass arise from uncertainties in the clusters' metallicities ($\Delta$[Fe/H] = 0.1 dex; Section \ref{CompSect}), ages (Section \ref{AgeSect}) and stellar properties. For the stellar properties, we consider uncertainties due to the choice of stellar evolution code, helium and $\alpha$-element abundances, and convective overshooting as described in Appendix \ref{TrackCompareSect}. Figure \ref{MinitErrorFig} shows the associated error arising from the parameters. Typically this combined error is $\sim$0.01--0.03 M$_\odot$, rising with metallicity.

\begin{figure}
\centerline{\includegraphics[height=0.47\textwidth,angle=-90]{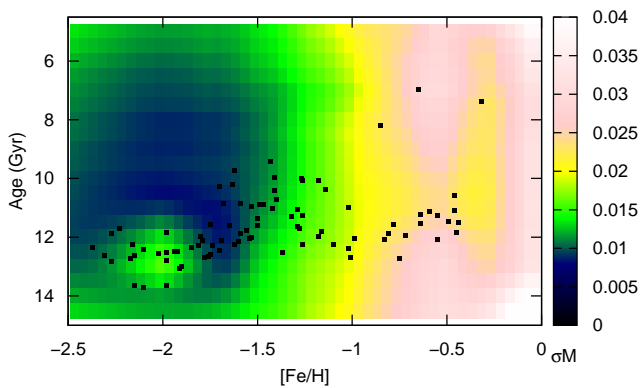}}
\caption{The error in initial mass arising from choice of stellar model, including contributions from convective overshooting, [$\alpha$/Fe] and natal helium abundance, as described in Appendix \ref{TrackCompareSect}.}
\label{MinitErrorFig}
\end{figure}


\section{Mass-loss efficiency on the RGB}
\label{AnalSect}

\subsection{Deriving $\eta$}
\label{EtaDeriveSect}


In this Section, we fit $\eta_{\rm R}$ and $\eta_{\rm SC}$ to match the mass lost by HB stars in each cluster between their birth (Section \ref{MassSect}) and their arrival on the ZAHB (GCB+10; Section \ref{HBMassSect}). This calculation is performed as follows. For each cluster, we identify the four $\eta_{\rm R} = 0.40$ {\sc mesa} models which bracket it in ZAHB age and [Fe/H]. For each model, we identify the mass lost between the star's birth and the ZAHB. Interpolating across the four models in age and [Fe/H], we find the mass loss appropriate for that cluster's age and [Fe/H]. We repeat this for the $\eta_{\rm R} = 0.50$ models. We construct a linear fit between the mass lost in the $\eta_{\rm R} = 0.40$ and 0.50 models, and identify where the observed mass loss falls on that linear fit, giving us an $\eta$ for that cluster.

This process is repeated for points at the extremities of the age and metallicity error range (Section \ref{MassSect}) to provide an error in $\eta_{\rm R}$ which is \emph{relative} compared to the values for the other clusters in this study. An additional \emph{systematic} error applies, arising from the choice of stellar evolution model and its parameters (also Section \ref{MassSect}), which is quoted separately.

The analysis described in this section is performed again using the law of SC05. The $\eta_{\rm SC}$ for each stellar evolution model is calculated as the value that provides the same amount of mass loss from the star's birth to the ZAHB. The calculation is then performed in the same way, although an additional error is added to account for the uncertainty created by adopting $\eta_{\rm SC}$ rather than $\eta_{\rm R}$ (Appendix \ref{ErrorEtaSCSect}).

\begin{figure*}
\centerline{\includegraphics[height=0.97\textwidth,angle=-90]{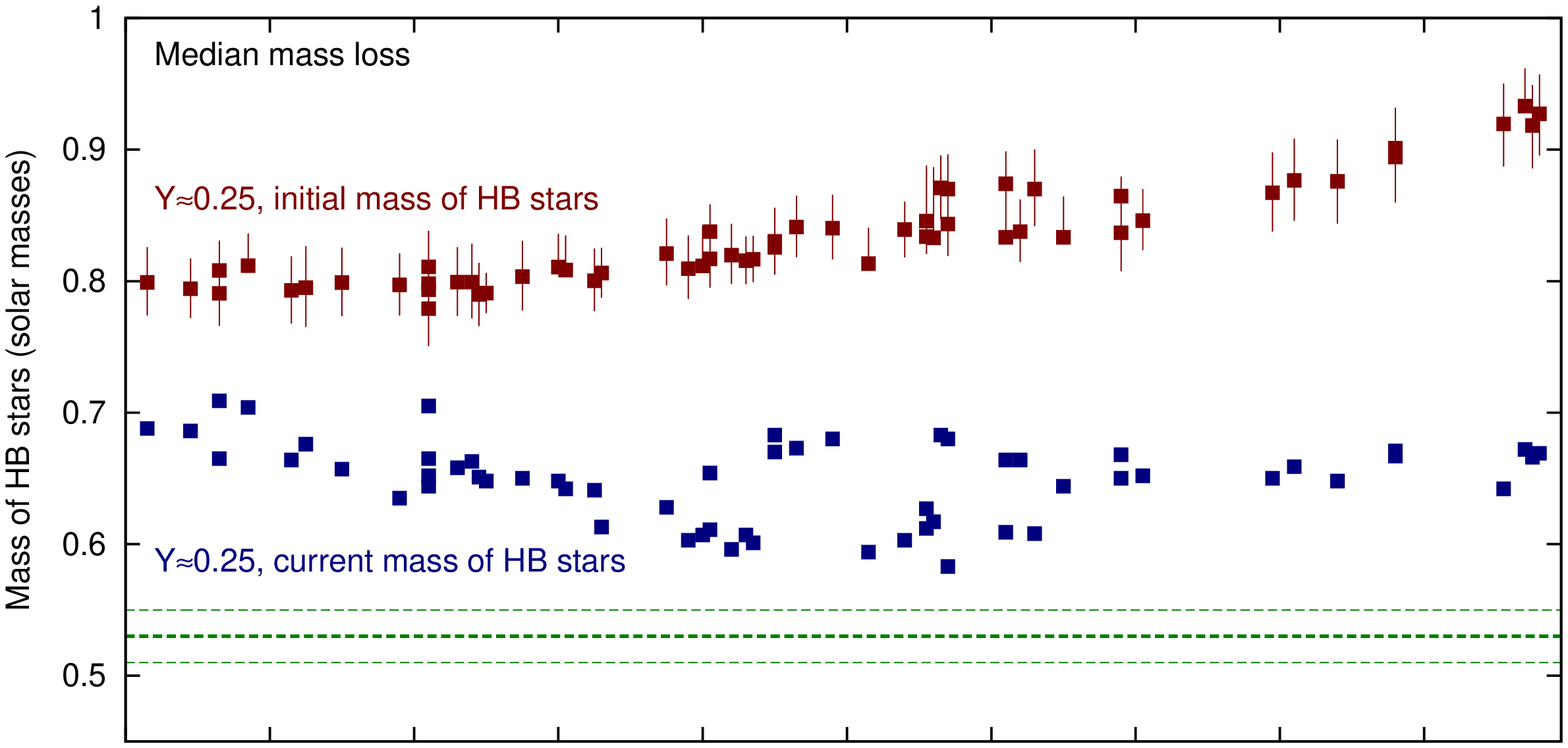}}
\centerline{\includegraphics[height=0.97\textwidth,angle=-90]{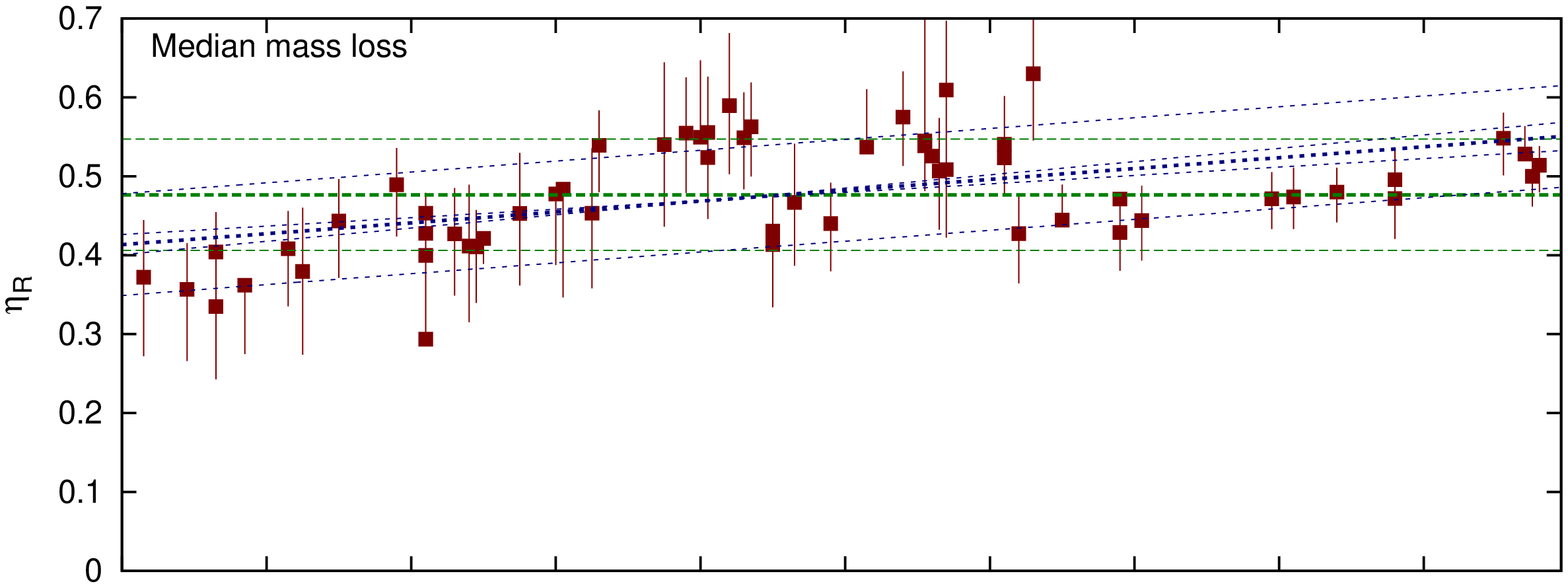}}
\centerline{\includegraphics[height=0.97\textwidth,angle=-90]{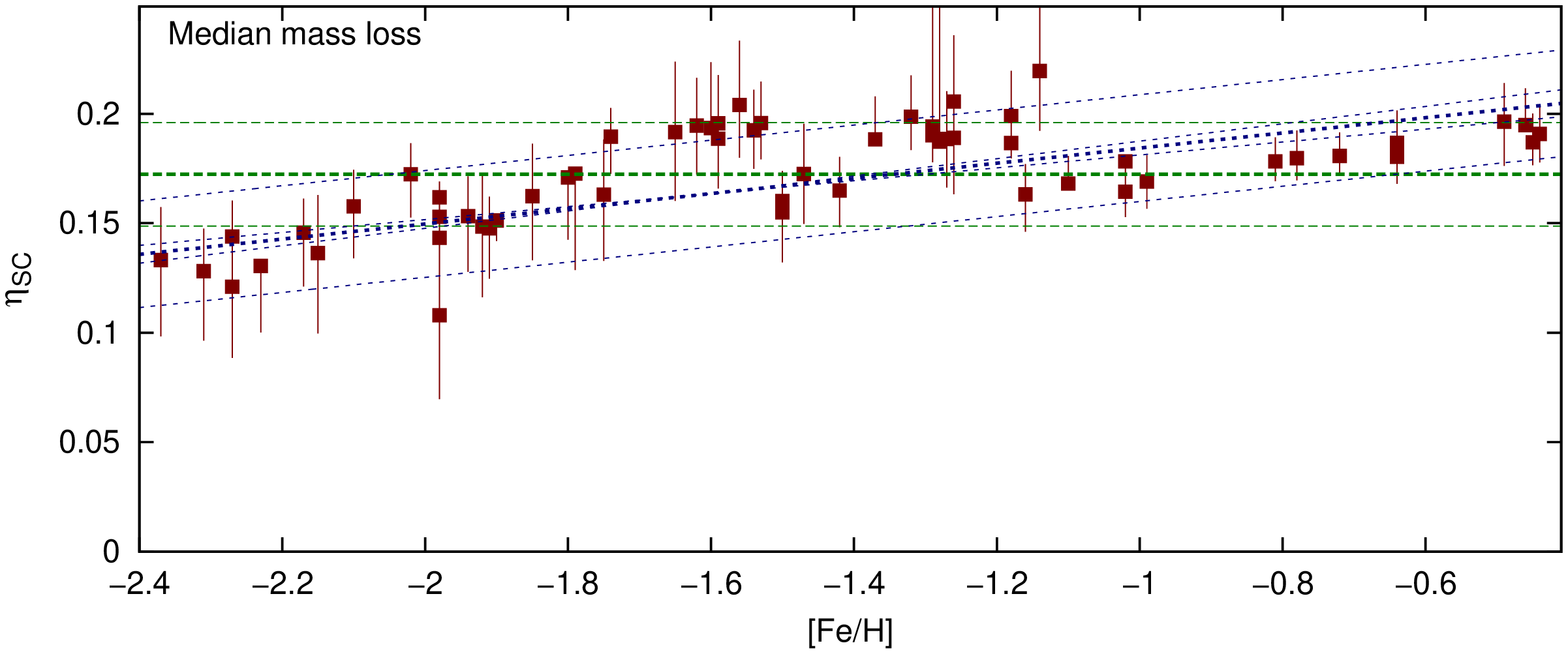}}
\caption{Top panel: initial masses of HB stars in each globular cluster (upper, red points with solid error bars) with the current median masses of those stars, from GCB+10 (lower, blue points, having negligible error bars). Green lines give the canonical white dwarf mass of 0.53 $\pm$ 0.02 M$_\odot$. Central and bottom panel: variations of the associated median mass-loss efficiencies ($\eta_{\rm R}$ and $\eta_{\rm SC}$). Red points show the observed values per cluster and their associated uncertainties \emph{under the assumption that there is no helium enrichment}. Green, dashed lines show the average value (thick line) and standard deviation (thin lines). Blue, dotted lines show a linear fit (thick line) along with the standard deviation from that fit (parallel thin lines) and error in the slope (other thin lines). Errors on clusters with [Fe/H] $>$ --0.64 are not always fully defined.}
\label{EtaFig}
\end{figure*}

Figure \ref{EtaFig} shows our derived values of $\eta$ for the median stars in each cluster.

\subsection{Application to the mimimum and maximum helium abundance}
\label{EtaHeSect}


As noted in Section \ref{HBMassSect}, GCB+10 provide a range of HB star masses for each cluster. However, the applicability of this range here is limited due to the uncertain helium abundance.

GCB+10 note a small (0.03--0.04 M$_\odot$), near-constant offset between the median and maximum masses in almost all cases, which translates to a change in $\eta_{\rm R}$ of 0.1, and a change in $\eta_{\rm SC}$ of 0.03. This would be indicative of a spread in $\eta$ within each cluster of $\sigma(\eta_{\rm R}) = \pm 0.05$ and $\sigma(\eta_{\rm R}) = \pm 0.015$. However, the correction applied by GCB+10 is too uncertain to say whether this represents an intrinsic spread in $\eta$ which might arise from a lower helium enrichment.

Conversely, the maximum $\eta$ can be a useful measure if the helium enrichment in that cluster is low. Anomalous enrichments of $\Delta Y \lesssim 0.01$ are seen by many authors (e.g.\ \citealt{CGSV+09,VPG09,VCAG+13}\footnote{However, note higher values of helium enrichment in M4: \citet{VGPG12}; \citet{DSF+13}.}) but severe enhancements of up to $Y \sim 0.4$ are estimated by other authors in the most massive clusters (e.g.\ NGC 2808, NGC 6441, $\omega$ Cen; \citealt{DABC+05,PVB+05,CDA07}).

Comparison of {\sc mesa} tracks of 0.90 M$_\odot$ stars at [Fe/H] = --0.60 dex show that a helium enrichment of $\Delta Y = 0.005$ will decrease in the timing of the RGB tip by 0.54 Gyr. This would decrease the initial mass of stars which today populate the HB of 0.011 M$_\odot$ (full details are presented in Appendix \ref{TrackHeSect}). This translates into a decrease of $\Delta\eta_{\rm R} \approx 0.022$ at [Fe/H] = --1 dex and $\Delta\eta_{\rm R} \approx 0.033$ at [Fe/H] = --2 dex.

GCB+10 model a large scatter in the difference between the median and minimum masses of HB stars, but the average across all metallicities is $\sim$0.057 M$_\odot$. If the change in HB mass is caused entirely be helium enrichment, this translates to an increase of $\Delta Y \sim 0.026$ between a cluster's median-mass and minimum-mass HB stars. Since GCB+10 model that 68 per cent of clusters have a maximum $Y$ of 0.293 or less, this is not an unreasonable result. Alternatively, if the change in HB mass is caused entirely by RGB mass-loss efficiency, this translates to an increase of $\eta_{\rm R} = 0.11$ at [Fe/H] = --1 dex and $\eta_{\rm R} = 0.16$ at [Fe/H] = --2 dex. This increase would be considerably greater than the spread between clusters.

Without accurate measures of helium abundance in globular clusters, it is not possible to differentiate between these two scenarios, hence we only further consider the $\eta$ that can be applied to each cluster's median-mass HB stars.

\subsection{Final average values of $\eta$}
\label{EtaHeSect}


From the data presented in Figure \ref{EtaFig} and Table \ref{DataTable}, we derive an average value across all clusters of $\eta_{\rm R} = 0.477 \pm 0.070$ and $\eta_{\rm SC} = 0.172 \pm 0.024$, where the quoted uncertainties refer to the standard deviation of best-estimate values for each cluster. However, the error budget for each cluster is dominated by the systematic uncertainties in determining the initial stellar mass and stellar evolution pathways, which derive from the uncertainties in the cluster ages, helium content and choice of isochrones. We adopt a systematic uncertainty in age of $\pm$0.7 Gyr, conservatively based on the standard deviation of ages derived for each cluster. Although we use similar sources, this can be compared to the difference in average age between \citet{RCGS14} and this work of 0.28 Gyr. This imparts a systematic uncertainty of $\Delta \eta_{\rm R} = 0.033$ and $\Delta \eta_{\rm SC} = 0.012$. Our estimated correction for helium enrichment adds a further downward uncertainty of $\Delta \eta_{\rm R} = 0.037$ and $\Delta \eta_{\rm SC} = 0.013$. The typical systematic uncertainty derived from the choice of isochrones is $\Delta \eta_{\rm R} = 0.037$ and $\Delta \eta_{\rm SC} = 0.014$. A final, additional error of $\Delta \eta_{\rm SC} = 0.003$ is applied to account for the fact we do not calculate our RGB isochrones using this mass-loss prescription (Appendix \ref{ErrorEtaSCSect}). Combining the systematic uncertainties in quadrature leads us to our final median values of $\eta$, which are:
\begin{eqnarray}
\eta_{\rm R} &=& 0.477 \pm 0.070 ^{+0.050}_{-0.062} \nonumber \\
\eta_{\rm SC} &=& 0.172 \pm 0.024 ^{+0.018}_{-0.023}, \nonumber
\end{eqnarray}
where we quote the standard deviation among clusters and the global systematic uncertainty, respectively.

There is little noticeable variation of $\eta$ among the Milky Way globular clusters, despite covering a significant range in parameter space, particularly in metallicity. However, we note a significantly larger spread in $\eta$ in the metal-intermediate clusters and a tail off to lower $\eta$ in the metal-poor clusters. We discuss these variations in Section \ref{EtaVarySect}.

\section{Discussion}
\label{DiscSect}

\subsection{Comparison with previous derivations of $\eta$}
\label{EtaDiscSect}

\begin{center}
\begin{table*}
\caption{Comparative values of $\eta$ from SC05. Their errors are implied from their text. Systematic errors have not been applied to our data here.}
\label{SCTable}
\begin{tabular}{ccccc}
    \hline \hline
Cluster & \multicolumn{2}{c}{SC05}	& \multicolumn{2}{c}{This work} \\
\ 	& $\eta_{\rm R}$ & $\eta_{\rm SC}$ & $\eta_{\rm R}$ & $\eta_{\rm SC}$ \\
    \hline
NGC 5904	& 0.6 ($\sim\pm0.1$) & 0.2 ($\sim\pm0.04$) & $0.545 ^{+0.065}_{-0.064}$ & $0.194 ^{+0.022}_{-0.016}$ \\
NGC 5927	& 0.5 ($\sim\pm0.1$) & 0.2 ($\sim\pm0.04$) & $0.548 ^{+0.032}_{-0.047}$ & $0.196 ^{+0.018}_{-0.020}$ \\
    \hline
\end{tabular}
\end{table*}
\end{center}

Our derived values of $\eta_{\rm R} = 0.477 \pm 0.070 ^{+0.050}_{-0.062}$ and $\eta_{\rm SC} = 0.172 \pm 0.024 ^{+0.018}_{-0.023}$ are in good agreement with previous literature. SC05 derive both $\eta_{\rm R}$ and $\eta_{\rm SC}$ for the clusters M5 (NGC 5904) and NGC 5927. Their results are summarised in Table \ref{SCTable}. Both clusters' values for $\eta_{\rm R}$ and $\eta_{\rm SC}$ are consistent within the combined error budgets. Further comparison can be made to field stars: \citet{CS11} model 47 Galactic stars, finding $\eta_{\rm SC} = 0.2125$ (with a presumed uncertainty of 0.0125): also consistent within the combined error budget.

The precise value of both $\eta_{\rm R}$ and $\eta_{\rm SC}$ one derives depends on the treatment of the uppermost parts of the red giant branch, where mass-loss rates reach $\sim$2 $\times 10^{-7}$ M$_{\odot}$ yr$^{-1}$. We have tried to incorporate reasonable errors by comparing several stellar evolution codes in our analysis, however it is not clear which stellar evolution model SC05 use.

Discussions brought up during the refereeing process of this work have highlighted the need for a proper treatment of stellar mass loss within the comparison code. While the uncertainties due to variation of giant branch mass loss within the stellar evolution codes are incorporated into our systematic error, their precise treatment can alter the value of $\eta$ significantly. For comparison, we have repeated our derivation of $\eta$ with the Dartmouth isochrones, which do not include mass loss while the stars are on the RGB. The Dartmouth-based $\eta$ is calculated as the substantially higher $\eta_{\rm R} = 0.550 \pm 0.062$ and $\eta_{\rm SC} = 0.261 \pm 0.024$ due to its hotter RGB tip and altered RGB lifetime (see Appendix \ref{MESACompareSect}).

\subsection{Empirical calibration of mass loss}
\label{RGBAGBSect}

The mass of a star can be calculated from observables using the formula:
\begin{equation}
M = \frac{gL}{4\pi\sigma T_{\rm eff}^4 G} .
\end{equation}
For measurements of individual stars, fractional uncertainties in both $L$ and $T^4$ can be as little as $\sim$3 per cent. However, purely spectroscopic derivations of $g$ are normally uncertain by a factor of three or more. This make it difficult to achieve sufficiently accurate masses for individual stars, particularly absolutely calibrated masses. However, computing the difference in mass between two populations of stars from the same observation is possible. This method has the advantage of being largely insensitive to the helium abundance of the star.

In collaboration with C.~I.~Johnson, we have previously adopted this approach to measure the difference between early AGB stars and RGB stars of similar luminosity in $\omega$ Centauri, measuring a 26 $\pm$ 4 per cent decrease in mass from the RGB to AGB \citep{MJZ11}. Assuming a 12-Gyr-old cluster at [Fe/H] = --1.62, this equates to 0.21 $\pm$ 0.03 M$_\odot$, which compares favourably with the 0.23 M$_\odot$ expected using our fit to Reimers' law with $\eta_{\rm R}$ = 0.477.

Similarly favourable results can be found in 47 Tuc, by combining the photometric temperatures and luminosities of \citet{MBvL+11} with the spectroscopic log($g$) measurements of RGB stars by \citet{CPJ+14} and of AGB stars by Johnson et al. (submitted). Although the errors from these more-direct measurements are still considerable, they constrain the possible range of $\eta_{\rm R}$ to between $\sim$0.30 to $\sim$0.60, re-enforcing the values we derive here.

\subsection{Implied evolution along the AGB}
\label{AGBSect}

The $\eta$ we have calculated above should also apply in later evolution, as the stars are thought to continue to lose mass via the same mechanisms well into their AGB evolution. In this section, we follow our stellar evolution models onto the AGB and compare their predictions to observations of AGB stars in globular clusters.

\subsubsection{Applying Reimers' mass-loss law to the AGB}
\label{AGBIsoSect}

\begin{figure}
\centerline{\includegraphics[height=0.47\textwidth,angle=-90]{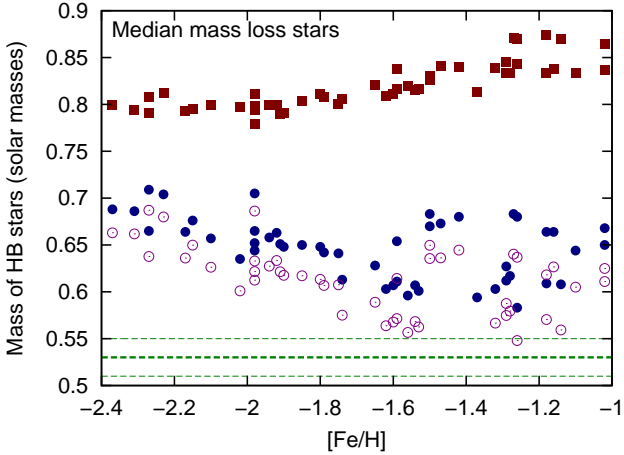}}
\caption{As Figure \ref{EtaFig}. Here, the addition of hollow magenta circles show the median masses predicted for stars reaching 1000 L$_\odot$ on the AGB, where dust production is expected to begin. 
Errors have been omitted for clarity, but are typically 0.025 M$_\odot$ on the initial masses, and 0.01 M$_\odot$ on the HB and at 1000 L$_\odot$.}
\label{MassEvolFig}
\end{figure}

\begin{figure*}
\centerline{\includegraphics[height=0.47\textwidth,angle=-90]{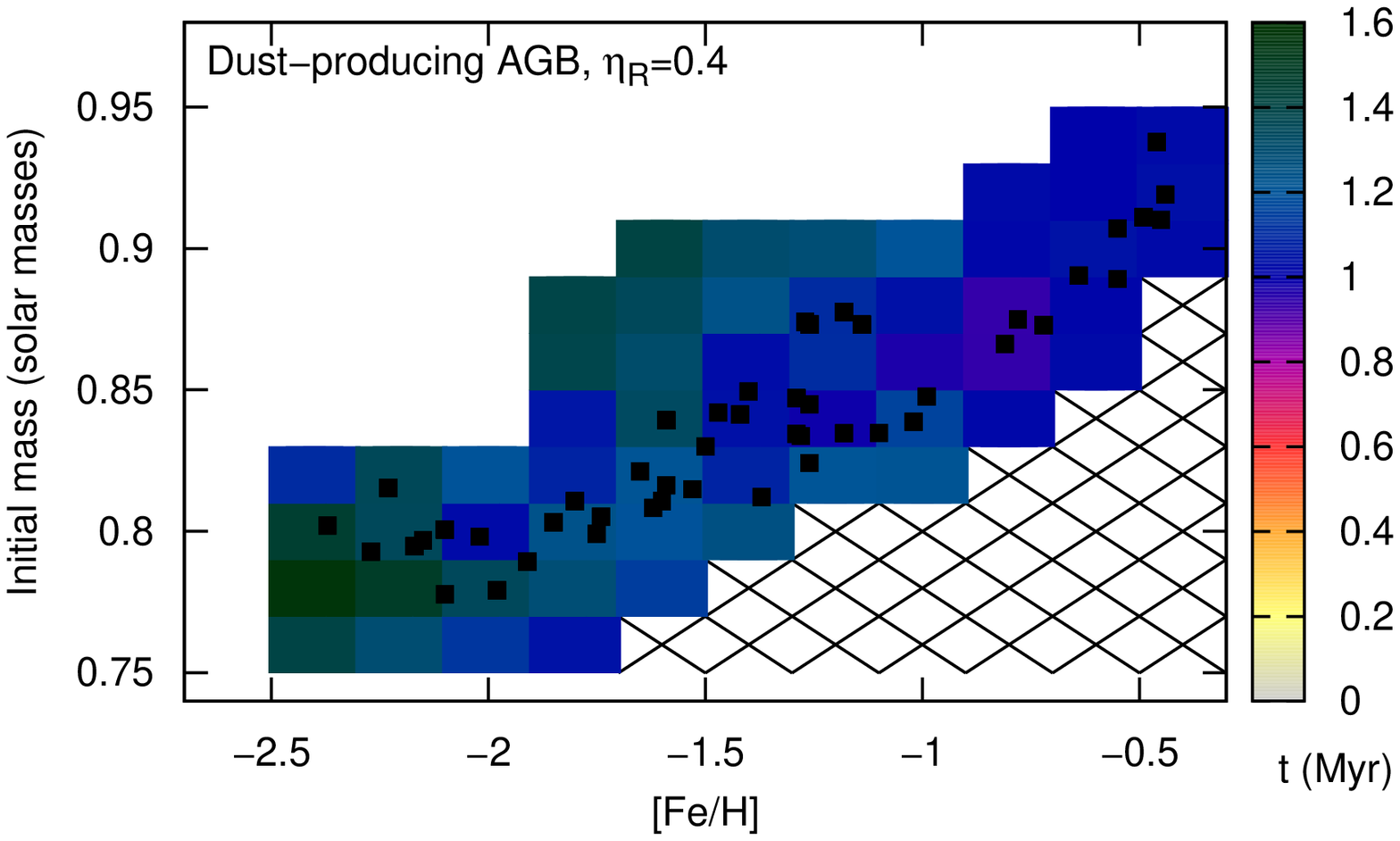} 
            \includegraphics[height=0.47\textwidth,angle=-90]{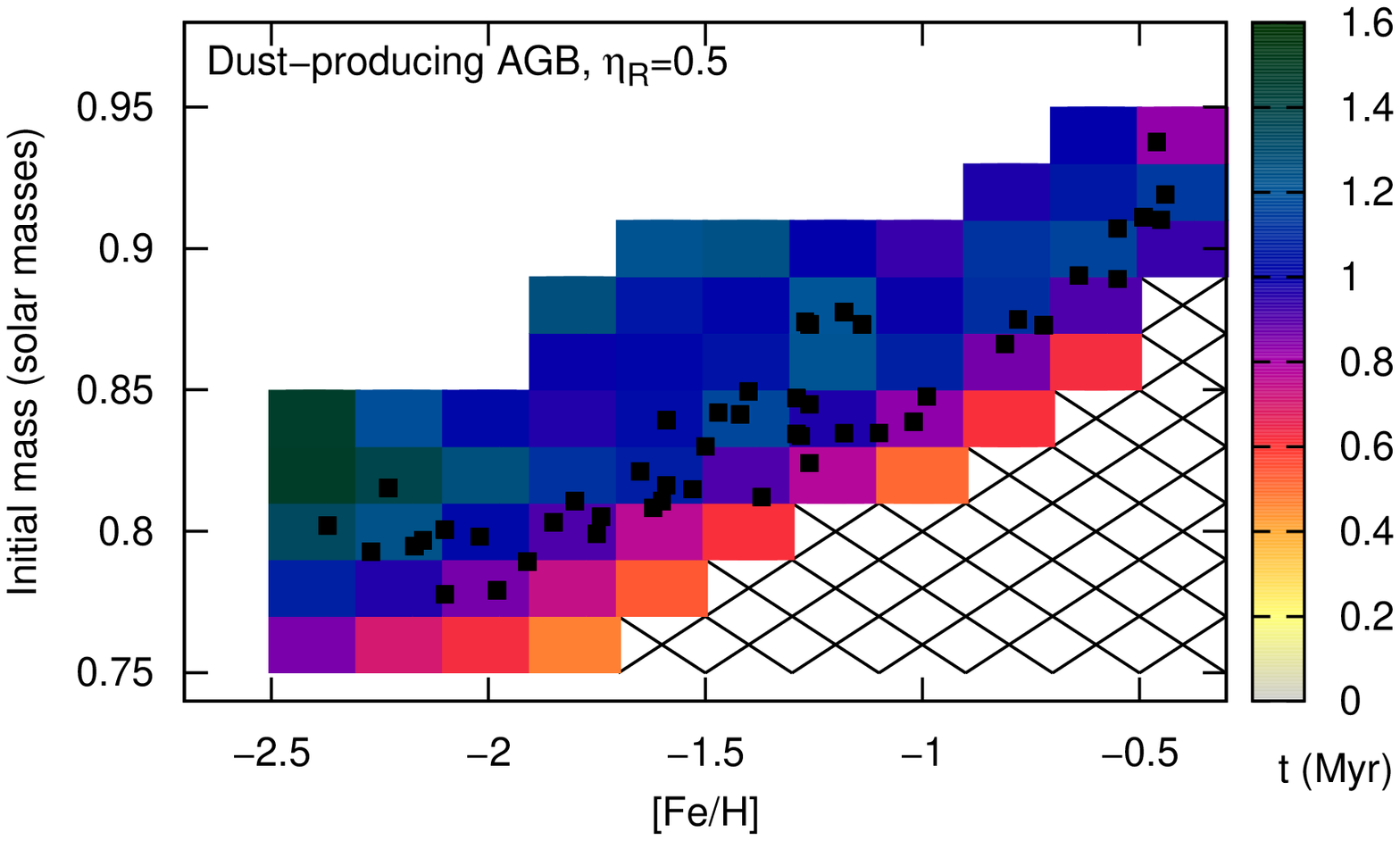}}
\centerline{\includegraphics[height=0.47\textwidth,angle=-90]{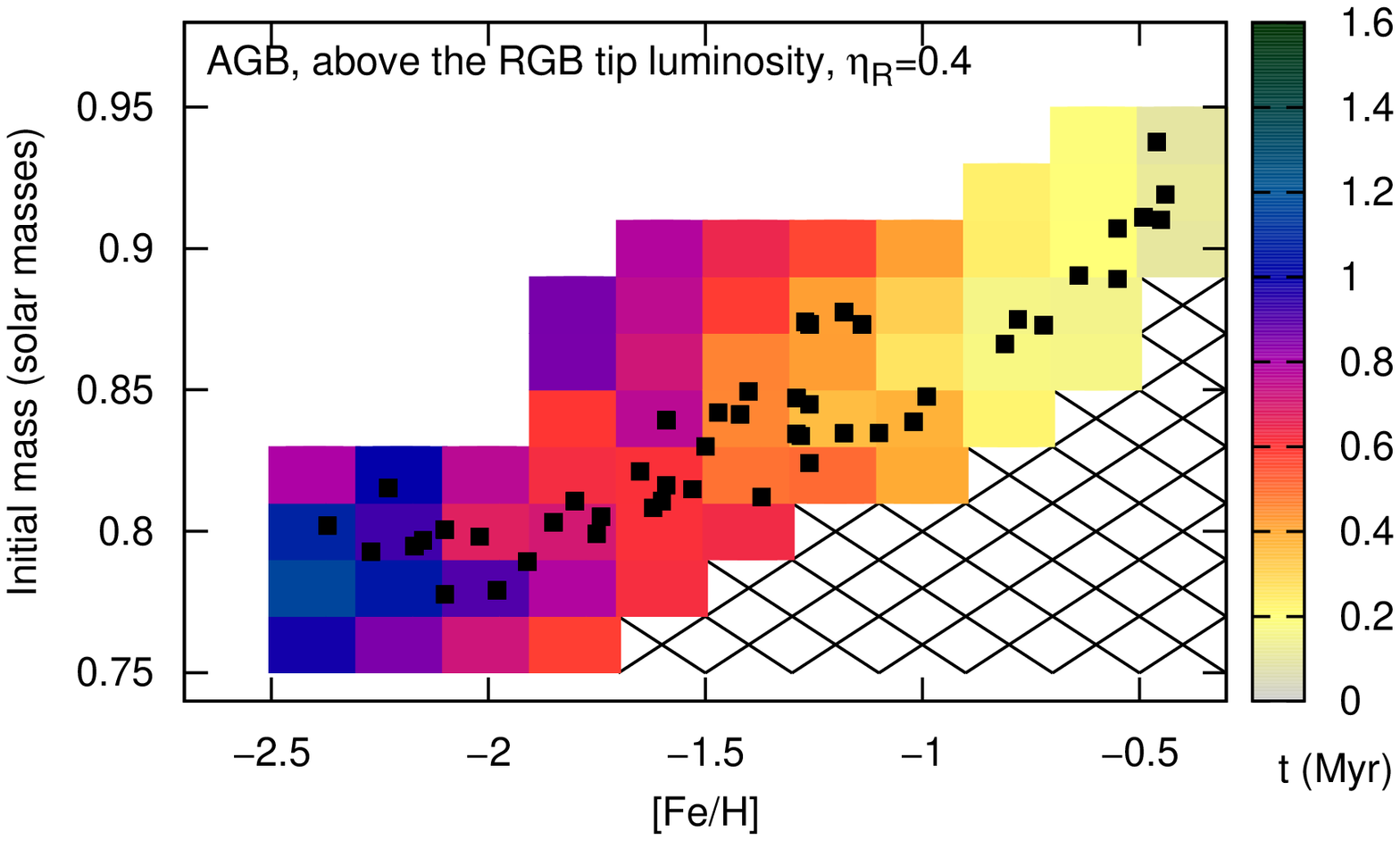} 
            \includegraphics[height=0.47\textwidth,angle=-90]{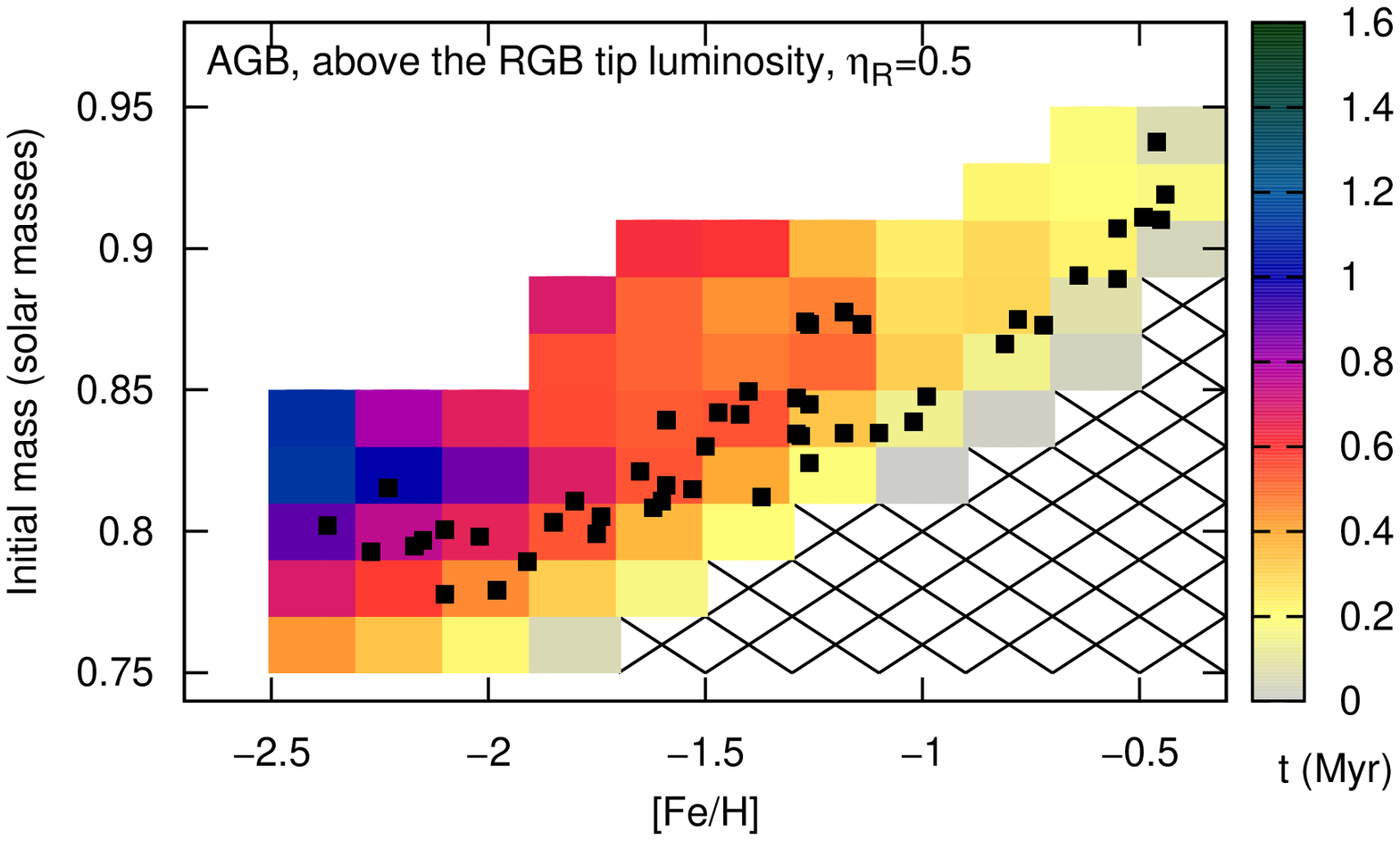}}
\caption{The predicted lifetime of stars (colour scale) on the dust-producing AGB ($L > 1000$ L$_\odot$) and as AGB stars visible above the RGB tip, for the two modelled values of $\eta$. Globular cluster initial masses, derived here, are shown as small black points. Grey colours indicate the star does not evolve as far as the indicated phase, clusters in yellow boxes may have some observable dusty/luminous AGB stars, clusters in darker (purple, blue or green boxes) should have observable dusty/luminous stars. Models were not computed in the hatched area, as evolution to the RGB tip takes longer than a Hubble time.}
\label{AGBFig}
\end{figure*}

Highly evolved stars are thought to lose mass primarily through pulsation-enhanced, dust-driven winds. Pulsations in the stellar atmosphere levitate material which can condense into dust; radiation pressure on this dust forces it from the star. However, the criteria for this to be the primary ejection mechanism are poorly determined, partly depending on the strength of the pulsation and the opacity of the dust (e.g.\ \citealt{Woitke06b,BHN+13}). In the early stages of this regime, where pulsation and dust production are visible but yet to become effective, we may still expect a Reimers-like law to be effective. However, if pulsation and dust driving enhance mass loss, the AGB should be truncated at a lower luminosity.

This process should be metallicity dependent. Metal-poor stars are hotter, typically exhibiting weaker, shorter-period, less effective pulsations. They also suffer from very low dust-to-gas ratios, meaning dust driving is less effective at accelerating all the ejected material. We can therefore expect that the AGBs of metal-rich clusters would be more curtailed than metal-poor clusters.

Dust production in globular clusters typically begins somewhere between 700 L$_\odot$ and 1500 L$_\odot$ (\citealt{MvLD+09,BMvL+09,MBvL+11,MBvLZ11,MvLS+11}; see also \citealt{MSS+12}). It is becoming clear that the average star starts to develop a dusty wind close to 1000 L$_\odot$. Our mass-loss efficiency can be used to predict the mass of a star reaching this 1000 L$_\odot$ point. By comparing this to the final white dwarf mass, we can determine how much more mass it should lose in a dusty wind.

However, accurate white dwarf masses for globular clusters are difficult to find. \citet{RFI+97} measured the mass of white dwarfs in M4 (NGC 6121) to be 0.51 $\pm$ 0.03 M$_\odot$, a mass later refined by \citet{KSDR+09} to 0.53 $\pm$ 0.01 M$_\odot$. \citet{MKZ+04} measured masses for white dwarfs in NGC 6752 at 0.53 $\pm$ 0.02 M$_\odot$. The initial--final mass function for stars at these masses is expected to be relatively flat \citep{KSDR+09,GZHS14}.

\subsubsection{Results from stellar evolution modelling}
\label{AGBModSect}

Finding the stellar mass at 1000 L$_\odot$ becomes a trivial matter of following each stellar evolution model until it reaches 1000 L$_\odot$, interpolating across both the $\eta_{\rm R} = 0.4$ and 0.5 tracks to find a mass appropriate to any particular value of $\eta$. To find the mass lost in a dusty wind, we also calculate the envelope mass of an AGB star at the luminosity of the RGB tip. These are presented in Figure \ref{MassEvolFig}.

We also estimate the final AGB luminosity, avoiding very short extensions to high luminosity during AGB thermal pulses. However, the lack of convergence during the final or second-to-final thermal pulse means that we can only give these approximately. Envelope masses at the final model step are typically $\sim$0.01 M$_\odot$, which will be lost in approximately $\sim$30\,000 years. Our computation of the time spent on the dust-producing AGB (above 1000 L$_\odot$; Figure \ref{AGBFig}) and the time spent as an AGB star with a luminosity greater than that attained on the RGB, will therefore be underestimated by $\sim$30\,000 years. Stars exceeding the RGB-tip luminosity are easily identifiable as AGB stars, as they stand brighter than the RGB tip in infrared colour--magnitude diagrams.

Clusters with metallicities above [Fe/H] = --1 dex are not shown in Figure \ref{MassEvolFig} as interpolation becomes difficult at higher metallicities. We stress these results are for the stars which have had median ZAHB mass: more- or less-massive stars will experience more or less dusty mass loss. Results from the mass-loss law of SC05 are not shown, as they have no clear terminus.

\subsubsection{Comparison to observed AGB tip luminosities}
\label{AGBTipSect}

Figures \ref{MassEvolFig} and \ref{AGBFig} imply that the median stars in clusters have between 0.01 and 0.15 M$_\odot$ to lose as dusty winds, with a typical value being closer to 0.10 M$_\odot$. Metal-poor clusters would produce proportionally more mass in dusty winds, although the presumably lower dust-to-gas ratios of those winds would mean less dust overall. Consequently, stars in metal-poor clusters should reach a higher luminosity on the AGB, producing dust for around $\sim$1.5 Myr and remaining above the RGB tip for $\sim$1 Myr. Conversely stars in metal-rich clusters should produce dust for $\lesssim$1 Myr. In many metal-rich clusters, there should not be a star above the AGB tip for much of the time.

Observationally, however, M15 (the most-metal-poor cluster) has an AGB which terminates at a luminosity below the RGB tip \citep{BWvL+06,MvLDB10}. It is likely that a lack of bright AGB stars is typical of most or all metal-poor clusters. AGB stars in clusters appear to just reach the RGB tip in $\omega$ Cen ([Fe/H] $\approx$ --1.6 dex; \citealt{MvLD+09}) and noticably exceed the RGB tip in luminosity by NGC 362 ([Fe/H] = --1.16 dex; \citealt{BMvL+09}). In 47 Tuc ([Fe/H] = --0.72 dex), stars spend $\sim$250\,000 years above the RGB tip, reaching nearly 5000 L$_\odot$ \citep{MBvL+11,LNH+14,MZ14}.

We therefore have a discrepancy whereby the prediction using Reimers' law is that metal-rich clusters have no bright, dusty AGB stars and metal-poor clusters have bright, dusty AGB stars, whereas the opposite is observed. The problem is compounded when one observes that metal-poor clusters may have a slightly lower $\eta_{\rm R}$ than metal-rich clusters, leading to longer-lived stars and a higher-luminosity AGB terminus (see Figure \ref{EtaFig}). A departure from a Reimers'-like law to a pulsation-enhanced, dust-driven wind would exacerbate this, therefore we do not think it likely that pulsation and dust production greatly affect the mass-loss rates of even the brightest globular cluster stars.

Adopting the SC05 law partially alleviates this discrepancy. The lack of a full treatment of the SC05 law means that the results are marginally less secure, however we can approximately identify the AGB terminus under the SC05 law by identifying when a star undergoing mass loss following the SC05 law will have lost its entire envelope. For metal-poor clusters ([Fe/H] $<$ --2 dex) the AGB terminates around 900\,000 years earlier for the SC05 law; for metal-rich clusters ([Fe/H] $>$ --1 dex) it is about 400\,000 years earlier. This is not enough to reverse the trend in AGB luminosity seen in Figure \ref{AGBFig}.


\subsection{The underlying cause of variations in $\eta$ among clusters}
\label{EtaVarySect}

The variance of $\eta$ among clusters is surprisingly small: the standard deviation among clusters is only 14 per cent. This is despite their variation by over 20 per cent in initial mass and two orders of magnitude in metallicity, with RGB-tip stars that are up to 1300 K warmer and a third fainter in metal-poor clusters than metal-rich clusters. That this works so well is a testament to the accuracy of both mass-loss formulisms studied here.

SC05 claim their formula should provide better accuracy than Reimers' due to the incorporation of physical princples, rather than a simple empirical fit. We note that both values for $\eta$ come with a 14 per cent standard deviation in this work, thus one would not appear any more accurate than the other. However, the median internal (cluster-to-cluster) error we derive for $\eta$ is also 14 per cent. This internal error is driven largely by the uncertainty in cluster age for metal-poor clusters, and a combination of age and metallicity for metal-rich clusters. It is not clear from our data that the SC05 law necessarily provides a much better fit than Reimers'.

Despite the relatively closeness in $\eta$ among clusters, there are still some trends that can be seen in Figure \ref{EtaFig}. Firstly, there is a much greater variation in $\eta$ for the metal-intermediate clusters (--1.7 $<$ [Fe/H] $<$ --1.1 dex). Outside of this range, there appear noticable gradients in the results with $\eta$ declining at the lowest metallicities. These are reflected in the predicted lifetimes of AGB stars in the dusty and super-RGB-tip phases described in Section \ref{AGBSect}, where metal-poor clusters have AGB tips higher than typically observed and metal-intermediate clusters have a range of expected AGB termini.

\subsubsection{The gradient in $\eta$}

To reconcile our Reimers' $\eta$ calculations with observed AGB populations for the metal-poor clusters, one or more adaptations could be made:
\begin{itemize}
\item A mass-loss law with a stronger temperature dependence could be used, possibly invoking a mechanism which more strongly affects hotter HB or early-AGB stars (e.g.\ \citealt{DSS09}). The temperature dependence would have to be stronger than that of SC05.
\item Slower evolution in metal-poor AGB stars in the $\sim$500--2000 L$_\odot$ range than predicted, allowing the wind to be lost over a greater amount of time.
\item The masses of the horizontal branch stars in metal-poor clusters could have been over-estimated by GCB+10. Figure \ref{EtaFig} shows a noticable upturn in the HB mass near [Fe/H] = --1.8 dex. If these stars can be made less massive, not only do the $\eta$ of these stars become similar to that of the other clusters: the AGB tip will also decrease in luminosity to the observed value.
\item The ages of the oldest globular clusters could be made younger. This would increase the initial mass of stars, increasing $\eta$ to values more consistent with metal-rich clusters and allowing more AGB mass loss to take place, truncating the AGB at the observed values.
\end{itemize}
The first two possibilities are relatively unlikely. HB mass loss would have to increase above predictions by a factor of $\sim$4 to prevent super-RGB-tip stars to become visible. Slower evolution would change the luminosity function around the RGB tip in younger environments (e.g.\ dwarf galaxies), but there is no noted drop in the number stars per magnitude over the AGB tip in these environments.

GCB+10 quote errors of only a few $\times$0.001 M$_\odot$ on their HB star masses. In practice, however, this neglects contributions from the uncertain metallicity of the clusters, and the stellar evolution codes they use to determine HB mass. The difference in metallicity quoted by GCB+10 and \citet{Harris10} for metal-poor clusters is $\sim$0.03 dex. A 0.1-dex uncertainty in an individual cluster's metallicity changes the HB mass by $\sim$0.01 M$_\odot$, and ($\eta_{\rm R}$ $\vert$ $\eta_{\rm SC}$) by $\sim$(0.03 $\vert$ 0.01). While this would remove a significant part of the gradient in the $\eta$ versus [Fe/H] relation, without a systematic change in many clusters, global [Fe/H] is not likely to be the underlying cause of the differences. Similarly, the contribution by helium is not expected to be significant (Appendix \ref{EtaHeSect}).

The absolute choice of metallicity ($Z$), however, may be an issue. GCB+10 use the metal-poor Pisa evolutionary models \citep{CDIC04}, which adopt [$Z$/$X$]$_\odot$ = 0.023, compared to the modern value of $Z$ = 0.0152 $\pm$ $\sim$0.0006 (see Appendix \ref{TrackHeScaleSect}). This amounts to a $\approx$0.083 dex difference in the zero (solar) point for metallicity, depending slightly on the helium abundance adopted. It is not immediately obvious how correction was made by GCB+10 from the solar-scaled Pisa models to the $\alpha$-enhanced globular clusters. \citet{CDIC04} provide a method for doing so, but it involves the adoption of a particular [$\alpha$/Fe], which is not quoted by GCB+10. Given the likely uncertainty in the average [$\alpha$/Fe] adopted ($\sim$0.1 dex), one can presume a further $\sim$0.06 dex uncertainty in $Z$, based on \citet[][their section 2]{CDIC04}. The combination of these uncertainties is sufficient to alter the derived horizontal branch masses significantly (we estimate by up to $\sim$0.02 M$_\odot$ for the metal-rich clusters). However, this alters the metal-rich end more than the metal-poor end, and increases rather than decreases the gradient with metallicity for both $\eta_{\rm R}$ and $\eta_{\rm SC}$ versus [Fe/H].

The choice of evolution model is also significant. The Pisa HB models for $Z = 0.0002$, $M = 0.65$ M$_\odot$ return a ZAHB temperature of $\sim$11\,700 K, whereas a corresponding {\sc mesa} model at [Fe/H] = --2.20 dex, $M_{\rm ZAHB} = 0.6455$ M$_\odot$ produces 10\,100 K. This corresponds to a difference of $(B-V) \approx 0.04$ and a difference in ZAHB mass as recorded by GCB+10 of $\sim$0.02 M$_\odot$. This difference should be in the correct sense, in that Pisa evolutionary tracks would record higher masses than the {\sc mesa} models. A consistent $\sim$0.02 M$_\odot$ difference across all clusters would serve to increase the average $\eta_{\rm R}$ by $\approx$0.05 (approximately the systematic uncertainty we have allotted for differences among evolutionary tracks) but would also serve to significantly flatten both distributions of $\eta$ versus [Fe/H]. It would remove the unobserved bright AGB stars in metal-poor clusters. However, it would also remove more of the bright AGB stars in metal-rich clusters, which are observed.

Reducing the age of the metal-poor globular clusters to bring their ages in line with MF+09 and D+10 flattens the age distribution of globular clusters, making them much more closely co-eval. The addition of $\sim$0.02 M$_\odot$ to the initial mass of metal-poor clusters and its subtraction from the metal-rich clusters flattens the $\eta$ distributions. The subsequent change in $\eta$ by $\sim\pm$0.05 at either end is sufficient to remove the bright AGB stars from all but the most-metal-poor clusters, while making the bright AGB stars from the metal-rich clusters more visible. A constant age does not entirely solve the AGB problem but it does alleviate it, plus removes the unexplained gradient in both $\eta_{\rm R}$ and $\eta_{\rm SC}$.

We suggest that a combination of these uncertainties could be sufficient to remove the gradient seen in $\eta$, therefore we cannot claim any metallicity dependence exists in $\eta$. An absence of a metallicity dependence would be in keeping with the magneto-acoustic driving mechanism proposed for these winds.

\subsubsection{The spread in $\eta$ for intermediate-metallicity clusters}

The noticeable spread in $\eta$ for intermediate-metallicity clusters (--1.8 $<$ [Fe/H] $<$ --1.0 dex) traces to the HB masses derived by GCB+10. The substantial spread in these, nearly 0.1 M$_\odot$, is not seen in the initial masses we derive (Figure \ref{EtaFig}). It also shows up as a significant spread in the RGB mass loss calculated by GCB+10 (their figure 11).

Generally speaking, the older clusters seem to have the lower HB masses and higher $\eta$, but the underlying causes in this spread are unclear. Well known HB pairs (M3 and M13, NGC 288 and 362) lie on either side of the divide in HB mass at 0.635 M$_\odot$. Clusters on the lower-HB-mass, higher-$\eta$ side average 23 per cent higher in mass but 39 per cent smaller in radius than clusters on the other side \citep{GZP+02}. Higher mass and central concentration are notable characteristics of clusters with multiple populations as they are more easily able to retain their ISM \citep{MZ14}. They also have significantly larger spreads in HB mass (GCB+10), providing evidence of their multiple populations. The median stars in these clusters could be significantly helium-enhanced, in which case our initial mass for the HB stars in these clusters would be over-estimated. If true, this would reduce the overall values of $\eta$ we derive, but still within the limits of the systematic errors we apply due to the uncertainty in helium abundance (Appendix \ref{EtaHeSect}).

\section{Conclusions}
\label{ConcSect}

In this paper, we have investigated the efficiency of mass loss experience by globular cluster stars and how that mass loss is distributed throughout the stars' evolution. Based on the ZAHB masses from GCB+10, we have assumed that the median star experiences negligible helium enrichment. We have combined these isochrones with evolutionary models to determine the following principle conclusions:
\begin{itemize}
\item Metal-rich globular cluster stars lose more mass on the RGB than metal-poor stars, simply due to their slower evolution.
\item For most giant branch stars in globular clusters, mass loss on the RGB dominates the total mass lost.
\item The mass-loss models of Reimers and Schr\"oder \& Cuntz show a relatively constant efficiency in converting (presumably) magneto-acoustic energy to stellar outflow, independent of metallicity. The modelled efficiencies are $\eta_{\rm R} = 0.477 \pm 0.070 ^{+0.050}_{-0.062}$ and $\eta_{\rm SC} = 0.172 \pm 0.024 ^{+0.018}_{-0.023}$. The quoted uncertainties denote the random spread among clusters and the overall systematic uncertainty (including a factor for helium enrichment), respectively. These values are towards the higher end of previous estimates in globular clusters, and may be somewhat higher than generally found in the field.
\item While a weak gradient with metallicity is possible, whereby metal-poor stars experience less-efficient mass loss, any gradient is likely to be within the systematic uncertainties of the underlying data.
\item Stars in globular clusters should generally lose $\sim$0.10 M$_\odot$ in their final AGB dusty wind. Some clusters may no longer regularly produce any dusty stars. A more precise value is difficult to determine given the remaining uncertainties in the data. Better empirical calibration of the AGB tip in globular clusters is recommended.
\end{itemize}

\section*{Acknowledgments}

We are grateful to Raphael Herschi and particularly Aaron Dotter for their help with using the {\sc mesa} code and checking its output. We are also grateful to Christian Johnson and the anonymous referee for their very helpful comments, which have greatly improved the quality and accuracy of this manuscript.


\appendix

\section{Information on literature studies of cluster ages}
\label{AgesSect}

SW02 built on methods employed in their earlier works \citep{SW97,SW98} to provide absolute ages for 50 clusters. Four metallicity bins were chosen, with breakpoints on the CG97 scale at [Fe/H] = --1.75, --1.3 and --0.9 dex. In each bin, a reference cluster was chosen (M15, M3, NGC 6171 and 47 Tuc) and its absolute age determined from the $V$-band magnitude difference between the red side of its HB and the main-sequence turn-off (MSTO) (taken from \citealt{RSPA99}). This is referred to in later literature as the ``vertical'' method. For each bin, relative ages were calculated corresponding to the difference in ($V-I$) or ($B-V$) colour between the MSTO and the base of the RGB. This is referred to as the ``horizontal'' method. Bespoke isochrones are used \citep{SW98}, assuming a helium fraction of $Y = 0.23 + 3Z$ (for metal fraction $Z$) and an alpha-enhancement of [$\alpha$/Fe] = +0.4 dex. This provides ages in good agreement with the earlier study of \citet{RSPA99}, which we do not include due to its relatively small number of clusters (35) and similar set of data to SW02. Ages are provided for both metallicity scales.

DA+05 calculated ages from $F439W$ and $F555W$ \emph{Hubble Space Telescope} (\emph{HST}) and ground-based $VI$ photometry of 55 clusters \citep{RPSA00,RASP00,PKD+02}. They follow the same approach as SW02, using five metallicity bins, broken at [Fe/H] = --1.8, --1.5, --1.3 and --1.1 dex, with templates NGC 4590, 5262, 5904, 1851 and 6362. Minor diferences exist in measuring the ZAHB magnitude for metal-rich clusters ([Fe/H] $>$ --1 dex, relying on the lower envelope of the HB) and metal-poor clusters (relying on matching to a template at that metallicity). They use the isochrones of \citet{PCSC04} and relative ages are provided for both metallicity scales.

MF+09 published ages for 64 globular clusters, on the basis of $F606W$ and $F814W$ $HST$ photometry \citep{SBC+07,ASB+08}. This is the same dataset and methodology which provides the space-based data for our horizontal branch masses, thus we invoke it as a primary age calibrator too. The depth of the $HST$ observations allow the authors to fit loci to the clusters' main sequences up to the luminosity of the HB. By overlaying them in colour--magnitude space, using the MS and RGB base as reference points, an intrinsic difference in MSTO magnitude can be found. This is analagous to the ``vertical'' method described above. Results are compared to the \citet{DCJ+07} isochrones for both metallicity scales, and to the evolutionary models of \citet{PCSC04}; \citet{BBC+94} and \citet{GBBC00} in the CG97 scale.

D+10 provide absolute ages for 61 globular clusters using the same \emph{HST} Advanced Camera for Surveys (ACS) data as MF+09. D+10 takes the ages of 55 of its clusters from fits to the \citet{DCJ+07} isochrones, which uses the ZW84 metallicity scale, but includes an extra six outer-halo globulars with ages calculated using the updated \citet{DCJ+08} models (AM-1, Eridanus, NGC 2419, and Palomar 3, 4 and 14). Absolute ages were calibrated by scaling to a nominal maximum age of 13.3 Gyr.

VBLC13 also use the \emph{HST} ACS data to derive ages for the original 55 globular clusters. Distances are set from the ZAHB magnitude, while ages are derived from isochrone fitting with appropriate metal mixtures \citep{VBD+12}. Particular care is given to the slopes of the sub-giant branch (SGB) and RGB in terms of the effects of chemistry, etc., though this was found to have little effect on the age derived as the MSTO is largely unaffected. Their approach, however, can be reduced to simply adopting versions of the ``vertical'' or ``horizontal'' (or both) methods described above.

\section{A comparison of stellar evolution models}
\label{ModelsSect}


\subsection{Existing stellar evolution models}

The choice of underlying stellar evolution model and its applicability to the globular clusters in question is a large source of systematic error in our work. In this section, we explore the differences between our {\sc mesa} models and five published different sets of isochrones. We will discuss the uncertainties the choice of stellar evolution models provides in estimating the initial mass of ZAHB stars in the clusters. These are namely:

{\it The Dartmouth stellar evolution database} \citep{DCJ+08}\footnote{http://stellar.dartmouth.edu/models/isolf\_new.html}. This is the original database generated for and calibrated against the \emph{HST} ACS data mentioned above (MF+09, D+10). Along with the usual parameters of age and metallicity, it allows the user to vary the helium abundance between a primordial value ($Y = 0.245 + 1.5Z$) and two fixed values ($Y = 0.33$, 0.40) and the $\alpha$-element abundance between [$\alpha$/Fe] = --0.2 to +0.8 dex in steps of 0.2 dex.

{\it The Padova database of stellar evolutionary tracks and isochrones}\footnote{http://stev.oapd.inaf.it/cgi-bin/cmd}. From this database, we use version 1.1 of the {\sc parsec} isochrones of \citet{BMG+12}. These isochrones assume a solar-scaled composition, with $Y = 0.2485 + 1.78Z$, based on $Z_\odot = 0.0152$. While different options for interstellar and circumstellar dust extinction and carbon stars exist, we do not need nor implement them here. No variation of [$\alpha$/Fe] is possible, but the {\sc parsec} isochrones do allow the user to change the RGB mass-loss rate via Reimers' $\eta$ (see Section \ref{EtaModelSect}). An alternative set of models is also available using the CMD\footnote{http://stev.oapd.inaf.it/cgi-bin/cmd} and YZVAR\footnote{http://stev.oapd.inaf.it/YZVAR/} interfaces, from \citet{MGB+08} and \citet{BGMN08}, respectively. A metallicity range from $Z = 0.0001$ to 0.07 is covered at $Y = 0.23$ to 0.46, with an interpolation tool provided on the website at a user-specified value of Reimers' $\eta$.

{\it The Pisa evolutionary library} \citep{CDIM+03,CDIC04}\footnote{http://astro.df.unipi.it/SAA/PEL/Z0.html}. This is an older set of models a grid of fixed metallicities and ages, with no provided interpolator. $Y = 0.23$ is assumed for $Z < 0.001$; $Y = 0.232$ for $Z = 0.001$; $Y = 0.237$ and 0.27 for $Z = 0.004$ and $Y = 0.25$ for $Z = 0.008$. Some variation in mixing length and convective overshooting is applied to the higher-metallicity models.

{\it The Victoria Regina models} \citep{VBD06}\footnote{http://www3.cadc-ccda.hia-iha.nrc-cnrc.gc.ca/community/{\linebreak}VictoriaReginaModels/}. A variety of sub-solar evolutionary tracks calculated at [$\alpha$/Fe] = 0.0, +0.3 and +0.6 dex, plus a set of solar-scaled models covering higher metallicities. An interpolation tool is provided, which has the ability to produce isochrones.

{\it A bag of stellar tracks and isochrones} (BaSTI, ver 5.0.1; \citealt{PCSC04}\footnote{http://albione.oa-teramo.inaf.it/}). Used by DA+05 and MF+09. A set of stellar isochrones from [Fe/H] = --2.27 to +0.40 dex, scaled with $Y = 0.245$ for $Z = 0$ and increasing to their adopted solar value of $Y = 0.273$ for $Z = 0.0198$. Both canonical and non-canonical mixing is included for both solar-scaled and $\alpha$-enhanced ([$\alpha$/Fe] = +0.35 dex) abundances. An interpolation tool is provided for these models. Additional models are available for increased helium abundances ($Y = 0.30$, 0.35 and 0.40) and for extreme C, N, O and Na abundances (at solar helium abundance with limited models at $Y = 0.28$, 0.35 and 0.40).

\subsection{Comparing of our {\sc mesa} tracks with existing models}
\label{MESACompareSect}

\begin{figure*}
\centerline{\includegraphics[height=0.95\textwidth,angle=-90]{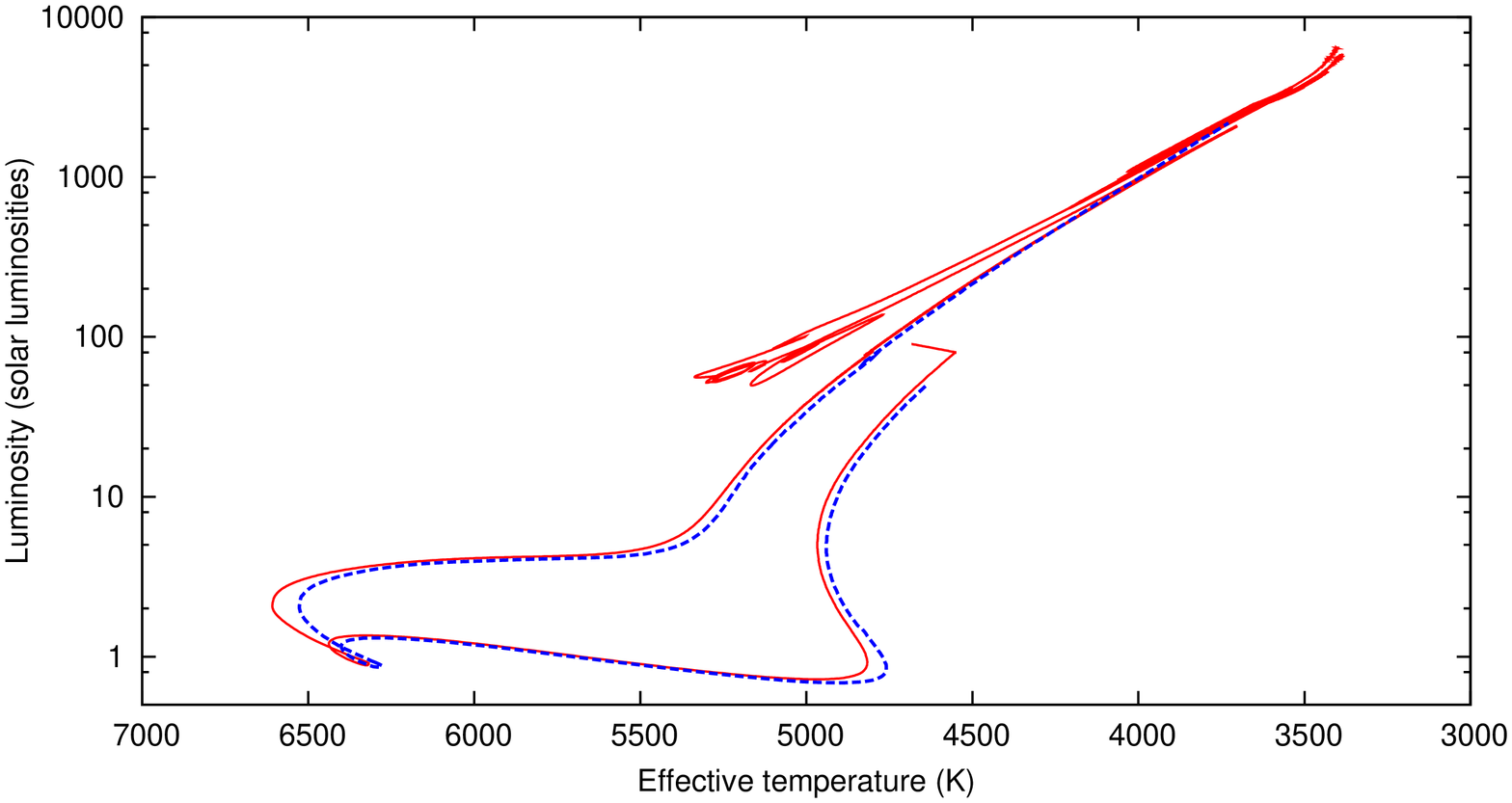}}
\centerline{\includegraphics[height=0.95\textwidth,angle=-90]{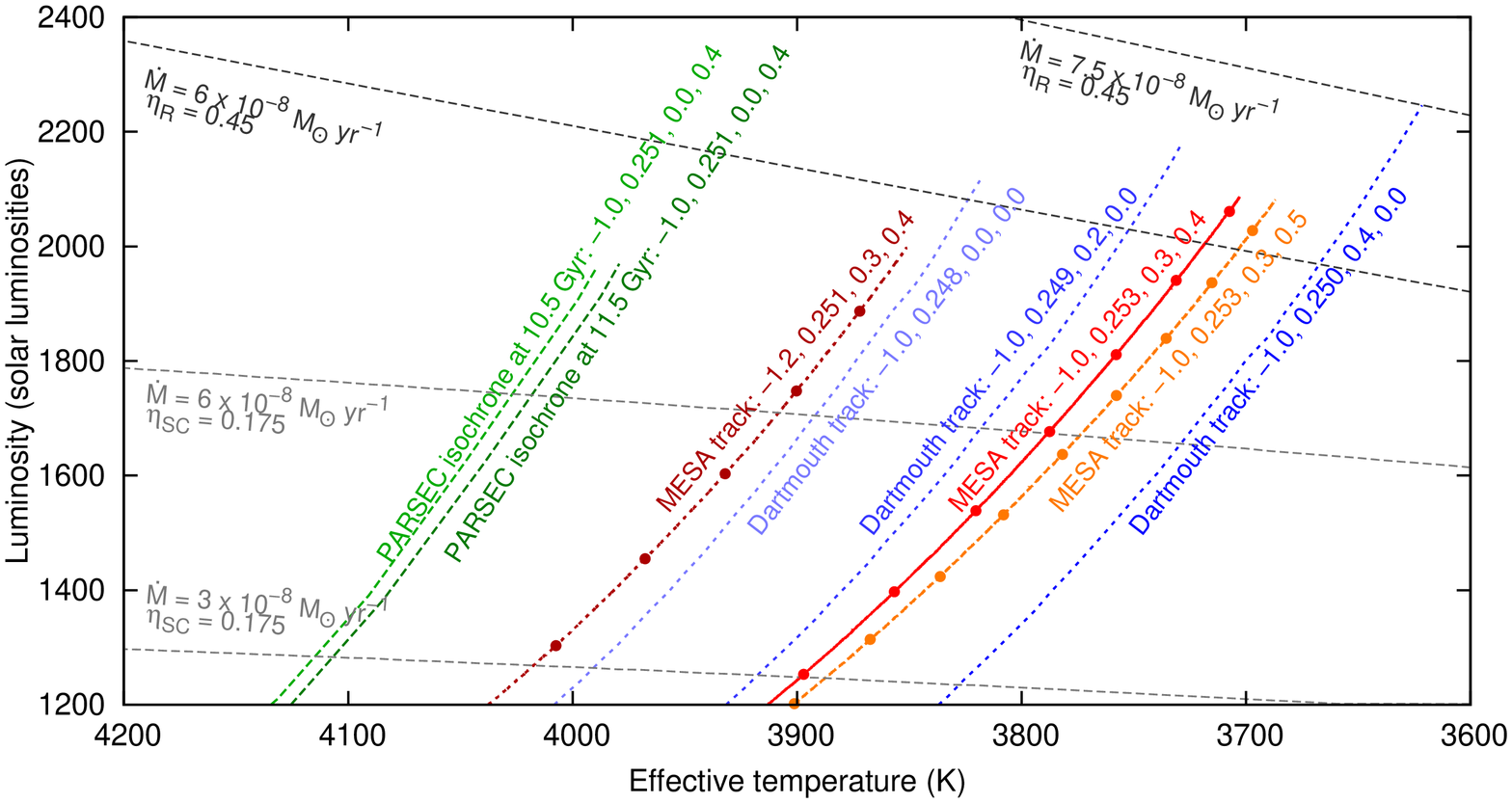}}
\caption{Top panel: stellar models on the Hertzsprung--Russell diagram. The solid, red line shows a {\sc mesa} track for a $M = 0.9$ M$_\odot$ star with [Fe/H] = --1 dex, losing mass with $\eta_{\rm R}$ = 0.2. A comparative Dartmouth track with [Fe/H] = --1 dex, [$\alpha$/Fe] = +0.4 dex, without mass loss, is shown as a dotted blue line. Bottom panel: a zoom into the RGB tip, showing a variety of {\sc mesa} and Dartmouth tracks and Padova isochrones. The format of the labels is: [Fe/H], $Y$ (helium fraction), [$\alpha$/Fe], $\eta_{\rm R}$. The effects of changing these three parameters can be seen by comparing the tracks. Also shown are lines of constant mass loss for the formulae of Reimers' and SC05. Each dot on the {\sc mesa} tracks denotes a loss of 0.01 M$_\odot$, the final dots being (from left to right) 0.75, 0.73 and 0.68 M$_\odot$.}
\label{MESAHRDFig}
\end{figure*}

Figure \ref{MESAHRDFig} compares our {\sc mesa} evolutionary tracks with the existing Dartmouth tracks. Evolutionary tracks for the {\sc parsec} models are not publicly available, but the range of masses on the upper RGB is sufficiently small that the {\sc parsec} isochrones are a reasonable substitute in this region.

The {\sc mesa} and Dartmouth isochrones follow each other reasonably well for most of the stellar evolution process. Minor differences exist, which are primarily due to the exact choices of helium content and detailed elemental abundances of the models (see \citet{DCJ+07} for a detailed discussion of these effects). Particularly important here is the oxygen abundance, which accounts for the difference in main-sequence turn-off temperature.

The lower panel of Figure \ref{MESAHRDFig} shows the upper part of the RGB of selected {\sc mesa} and other evolutionary models, revealing the sensitivity of the RGB tip to a variety of different parameters. The adoption of an $\alpha$-enhanced model has moved our RGB tip to cooler temperatures (higher mass-loss rates) than the Padova models, though the proportional increase in helium abundance has partly compensated for this. The additional helium abundance compared to the Dartmouth evolution tracks has meant that the tracks evolve faster and are truncated at lower luminosities.

The inclusion of mass loss also shifts the tracks to cooler temperatures, increasing the mass-loss rate, but decreasing the luminosity of the RGB tip. Faster mass loss also leads to marginally longer RGB evolution, with a difference of 9.6 Myr between the $\eta = 0.4$ and 0.5 {\sc mesa} models. The choice of $\eta$ in the stellar models therefore has an impact on the calculated $\eta$ we derive: a change in the stellar evolution model of $\Delta\eta_{\rm R} = 0.1$ results in a change in the derived value of $\Delta\eta_{\rm R} = 0.0175$. It is on this basis that we chose to calculate models with two different values of $\eta_{\rm R}$: one with $\eta_{\rm R} = 0.4$ and one with $\eta_{\rm R} = 0.5$.

\subsection{Comparing of our {\sc mesa} tracks with observations}
\label{MESADataSect}

\begin{figure*}
\centerline{\includegraphics[height=0.95\textwidth,angle=-90]{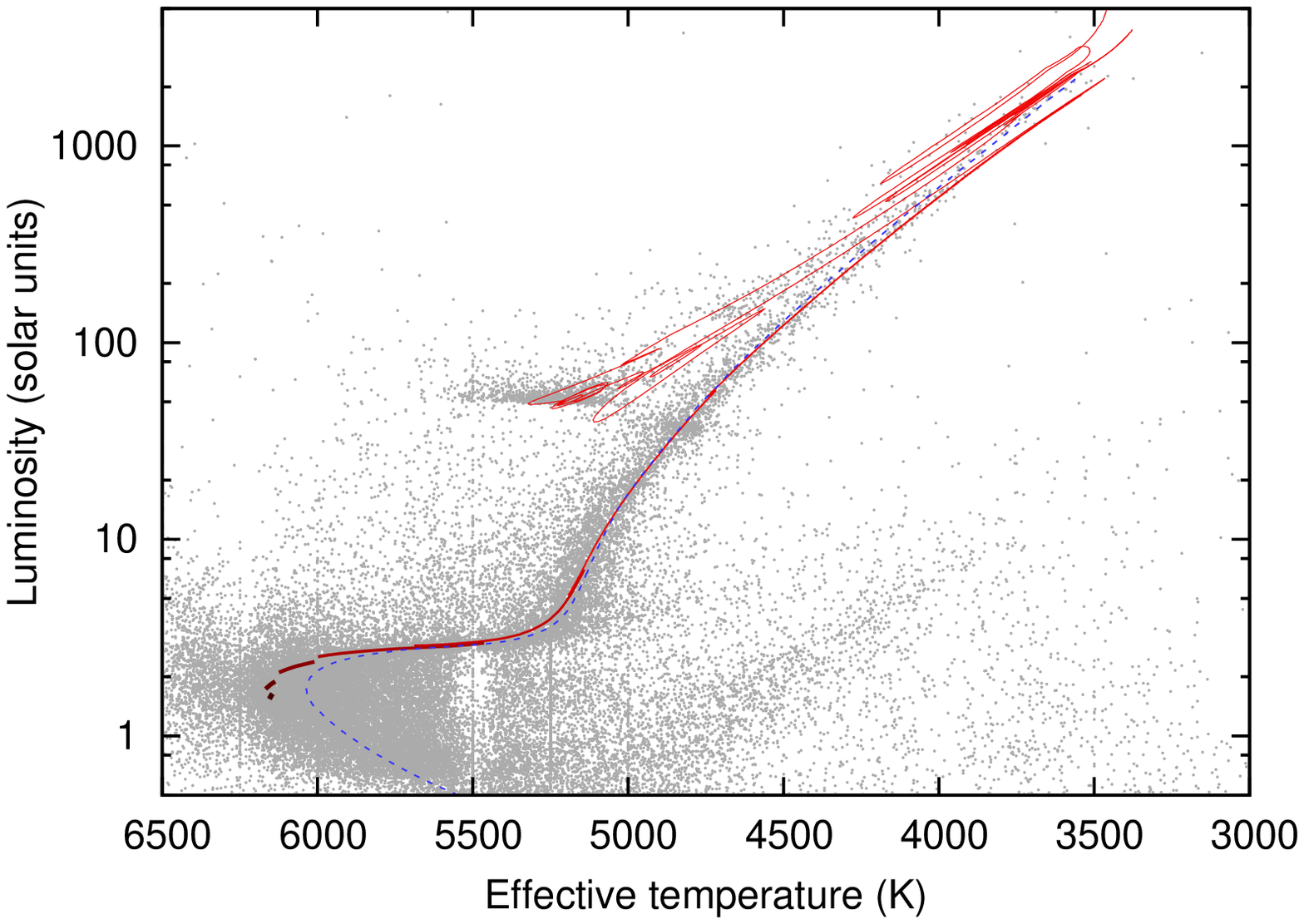}}
\caption{A Hertzsprung--Russell diagram of 47 Tucanae (grey points from \citealt{MBvL+11}). A 12-Gyr, $Y$ = 0.25, [$\alpha$/Fe] = +0.2 dex, [Fe/H] = --0.75 Dartmouth isochrone is shown as a dotted blue line. Stellar evolution tracks from {\sc mesa} are shown in red. Tracks at several different masses are shown between 11.6 Gyr and 12.119 Gyr in order to approximate a stellar isochrone appropriate for 47 Tuc. From bottom (dark) to top (light), the tracks are for masses 0.83, 0.84, 0.85, 0.86, 0.867 and 0.875 M$_\odot$.}
\label{MESA47TucFig}
\end{figure*}

Figure \ref{MESA47TucFig} shows our {\sc mesa} stellar evolution model applied to the globular cluster 47 Tucanae (NGC 104). We have chosen this cluster because it has one of the most homogeneous populations of all the populous clusters (GCB+10). The stellar abundances are well determined by recent publications and closely follow those we have assumed for the {\sc mesa} evolutionary tracks \citep[][Johnson et al., submitted]{GLS+13,CPJ+14,CKB+14,DKB+14,LML+14,TSA+14}. Temperatures of all stars above the MSTO have been derived in \citet{MBvL+11}. These temperatures have been confirmed spectroscopically on the RGB and AGB by \citet{CPJ+14} and Johnson et al.\ (submitted) and should therefore have an absolute accuracy of $<$30 K between 4200 and 4700 K. This spectroscopic calibration ensures we have appropriately addresssed the small reddening correction towards the cluster and that we have a model with the correct chemical abundances. This allows us to confidently adopt parameters in {\sc mesa} that fit the observed Hertzsprung--Russell diagram.

The {\sc mesa} tracks shown in Figure \ref{MESA47TucFig} have been evolved for 12.119 Gyr, and the last 519 Myr of each track is shown. They are computed for our standard elemental mixture, scaled to [Fe/H] = --0.72 dex as appropriate for 47 Tuc \citep[][and references above]{Harris10}, and adopt $\eta_{\rm R} = 0.45$ for their entire evolution. The {\sc mesa} isochrone is marginally better at reproducing the main-sequence turnoff than the equivalent Dartmouth isochrone without affecting the luminosity of the sub-giant branch, although the accuracy of both is within the uncertainties imparted by cluster and and distance. Both isochrones reproduce the location of the RGB tolerably well. The {\sc mesa} isochrone over-estimates the luminosity of the RGB bump, which is due to our simplistic treatment of mixing in these stars (cf.\ \citealt{CSB+02}). However, this does not have an important effect on later stellar evolution for the purpose of this paper.

Further up the giant branch, at cooler temperatures, the Dartmouth isochrone and {\sc mesa} track diverge, primarily due to the mass loss experienced by the {\sc mesa} model. Qualitatively, the Dartmouth model appears to better represent the upper RGB. However, non-LTE and dynamical effects become important as the stars are pulsating. Detailed treatment of molecular bands becomes important below 4000 K. Dust production also artificially lowers the computed stellar temperature in the brightest stars \citep{MBvLZ11}. As the temperature in Figure \ref{MESA47TucFig} are computed with static atmospheres under the assumption of LTE and no dust, temperatures below 4000 K may be less trustworthy. A more refined treatment by \citet{LNH+14} suggests that the brightest, pulsating stars may have temperatures slightly ($\sim$100 K) lower than predicted, although there are considerable systematic uncertainties preventing us stating this with certainty.

The post-RGB evolution is largely controlled by the mass-loss treatment on the RGB. Qualtitatively, our $\eta = 0.45$ model provides the appropriate temperature and luminosity of the horizontal branch in 47 Tuc. The start of the AGB, at $\sim$135 L$_\odot$ is not well reproduced by the {\sc mesa} track (80 L$_\odot$), which is largely also due to our treatment of mixing. The model fails to converge after three thermal pulses, at which point it has a mass of 0.567 M$_\odot$. A fourth thermal pulse could be expected, which is not expected to be enough to form a carbon star. Carbon stars are not observed in 47 Tucanae.

\section{Uncertainties in $\eta$}
\label{TrackCompareSect}

Calculation of the formal systematic uncertainty in $\eta$ is difficult. Differences in the treatment of stellar evolution are important, but the main source of systematic uncertainty in $\eta$ is caused by assumptions about the stellar parameters themselves. An exhaustive study, requiring many self-consistent grids of stellar evolution models, is computationally prohibitive. For the purposes of this study, an adequate approximation can be made by comparing pre-existing grids of stellar evolution models and test cases of {\sc mesa} evolution models.

Since $\eta$ is measured by following stellar evolution models, comparing an initial mass with a current mass, we have divided our error analysis into these three categories.

\begin{figure}
\centerline{\includegraphics[height=0.47\textwidth,angle=-90]{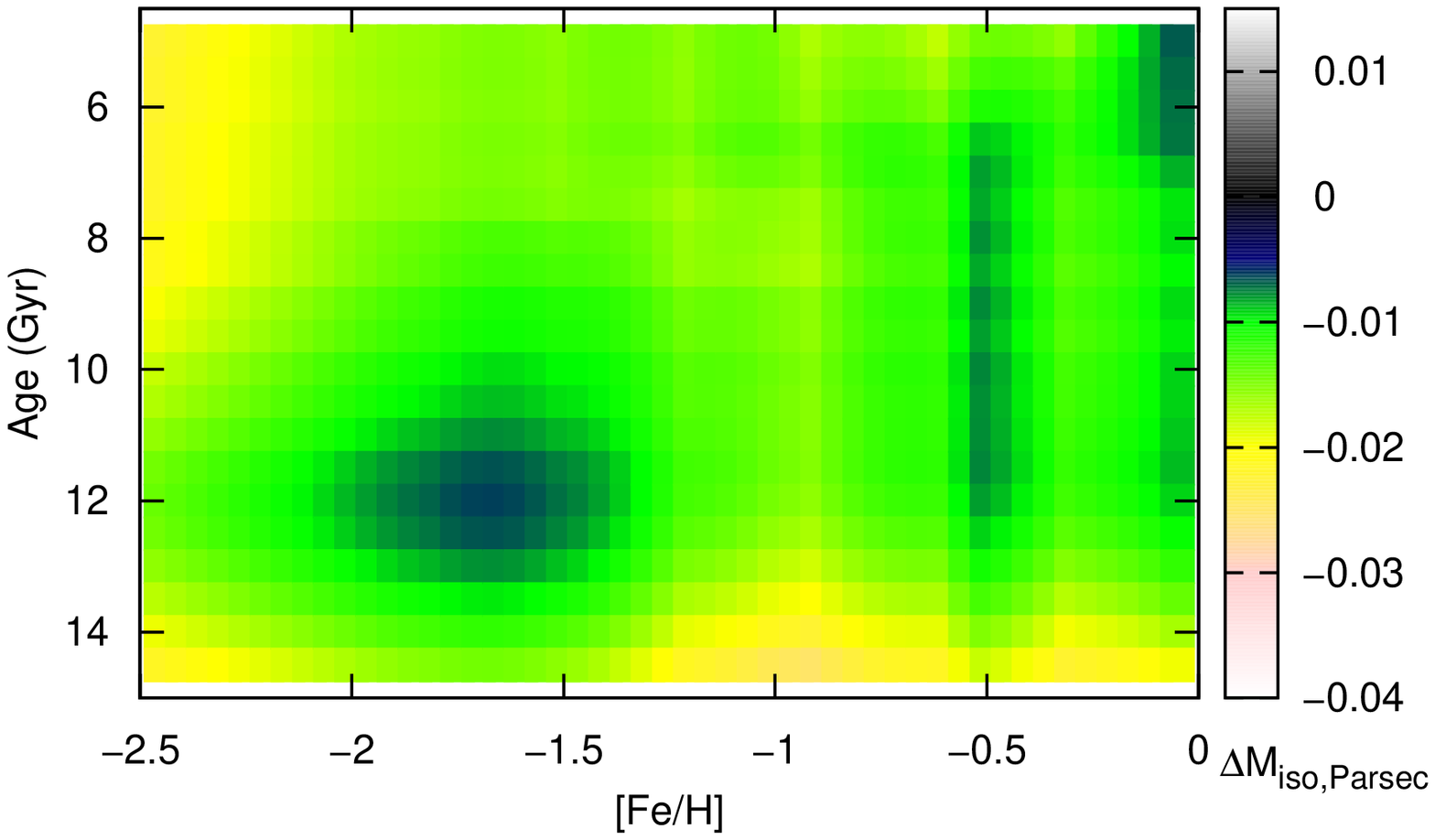}}
\centerline{\includegraphics[height=0.47\textwidth,angle=-90]{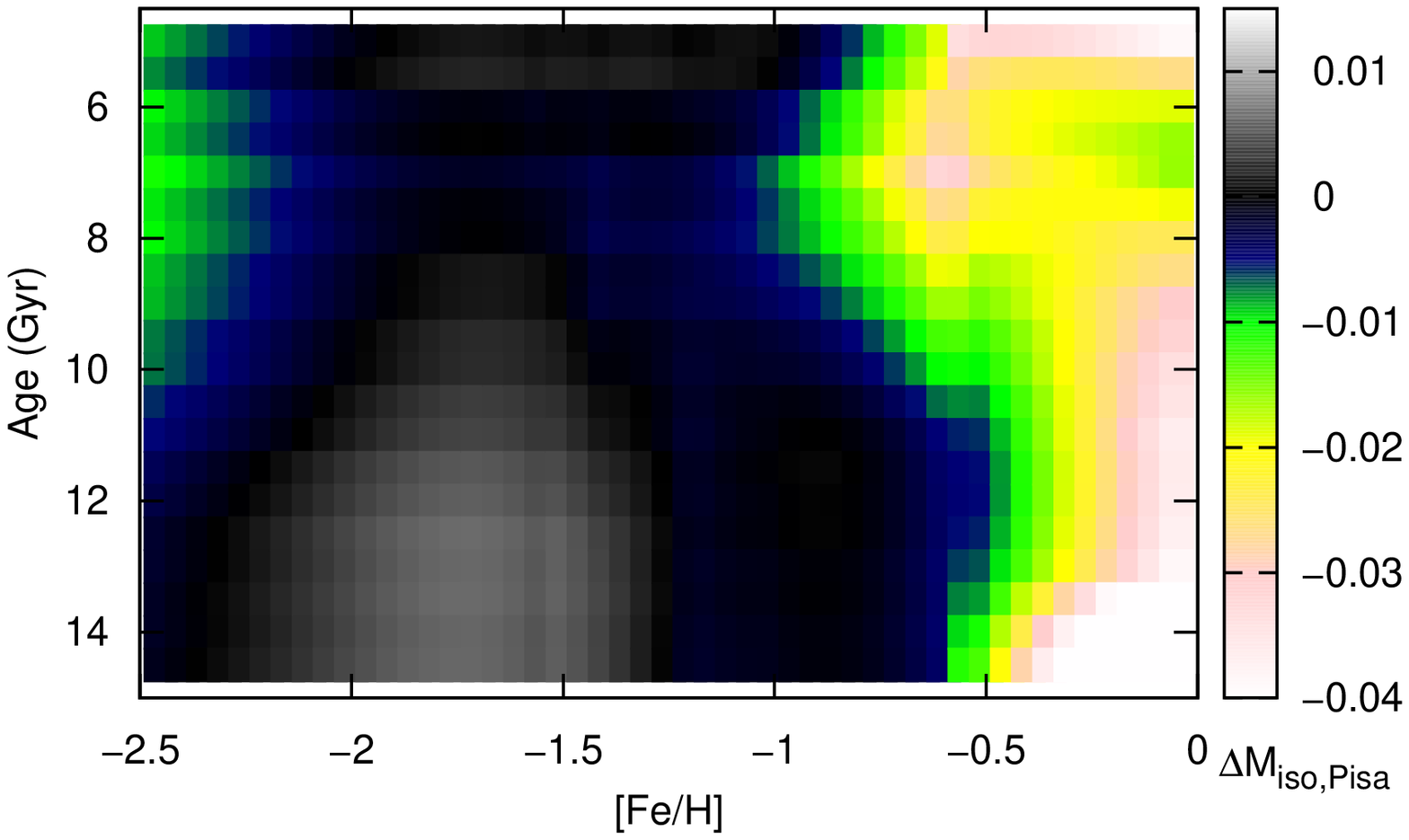}}
\centerline{\includegraphics[height=0.47\textwidth,angle=-90]{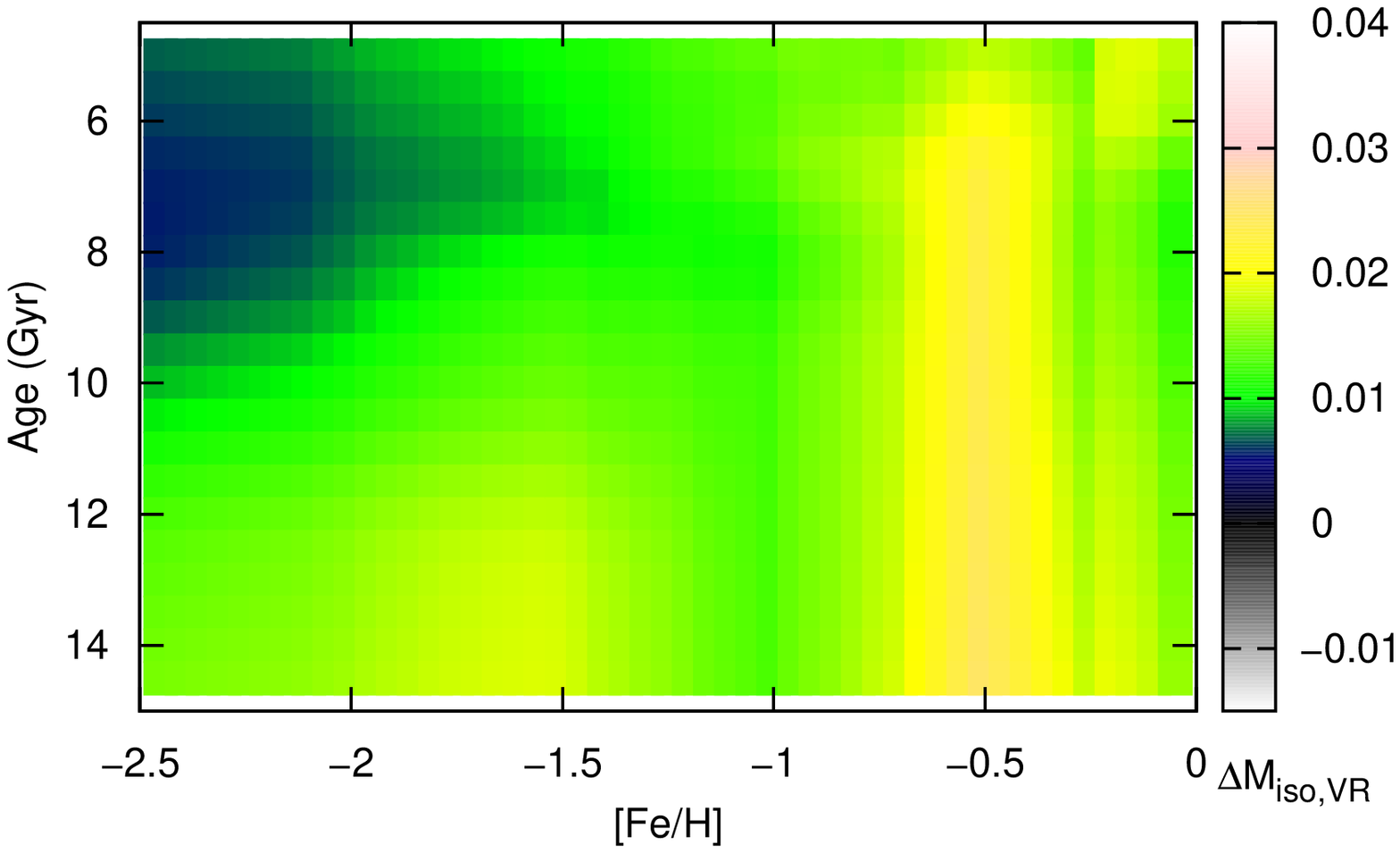}}
\centerline{\includegraphics[height=0.47\textwidth,angle=-90]{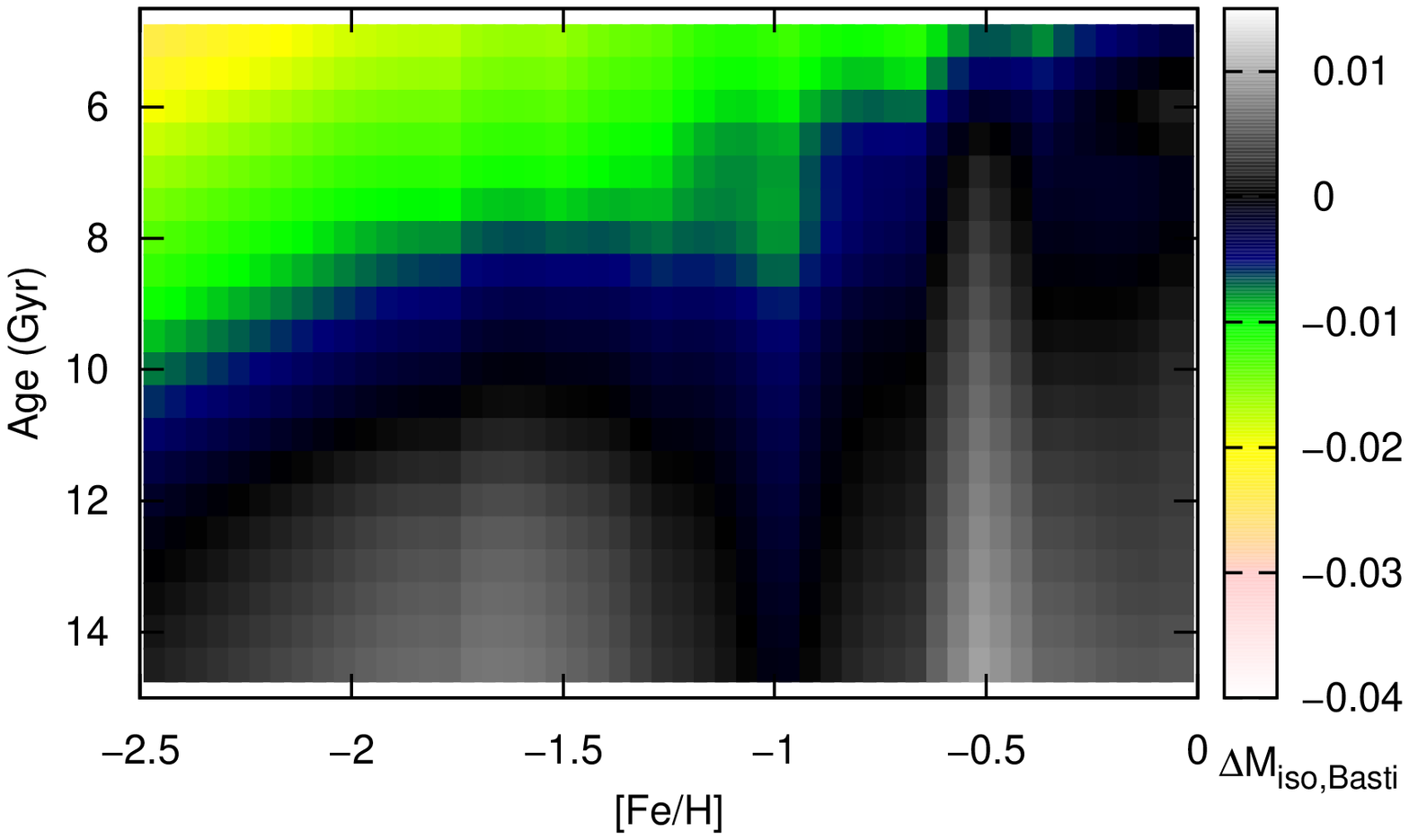}}
\caption{Difference between the Dartmouth RGB tip masses and other isochrones, for [$\alpha$/Fe] = 0 isochrones with solar-scaled helium abundances. The models are (top to bottom): {\sc parsec}, Pisa, Victoria Regina and BaSTI.}
\label{ModelDiffFig}
\end{figure}
\begin{figure}
\centerline{\includegraphics[height=0.47\textwidth,angle=-90]{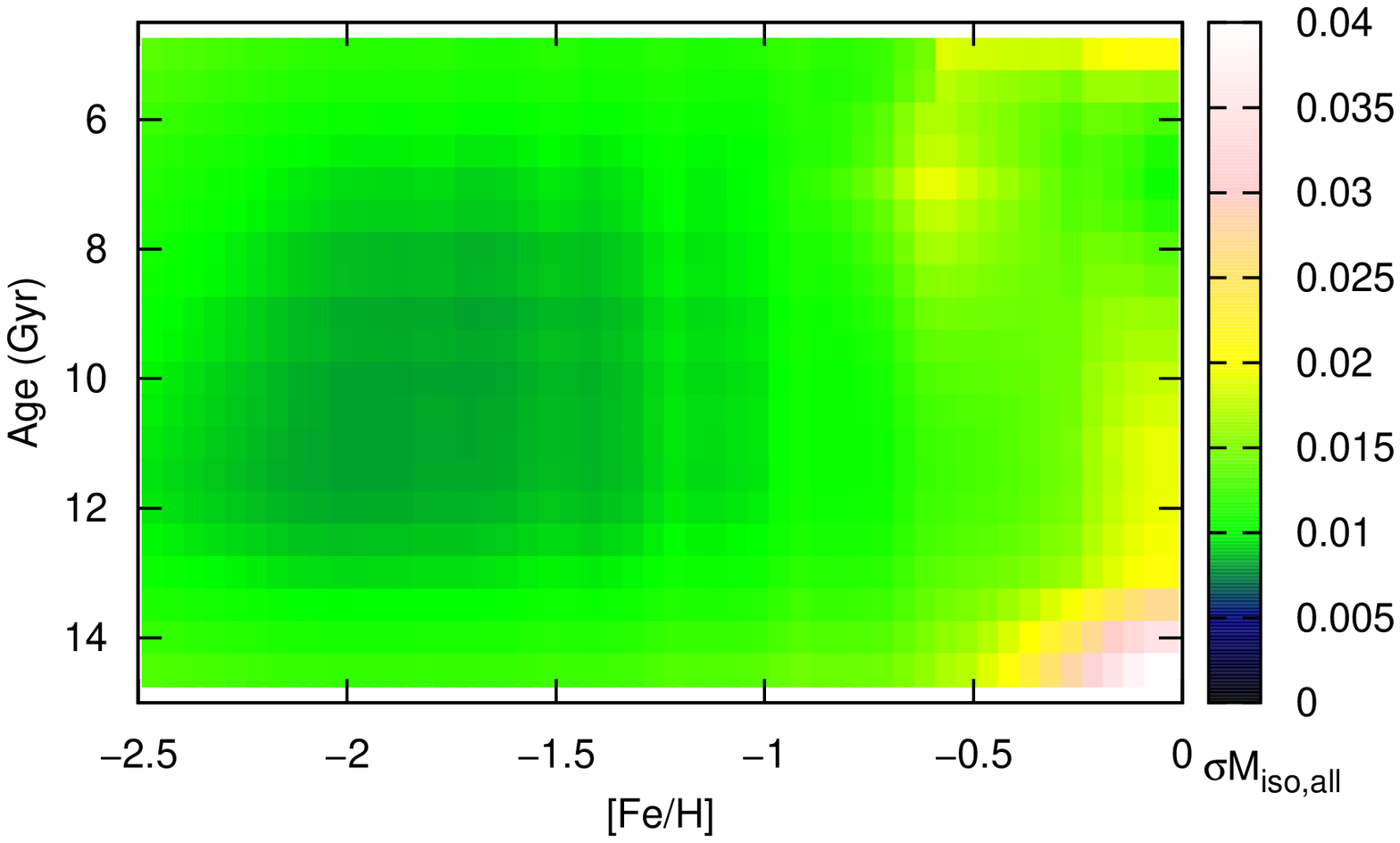}}
\centerline{\includegraphics[height=0.47\textwidth,angle=-90]{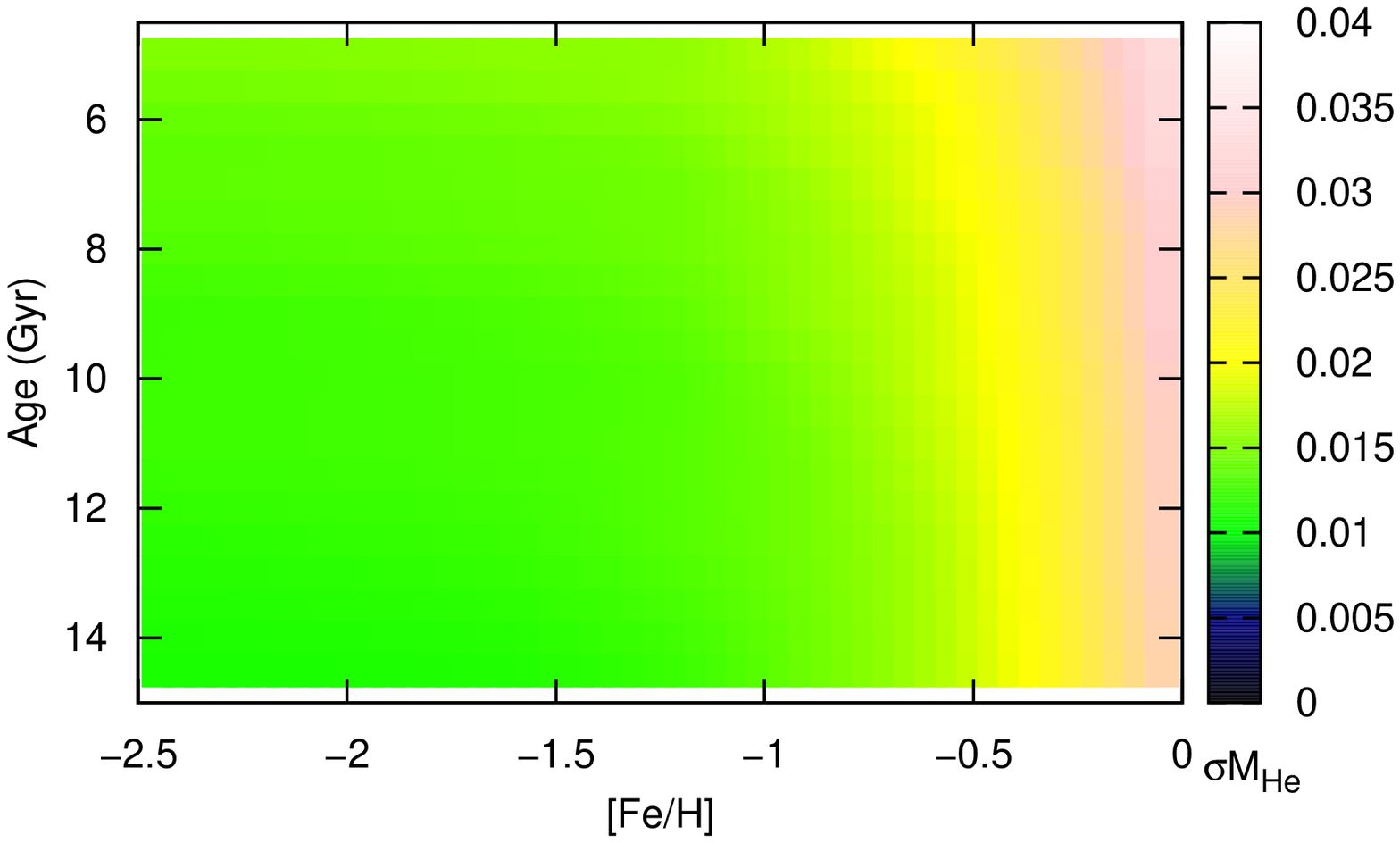}}
\centerline{\includegraphics[height=0.47\textwidth,angle=-90]{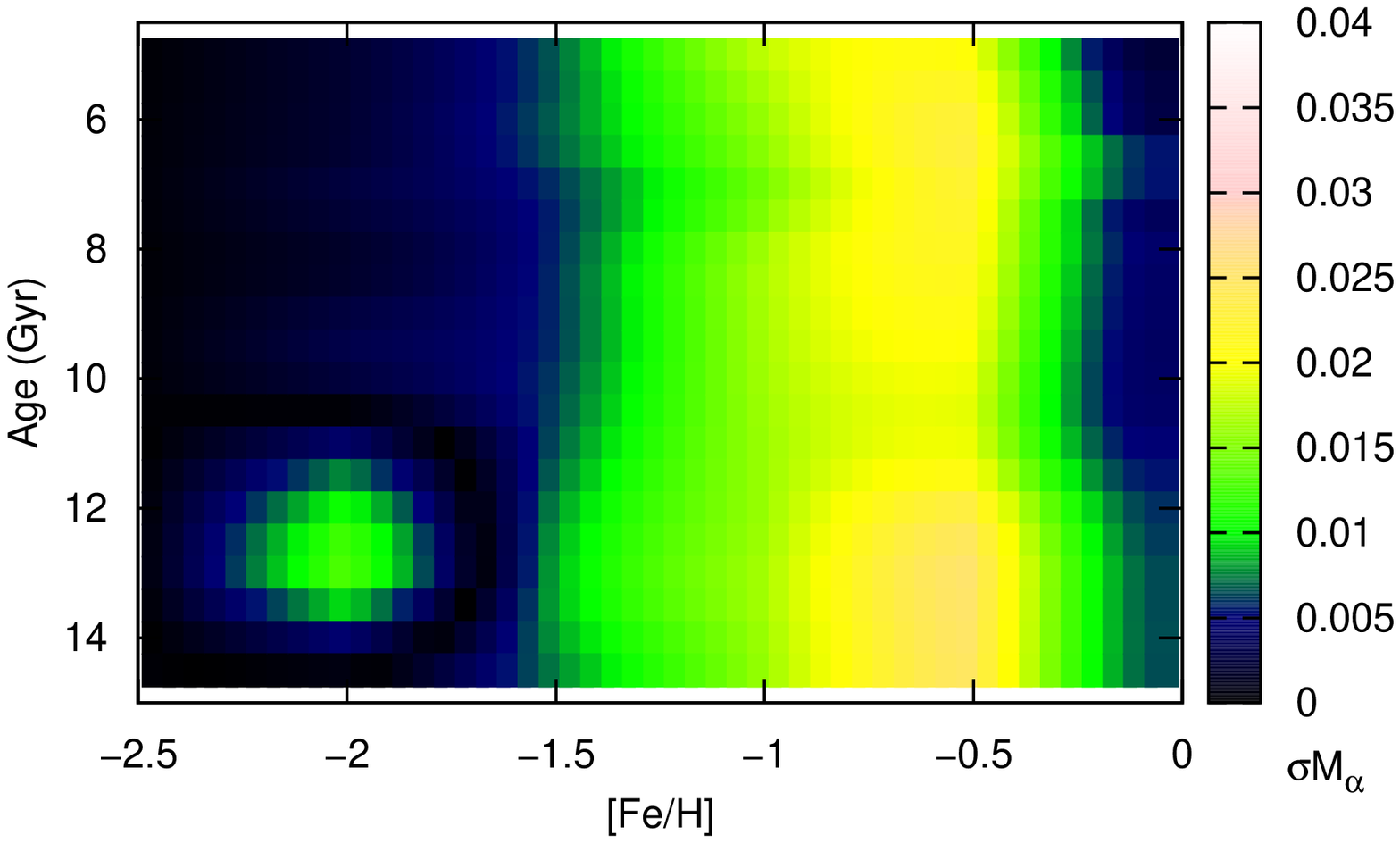}}
\centerline{\includegraphics[height=0.47\textwidth,angle=-90]{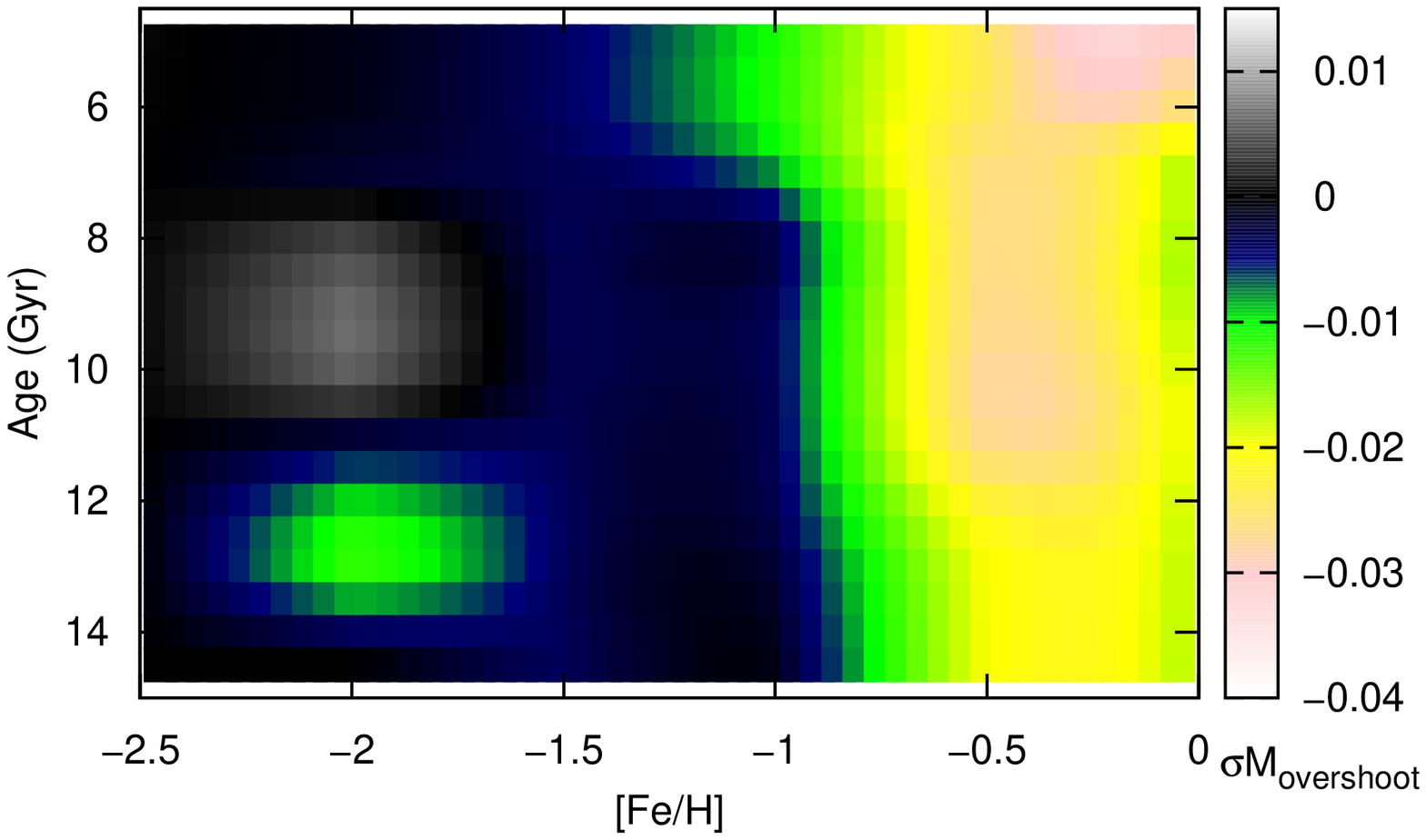}}
\caption{The major sources of error contributing to determination of the the initial mass of RGB tip stars. Top panel: the error associated with choice of stellar evolution model. Second panel: combined random and systematic errors  associated with the natal helium abundance. Third panel: the absolute difference between [$\alpha$/Fe] = +0.2 and +0.4 dex models, indicating the uncertainty due to a varying [$\alpha$/Fe]. Bottom panel: from the BaSTI models, the error due to convective overshooting prescription.}
\label{ErrorFig}
\end{figure}

\subsection{Uncertainties arising from choice of evolution model}

Errors associated with the choice of evolution model can be approximated by choosing models with the same parameters but from different evolution codes. Figure \ref{ModelDiffFig} shows the difference in initial mass of stars at the RGB tip between each set of isochrones and the Dartmouth models. To arrive at an accurate comparison, we have had to interpolate over both age and metallicity. We do this by two-dimensional linear interpolation between points in age and [Fe/H], interpolating past the end of existing models if necessary. Note that many of the vertical features in the diagrams may arise from this interpolation as some models are sparsely sampled in metallicity. Where [Fe/H] is not directly given, we have calculated it from the metal abundance assuming $Z_\odot = 0.0152$.

The systematic error caused by the choice of stellar evolution model is taken to be the standard deviation of the RGB tip masses of all five evolutionary codes for a particular age and metallicity. It is shown as the top panel in Figure \ref{ErrorFig}. For the age--metallicity range covered by globular clusters, the error is typically $\sim$0.01 M$_\odot$, which imparts a systematic error of $\sim$0.05 to $\eta$. Note that exact values are used in the derivation of the systematic error in the main text.

\subsection{Uncertainties in initial mass}
\label{TrackCompareMassSect}

Once the evolutionary model is set, the primary uncertainties arise from the stellar physics and chemistry included in the model. These include the helium abundance, [$\alpha$/Fe] ratio and treatment of convective overshooting. Figure \ref{ErrorFig} shows our adopted uncertainties for these effects. The quadrature sum of these four error components (evolution model, helium abundance, [$\alpha$/Fe] and convective overshooting) is taken as the total systematic error in the age--initial-mass relation. This error is shown in Figure \ref{MinitErrorFig} in the main text (Section \ref{MassSect}). An additional error is present on $\eta_{\rm SC}$ that is not present on $\eta_{\rm R}$ due to the fact our model grid is calculated using Reimers' mass-loss law. This is explored in more detail in Section \ref{ErrorEtaSCSect}.

\subsubsection{Helium abundance in evolution models}
\label{TrackHeSect}

\begin{figure}
\centerline{\includegraphics[height=0.47\textwidth,angle=-90]{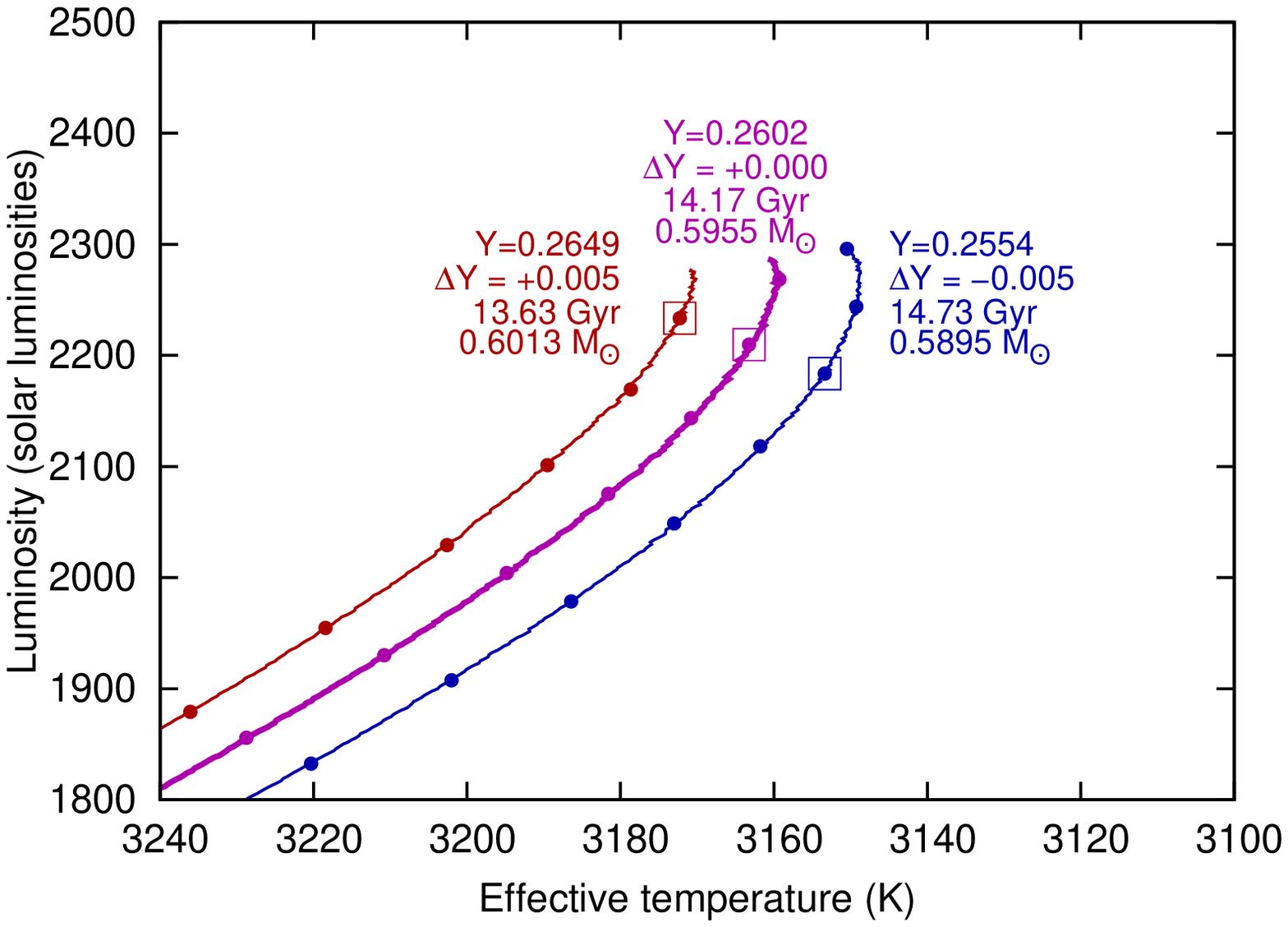}}
\caption{The RGB tip as a function of helium abundance for [Fe/H] = --0.60 dex models at 0.90 M$_\odot$ with $\eta_{R} = 0.45$. See Figure \ref{MESAHRDFig} for details of the symbols. The open squares mark a mass of 0.61 M$_\odot$.}
\label{MESAHeFig}
\end{figure}

The helium abundance has a relatively small effect on the Hertzsprung--Russell diagram, resulting in a marginally hotter RGB and a less-luminous RGB tip. A good comparison can be made between two Dartmouth isochrones at 11.5 Gyr, [Fe/H] = --1.0 dex and [$\alpha$/Fe] = +0.4 dex, for two different helium abundances. The isochrones at $Y = 0.25$ and $Y = 0.33$ ($\Delta Y = 0.08$) show a difference in the RGB tip position of 40 K and 70 L$_\odot$. By far the stronger factor is the rate of stellar evolution. The same isochrones provide initial stellar masses of 0.877 and 0.750 M$_\odot$ at the RGB tip. For a typical HB mass of 0.65 M$_\odot$, the helium-rich model produces an $\eta$ that is only 44 per cent of the base model.

This is also illustratable in the {\sc mesa} models. Figure \ref{MESAHeFig} shows the RGB tip in the Hertzsprung--Russell diagram for three evolutionary tracks with marginally different helium abundances, each with $M = 0.90$ M$_\odot$. Sodium and magnesium abundances have been increased by $5\Delta Y$ and oxygen and aluminium abundances decreased by the same amount to account for the observed Na--O and Mg--Al anticorrelations that are thought to be associated with helium enrichment. The stellar parameters near the RGB are negligibly different for the three different cases, but a difference of $\pm$0.005 in helium fraction results in a difference of $^{-0.54}_{+0.56}$ Gyr in stellar age and $^{+0.0058}_{-0.0060}$ M$_\odot$ in RGB tip mass.

Since we observe clusters at a known age, rather than a known mass, we must convert this into a mass differential to understand its effect on $\eta$. A change in the same model of $\pm$0.02 M$_\odot$ in initial mass produces a change in stellar age of $^{-0.94}_{+1.02}$ Gyr. We can therefore assume that a change of $\Delta Y$ = 0.005 produces a change in initial mass at fixed age of $\sim\pm$0.011 M$_\odot$.

\subsubsection{Error introduced by the uncertain helium abundance}
\label{TrackHeScaleSect}

GCB+10 model that most clusters have a median helium richness close to the primordial value (68 per cent have $Y_{\rm med} \leq 0.254$, compared to an assumed average of $Y = 0.247$). We can therefore adopt an uncertainty in each cluster's helium enhancement of $\Delta Y = 0.007$. As this is purely due to helium \emph{enrichment}, not depletion, these uncertainties make the overall error asymmetric, reducing $\Delta M$ and $\eta$.

For our adopted error on each cluster, we use the difference between the $Y$ = 0.23 and 0.26 Padova models \citep{BGMN08}, scaled by 0.23$\times$ as appropriate for our uncertainty of $\Delta Y = 0.007$. The resulting error is around 0.01 M$_\odot$ in most cases, rising to $\sim$0.02 M$_\odot$ for the highest-metallicity cluster. The corresponding values for test cases mentioned in the previous section are $\sim$0.011 M$_\odot$ for the metal-poor Dartmouth test case and $\sim$0.015 M$_\odot$ for the metal-rich {\sc mesa} test case, i.e.\ very similar uncertainties are found among all three models.

In addition to the individual errors, there in a substantial systematic error in the helium abundance attributable to the relative enrichment of helium with metallicity. Following the {\sc parsec} models, we have adopted $\Delta Y / \Delta Z$ = 1.78 for our evolutionary tracks, with a primordial $Y = 0.2485$. Primordial helium abundances scatter by about $\pm$0.001, while modern scalings of the initial solar helium abundance tend to vary by $\pm$0.004 \citep{AGSS09,LPG09,SB10,CUV13,AOPS13,Planck-2013-XVI}. Meanwhile, the same sources give $Z_\odot$ to be uncertain by around 4 per cent (0.0152 $\pm$ 0.0006). The uncertainty in primordial helium fraction is negligible compared to other sources of uncertainty, but the scaling of the helium enrichment with $Z$ becomes important at higher metallicities. 

We here assume that the first generation of globular cluster stars have experienced a helium enrichment which is proportionally similar to the Sun. This may not be the case, but the uncertainty in the relationship is sufficient to take into account the variations imposed by GCB+10. Based on the primordial and solar helium abundances, we can then derive an uncertainty in the helium content of these first-generation stars as:
\begin{equation}
\Delta Y_{\rm systematic} = 0.001 + 0.263 Z .
\end{equation}
Based on the calibration found in the last section, this can be translated to an systematic uncertainty in initial mass of:
\begin{equation}
\Delta M_{\rm init,\ systematic} = 0.0022 + 0.0090 Z / Z_\odot .
\end{equation}
Following the abundances we adopted in our {\sc mesa} models, this then becomes:
\begin{equation}
\Delta M_{\rm init,\ systematic} = 0.0022 + 0.0158 \times 10^{\rm [Fe/H]} .
\end{equation}

\subsubsection{Uncertainties arising from other chemical and physical choices}

To estimate the uncertainty introduced by assuming a fixed $\alpha$-enhancement ([$\alpha$/Fe] = +0.3 dex), we use the difference between the two Dartmouth models at +0.2 and +0.4 dex. The range of $\pm$0.1 dex is close to the standard deviation in globular clusters' [$\alpha$/Fe] of 0.11 dex \citep{RCGS13}. The uncertainty arising from changes in [$\alpha$/Fe] dominates the uncertainty budget for clusters around [Fe/H] = --2 dex and between [Fe/H] $\approx$ --1 and --0.4 dex. Outside these areas, the choice is dominated by the choice of evolutionary model.

Finally, we include the uncertainty due to convective over-shooting by taking the difference between BaSTI models including two different levels of over-shooting. Again the effects are strongest at higher metallicities, and particularly become important above [Fe/H] = --1.

For the most-metal-poor clusters, the error in initial mass is set by the choice of stellar evolution model. At [Fe/H] $\approx$ --2 and $\approx$ --1.2, the [$\alpha$/Fe] and choice of evolution model dominate. Beyond [Fe/H] = --1, the [$\alpha$/Fe] and convective overshooting become more important and, in the most-metal-rich clusters, the convective overshooting and helium abundance dominate the errors.

To summarise, we have adopted the following random uncertainties in our calculation of initial mass:
\begin{itemize}
\item Age: random error following variations among literature derivations, as prescribed in the main text;
\item {[Fe/H]}: total error of $\pm$0.1 dex (see main text), to account for uncertainties in derivations for individual clusters;
\item Helium abundance ($Y$): fixed ($^{+0.000}_{-0.016}$ M$_\odot$);
\item {[$\alpha$/Fe]}: variable (up to $\sim$0.01 M$_\odot$);
\end{itemize}
and the following systematic uncertainties:
\begin{itemize}
\item {[Fe/H]}: an error due to calibrating an absolute metallicity scale incorporated in random error;
\item Choice of evolutionary model: variable ($\sim$0.01 M$_\odot$);
\item Helium abundance ($Y$): variable ($0.0022 + 0.0158 \times 10^{\rm [Fe/H]}$);
\item Convective overshooting: variable (up to $\sim$0.02 M$_\odot$).
\end{itemize}

\subsection{Uncertainties in horizontal branch mass}
\label{EtaHeSect}

\begin{figure}
\centerline{\includegraphics[height=0.47\textwidth,angle=-90]{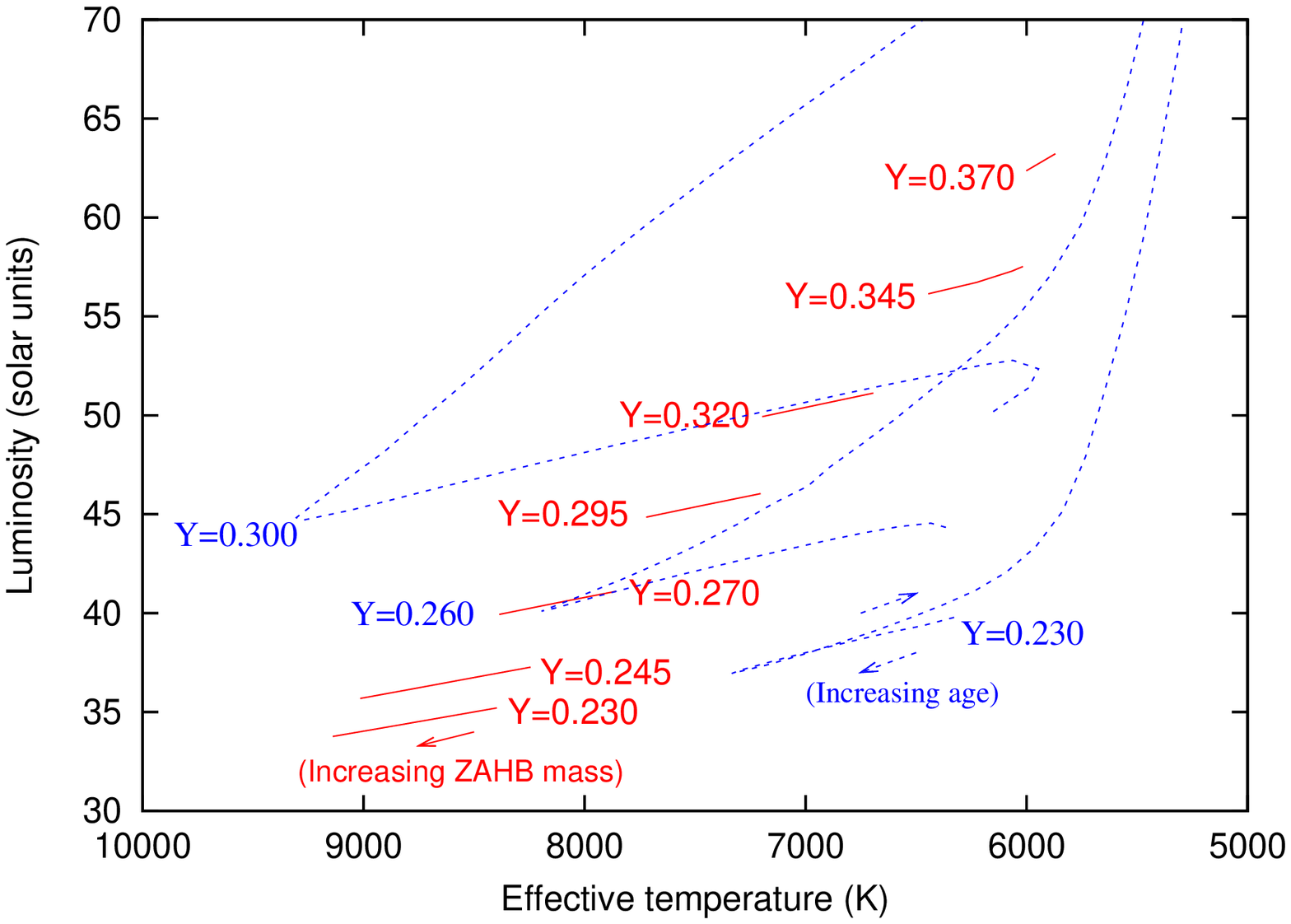}}
\caption{Solid red lines show zero-age HB loci for stars between 0.60 M$_\odot$ and 0.61 M$_\odot$ (increasing as shown) for different helium fractions ($Y$). Data are from the PGPUC models for [Fe/H] = --1.25 dex, [$\alpha$/Fe] = +0.3 dex. Blue dotted lines show HB and post-HB evolutionary tracks for $Z = 0.001$, $M = 0.60$ M$_\odot$ stars from the Padova isochrones, with time increasing as shown.}
\label{HeZAHBFig}
\end{figure}

Figure \ref{HeZAHBFig} shows the effects of helium abundance on ZAHB loci and HB evolutionary models using data from two evolutionary models. The PGPUC ZAHB models \citep{VCS12,VCDM13} use a fixed initial mass, leading to differences in the core mass. The Padova synthetic HB tracks use a fixed core and total mass. For our instantaneous view of a globular cluster, the PGPUC models give the more accurate representation of where stars will land on the ZAHB, while the Padova models can be used as a representation of the directions in which those stars will evolve.

The derivation of HB mass in GCB+10 uses purely colour terms (their equations (3) through (8)), which can be translated almost directly into effective temperatures. The length of the solid, red PGPUC model lines in Figure \ref{HeZAHBFig} represents a 0.01 M$_\odot$ change in ZAHB mass. The models are separated by 1.5--2.5 per cent in $Y$, with the lines almost overlapping in effective temperature. For modest increases in $Y$, up to about $Y = 0.270$, the lines continue to overlap. Thus, an increase of $Y$ up to this value will have a $<$0.01 M$_\odot$ effect on the ZAHB mass derived by GCB+10. This increase, $\Delta Y \approx 0.03$, is much more than either the typical random variation among clusters of $\Delta Y \approx 0.007$ found by GCB+10, or the projected uncertainty in helium enrichment with metallicity. We therefore estimate that the uncertainty in ZAHB mass derived by GCB+10 is $\ll$0.01 M$_\odot$.

An additional, second-order effect can be seen in the Padova models, where helium-rich stars are shown evolving to hotter temperatures while on the HB (Figure \ref{HeZAHBFig}). HB stars spend most of this period near the ZAHB locus, so the error in derived mass caused by changes to the HB evolution should be smaller than that caused by the change in ZAHB locus. We therefore combine both these errors to estimate that the ZAHB mass of the clusters as measured by GCB+10 is $\sim$0.01 M$_\odot$.

\subsection{An additional error on $\eta_{\rm SC}$}
\label{ErrorEtaSCSect}

\begin{figure}
\centerline{\includegraphics[height=0.47\textwidth,angle=-90]{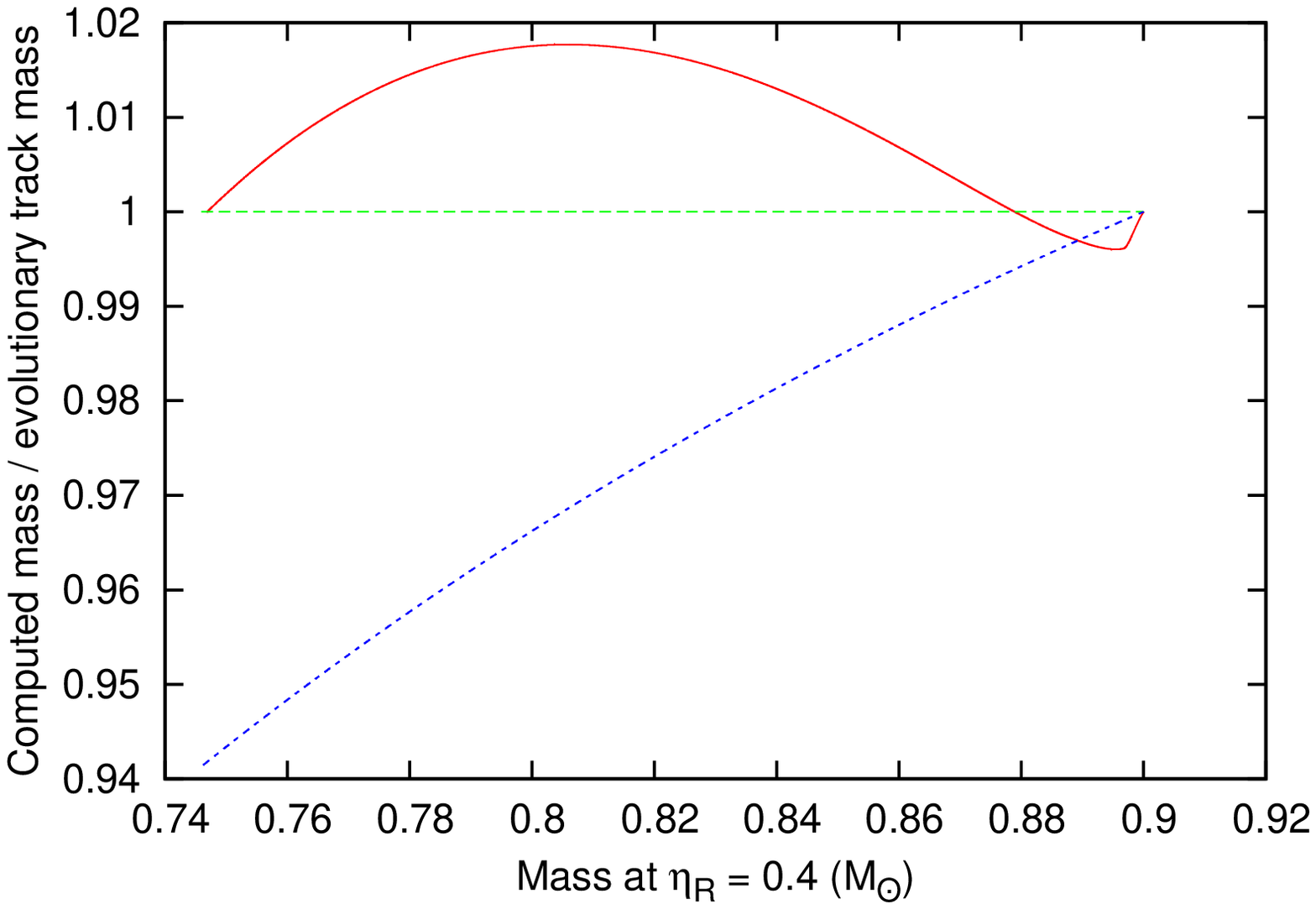}}
\caption{Error in mass caused by improper treatment of RGB mass loss. Evolutionary time runs right to left. The underlying evolutionary track is a model at [Fe/H] = --1, $M$ = 0.90 M$_\odot$. The dashed blue line shows the mass attained using $\eta_{\rm R} = 0.4$, applied to a $\eta_{\rm R} = 0.5$ track. The solid red line shows the mass attained using $\eta_{\rm SC} = 0.158$, applied to a $\eta_{\rm R} = 0.4$ track.}
\label{EtaSCErrorFig}
\end{figure}

Our evolutionary tracks are calculated based on Reimers' $\eta$. Applications to other mass-loss laws (not implemented in {\sc mesa}) acquire an additional uncertainty. That uncertainty is difficult to assess properly, however it can be estimated using a comparison between our $\eta_{\rm R} = 0.4$ and $\eta_{\rm R} = 0.5$ tracks.

Figure \ref{EtaSCErrorFig} shows the effect of applying the wrong value of $\eta$ to a track. Computing the stellar mass using $\eta_{\rm R} = 0.4$, applied to an evolutionary track computed for $\eta_{\rm R} = 0.5$ results in an RGB tip mass that is 6 per cent too low. Since the associated change in $L$ and $R$ are comparatively small, this translates into a fractional error of 6 per cent on $LR/M$, hence also on the value of $\eta_{\rm R}$ derived. For a derived $\eta = 0.5$, for example, the associated additional uncertainty would be 0.03.

Figure \ref{EtaSCErrorFig} also shows the difference in mass found when applying $\eta_{\rm SC} = 0.158$ to a $\eta_{\rm R} = 0.4$ model. The value of $\eta_{\rm SC}$ here is chosen to make the RGB tip masses identical. The difference in mass is always less than 2 per cent. Since the difference in $L$, $R$, $T$ and $g$ are comparatively small, we estimate that the additional fractional uncertainty associated with applying the SC05 law to a track computed for Reimers' law is $<$2 per cent. To be conservative, we adopt a 2 per cent error in the main text.


\begin{thebibliography}{98}
\expandafter\ifx\csname natexlab\endcsname\relax\def\natexlab#1{#1}\fi

\bibitem[{{Anderson} {et~al.}(2008){Anderson}, {Sarajedini}, {Bedin}, {King},
  {Piotto}, {Reid}, {Siegel}, {Majewski}, {Paust}, {Aparicio}, {Milone},
  {Chaboyer}, \& {Rosenberg}}]{ASB+08}
{Anderson} J., {Sarajedini} A., {Bedin} L.~R., {King} I.~R., {Piotto} G.,
  {Reid} I.~N., {Siegel} M., {Majewski} S.~R., {Paust} N.~E.~Q., {Aparicio} A.,
  {Milone} A.~P., {Chaboyer} B., {Rosenberg} A., 2008, AJ, 135, 2055

\bibitem[{{Asplund} {et~al.}(2009){Asplund}, {Grevesse}, {Sauval}, \&
  {Scott}}]{AGSS09}
{Asplund} M., {Grevesse} N., {Sauval} A.~J., {Scott} P., 2009, ARA\&A, 47, 481

\bibitem[{{Aver} {et~al.}(2013){Aver}, {Olive}, {Porter}, \&
  {Skillman}}]{AOPS13}
{Aver} E., {Olive} K.~A., {Porter} R.~L., {Skillman} E.~D., 2013, 11, 17

\bibitem[{{Bertelli} {et~al.}(1994){Bertelli}, {Bressan}, {Chiosi}, {Fagotto},
  \& {Nasi}}]{BBC+94}
{Bertelli} G., {Bressan} A., {Chiosi} C., {Fagotto} F., {Nasi} E., 1994, A\&AS,
  106, 275

\bibitem[{{Bertelli} {et~al.}(2008){Bertelli}, {Girardi}, {Marigo}, \&
  {Nasi}}]{BGMN08}
{Bertelli} G., {Girardi} L., {Marigo} P., {Nasi} E., 2008, A\&A, 484, 815

\bibitem[{{Bladh} {et~al.}(2013){Bladh}, {H{\"o}fner}, {Nowotny}, {Aringer}, \&
  {Eriksson}}]{BHN+13}
{Bladh} S., {H{\"o}fner} S., {Nowotny} W., {Aringer} B., {Eriksson} K., 2013,
  A\&A, 553, A20

\bibitem[{{Boyer} {et~al.}(2009){Boyer}, {McDonald}, {van Loon}, {Gordon},
  {Babler}, {Block}, {Bracker}, {Engelbracht}, {Hora}, {Indebetouw}, {Meade},
  {Meixner}, {Misselt}, {Oliveira}, {Sewilo}, {Shiao}, \& {Whitney}}]{BMvL+09}
{Boyer} M.~L., {McDonald} I., {van Loon} J.~T., {Gordon} K.~D., {Babler} B.,
  {Block} M., {Bracker} S., {Engelbracht} C., {Hora} J., {Indebetouw} R.,
  {Meade} M., {Meixner} M., {Misselt} K., {Oliveira} J.~M., {Sewilo} M.,
  {Shiao} B., {Whitney} B., 2009, ApJ, 705, 746

\bibitem[{{Boyer} {et~al.}(2006){Boyer}, {Woodward}, {van Loon}, {Gordon},
  {Evans}, {Gehrz}, {Helton}, \& {Polomski}}]{BWvL+06}
{Boyer} M.~L., {Woodward} C.~E., {van Loon} J.~T., {Gordon} K.~D., {Evans} A.,
  {Gehrz} R.~D., {Helton} L.~A., {Polomski} E.~F., 2006, AJ, 132, 1415

\bibitem[{{Bressan} {et~al.}(2012){Bressan}, {Marigo}, {Girardi}, {Salasnich},
  {Dal Cero}, {Rubele}, \& {Nanni}}]{BMG+12}
{Bressan} A., {Marigo} P., {Girardi} L., {Salasnich} B., {Dal Cero} C.,
  {Rubele} S., {Nanni} A., 2012, MNRAS, 427, 127

\bibitem[{{Caloi} \& {D'Antona}(2007)}]{CDA07}
{Caloi} V., {D'Antona} F., 2007, A\&A, 463, 949

\bibitem[{{Cariulo} {et~al.}(2004){Cariulo}, {Degl'Innocenti}, \&
  {Castellani}}]{CDIC04}
{Cariulo} P., {Degl'Innocenti} S., {Castellani} V., 2004, A\&A, 421, 1121

\bibitem[{{Carney}(1996)}]{Carney96}
{Carney} B.~W., 1996, PASP, 108, 900

\bibitem[{{Carretta} {et~al.}(2009){Carretta}, {Bragaglia}, {Gratton},
  {D'Orazi}, \& {Lucatello}}]{CBG+09b}
{Carretta} E., {Bragaglia} A., {Gratton} R., {D'Orazi} V., {Lucatello} S.,
  2009, A\&A, 508, 695

\bibitem[{{Carretta} \& {Gratton}(1997)}]{CG97}
{Carretta} E., {Gratton} R.~G., 1997, A\&AS, 121, 95

\bibitem[{{Cassisi} {et~al.}(2002){Cassisi}, {Salaris}, \& {Bono}}]{CSB+02}
{Cassisi} S., {Salaris} M., {Bono} G., 2002, ApJ, 565, 1231

\bibitem[{{Castellani} {et~al.}(2003){Castellani}, {Degl'Innocenti}, {Marconi},
  {Prada Moroni}, \& {Sestito}}]{CDIM+03}
{Castellani} V., {Degl'Innocenti} S., {Marconi} M., {Prada Moroni} P.~G.,
  {Sestito} P., 2003, A\&A, 404, 645

\bibitem[{{Catelan} {et~al.}(2009){Catelan}, {Grundahl}, {Sweigart},
  {Valcarce}, \& {Cort{\'e}s}}]{CGSV+09}
{Catelan} M., {Grundahl} F., {Sweigart} A.~V., {Valcarce} A.~A.~R.,
  {Cort{\'e}s} C., 2009, ApJ, 695, L97

\bibitem[{{Coc} {et~al.}(2013){Coc}, {Uzan}, \& {Vangioni}}]{CUV13}
{Coc} A., {Uzan} J.-P., {Vangioni} E., 2013, ArXiv e-prints

\bibitem[{{Cordero} {et~al.}(2014){Cordero}, {Pilachowski}, {Johnson},
  {McDonald}, {Zijlstra}, \& {Simmerer}}]{CPJ+14}
{Cordero} M.~J., {Pilachowski} C.~A., {Johnson} C.~I., {McDonald} I.,
  {Zijlstra} A.~A., {Simmerer} J., 2014, ApJ, 780, 94

\bibitem[{{Cranmer} \& {Saar}(2011)}]{CS11}
{Cranmer} S.~R., {Saar} S.~H., 2011, ApJ, 741, 54

\bibitem[{{Dalessandro} {et~al.}(2013){Dalessandro}, {Salaris}, {Ferraro},
  {Mucciarelli}, \& {Cassisi}}]{DSF+13}
{Dalessandro} E., {Salaris} M., {Ferraro} F.~R., {Mucciarelli} A., {Cassisi}
  S., 2013, MNRAS, 430, 459

\bibitem[{{d'Antona} {et~al.}(2005){d'Antona}, {Bellazzini}, {Caloi}, {Pecci},
  {Galleti}, \& {Rood}}]{DABC+05}
{d'Antona} F., {Bellazzini} M., {Caloi} V., {Pecci} F.~F., {Galleti} S., {Rood}
  R.~T., 2005, ApJ, 631, 868

\bibitem[{{de Angeli} {et~al.}(2005){de Angeli}, {Piotto}, {Cassisi}, {Busso},
  {Recio-Blanco}, {Salaris}, {Aparicio}, \& {Rosenberg}}]{DAPC+05}
{de Angeli} F., {Piotto} G., {Cassisi} S., {Busso} G., {Recio-Blanco} A.,
  {Salaris} M., {Aparicio} A., {Rosenberg} A., 2005, AJ, 130, 116

\bibitem[{{Dobrovolskas} {et~al.}(2014){Dobrovolskas}, {Ku{\v c}inskas},
  {Bonifacio}, {Korotin}, {Steffen}, {Sbordone}, {Caffau}, {Ludwig}, {Royer},
  \& {Prakapavi{\v c}ius}}]{DKB+14}
{Dobrovolskas} V., {Ku{\v c}inskas} A., {Bonifacio} P., {Korotin} S.~A.,
  {Steffen} M., {Sbordone} L., {Caffau} E., {Ludwig} H.-G., {Royer} F.,
  {Prakapavi{\v c}ius} D., 2014, A\&A, 565, A121

\bibitem[{{Dotter} {et~al.}(2007){Dotter}, {Chaboyer}, {Jevremovi{\'c}},
  {Baron}, {Ferguson}, {Sarajedini}, \& {Anderson}}]{DCJ+07}
{Dotter} A., {Chaboyer} B., {Jevremovi{\'c}} D., {Baron} E., {Ferguson} J.~W.,
  {Sarajedini} A., {Anderson} J., 2007, AJ, 134, 376

\bibitem[{{Dotter} {et~al.}(2008){Dotter}, {Chaboyer}, {Jevremovi{\'c}},
  {Kostov}, {Baron}, \& {Ferguson}}]{DCJ+08}
{Dotter} A., {Chaboyer} B., {Jevremovi{\'c}} D., {Kostov} V., {Baron} E.,
  {Ferguson} J.~W., 2008, ApJS, 178, 89

\bibitem[{{Dotter} {et~al.}(2010){Dotter}, {Sarajedini}, {Anderson},
  {Aparicio}, {Bedin}, {Chaboyer}, {Majewski}, {Mar{\'{\i}}n-Franch}, {Milone},
  {Paust}, {Piotto}, {Reid}, {Rosenberg}, \& {Siegel}}]{DSA+10}
{Dotter} A., {Sarajedini} A., {Anderson} J., {Aparicio} A., {Bedin} L.~R.,
  {Chaboyer} B., {Majewski} S., {Mar{\'{\i}}n-Franch} A., {Milone} A., {Paust}
  N., {Piotto} G., {Reid} I.~N., {Rosenberg} A., {Siegel} M., 2010, ApJ, 708,
  698

\bibitem[{{Dupree} \& {Avrett}(2013)}]{DA13}
{Dupree} A.~K., {Avrett} E.~H., 2013, ArXiv e-prints

\bibitem[{{Dupree} {et~al.}(2009){Dupree}, {Smith}, \& {Strader}}]{DSS09}
{Dupree} A.~K., {Smith} G.~H., {Strader} J., 2009, AJ, 138, 1485

\bibitem[{{Fenner} {et~al.}(2004){Fenner}, {Campbell}, {Karakas}, {Lattanzio},
  \& {Gibson}}]{FCK+04}
{Fenner} Y., {Campbell} S., {Karakas} A.~I., {Lattanzio} J.~C., {Gibson} B.~K.,
  2004, MNRAS, 353, 789

\bibitem[{{Fusi Pecci} \& {Bellazzini}(1997)}]{FPB97}
{Fusi Pecci} F., {Bellazzini} M., 1997, in The Third Conference on Faint Blue
  Stars, {Philip} A.~G.~D., {Liebert} J., {Saffer} R., {Hayes} D.~S., eds., p.
  255

\bibitem[{{Gesicki} {et~al.}(2014){Gesicki}, {Zijlstra}, {Hajduk}, \&
  {Szyszka}}]{GZHS14}
{Gesicki} K., {Zijlstra} A.~A., {Hajduk} M., {Szyszka} C., 2014, A\&A, 566, A48

\bibitem[{{Girardi} {et~al.}(2000){Girardi}, {Bressan}, {Bertelli}, \&
  {Chiosi}}]{GBBC00}
{Girardi} L., {Bressan} A., {Bertelli} G., {Chiosi} C., 2000, A\&AS, 141, 371

\bibitem[{{Gnedin} {et~al.}(2002){Gnedin}, {Zhao}, {Pringle}, {Fall}, {Livio},
  \& {Meylan}}]{GZP+02}
{Gnedin} O.~Y., {Zhao} H., {Pringle} J.~E., {Fall} S.~M., {Livio} M., {Meylan}
  G., 2002, ApJ, 568, L23

\bibitem[{{Gratton} {et~al.}(2010{\natexlab{a}}){Gratton}, {Carretta},
  {Bragaglia}, {Lucatello}, \& {D'Orazi}}]{GCB+10}
{Gratton} R.~G., {Carretta} E., {Bragaglia} A., {Lucatello} S., {D'Orazi} V.,
  2010{\natexlab{a}}, A\&A, 517, A81

\bibitem[{{Gratton} {et~al.}(2010{\natexlab{b}}){Gratton}, {D'Orazi},
  {Bragaglia}, {Carretta}, \& {Lucatello}}]{GDOB+10}
{Gratton} R.~G., {D'Orazi} V., {Bragaglia} A., {Carretta} E., {Lucatello} S.,
  2010{\natexlab{b}}, A\&A, 522, A77

\bibitem[{{Gratton} {et~al.}(2013){Gratton}, {Lucatello}, {Sollima},
  {Carretta}, {Bragaglia}, {Momany}, {D'Orazi}, {Cassisi}, {Pietrinferni}, \&
  {Salaris}}]{GLS+13}
{Gratton} R.~G., {Lucatello} S., {Sollima} A., {Carretta} E., {Bragaglia} A.,
  {Momany} Y., {D'Orazi} V., {Cassisi} S., {Pietrinferni} A., {Salaris} M.,
  2013, A\&A, 549, A41

\bibitem[{{Grevesse} \& {Sauval}(1998)}]{GS98}
{Grevesse} N., {Sauval} A.~J., 1998, Space Science Reviews, 85, 161

\bibitem[{{Groenewegen}(2014)}]{Groenewegen14}
{Groenewegen} M.~A.~T., 2014, A\&A, 561, L11

\bibitem[{{Habgood}(2001)}]{Habgood01}
{Habgood} M.-J.~J., 2001, PhD thesis, The University of North Carolina at
  Chapel Hill

\bibitem[{Harris(1996)}]{Harris96}
Harris W.~E., 1996, ApJ, 112, 1487

\bibitem[{{Harris}(2010)}]{Harris10}
{Harris} W.~E., 2010, ArXiv e-prints

\bibitem[{{Henyey} {et~al.}(1965){Henyey}, {Vardya}, \& {Bodenheimer}}]{HVB65}
{Henyey} L., {Vardya} M.~S., {Bodenheimer} P., 1965, ApJ, 142, 841

\bibitem[{{Kalirai} {et~al.}(2009){Kalirai}, {Saul Davis}, {Richer},
  {Bergeron}, {Catelan}, {Hansen}, \& {Rich}}]{KSDR+09}
{Kalirai} J.~S., {Saul Davis} D., {Richer} H.~B., {Bergeron} P., {Catelan} M.,
  {Hansen} B.~M.~S., {Rich} R.~M., 2009, ApJ, 705, 408

\bibitem[{{Karakas} {et~al.}(2006){Karakas}, {Fenner}, {Sills}, {Campbell}, \&
  {Lattanzio}}]{KFS+06}
{Karakas} A., {Fenner} Y., {Sills} A., {Campbell} S.~W., {Lattanzio} J.~C.,
  2006, Memorie della Societa Astronomica Italiana, 77, 858

\bibitem[{{Lapenna} {et~al.}(2014){Lapenna}, {Mucciarelli}, {Lanzoni}, {Rosario
  Ferraro}, {Dalessandro}, {origlia}, \& {Massari}}]{LML+14}
{Lapenna} E., {Mucciarelli} A., {Lanzoni} B., {Rosario Ferraro} F.,
  {Dalessandro} E., {origlia} L., {Massari} D., 2014, ArXiv e-prints

\bibitem[{{Lebzelter} {et~al.}(2014){Lebzelter}, {Nowotny}, {Hinkle},
  {H{\"o}fner}, \& {Aringer}}]{LNH+14}
{Lebzelter} T., {Nowotny} W., {Hinkle} K.~H., {H{\"o}fner} S., {Aringer} B.,
  2014, A\&A, 567, A143

\bibitem[{{Lodders} {et~al.}(2009){Lodders}, {Palme}, \& {Gail}}]{LPG09}
{Lodders} K., {Palme} H., {Gail} H.-P., 2009, Landolt B{\"o}rnstein, 44

\bibitem[{{Marigo} {et~al.}(2008){Marigo}, {Girardi}, {Bressan}, {Groenewegen},
  {Silva}, \& {Granato}}]{MGB+08}
{Marigo} P., {Girardi} L., {Bressan} A., {Groenewegen} M.~A.~T., {Silva} L.,
  {Granato} G.~L., 2008, A\&A, 482, 883

\bibitem[{{Mar{\'{\i}}n-Franch} {et~al.}(2009){Mar{\'{\i}}n-Franch},
  {Aparicio}, {Piotto}, {Rosenberg}, {Chaboyer}, {Sarajedini}, {Siegel},
  {Anderson}, {Bedin}, {Dotter}, {Hempel}, {King}, {Majewski}, {Milone},
  {Paust}, \& {Reid}}]{MFAP+09}
{Mar{\'{\i}}n-Franch} A., {Aparicio} A., {Piotto} G., {Rosenberg} A.,
  {Chaboyer} B., {Sarajedini} A., {Siegel} M., {Anderson} J., {Bedin} L.~R.,
  {Dotter} A., {Hempel} M., {King} I., {Majewski} S., {Milone} A.~P., {Paust}
  N., {Reid} I.~N., 2009, ApJ, 694, 1498

\bibitem[{{McDonald} {et~al.}(2011{\natexlab{a}}){McDonald}, {Boyer}, {van
  Loon}, \& {Zijlstra}}]{MBvLZ11}
{McDonald} I., {Boyer} M.~L., {van Loon} J.~T., {Zijlstra} A.~A.,
  2011{\natexlab{a}}, ApJ, 730, 71

\bibitem[{{McDonald} {et~al.}(2011{\natexlab{b}}){McDonald}, {Boyer}, {van
  Loon}, {Zijlstra}, {Hora}, {Babler}, {Block}, {Gordon}, {Meade}, {Meixner},
  {Misselt}, {Robitaille}, {Sewi{\l}o}, {Shiao}, \& {Whitney}}]{MBvL+11}
{McDonald} I., {Boyer} M.~L., {van Loon} J.~T., {Zijlstra} A.~A., {Hora} J.~L.,
  {Babler} B., {Block} M., {Gordon} K., {Meade} M., {Meixner} M., {Misselt} K.,
  {Robitaille} T., {Sewi{\l}o} M., {Shiao} B., {Whitney} B.,
  2011{\natexlab{b}}, ApJS, 193, 23

\bibitem[{{McDonald} {et~al.}(2011{\natexlab{c}}){McDonald}, {Johnson}, \&
  {Zijlstra}}]{MJZ11}
{McDonald} I., {Johnson} C.~I., {Zijlstra} A.~A., 2011{\natexlab{c}}, MNRAS,
  416, L6

\bibitem[{{McDonald} \& {van Loon}(2007)}]{MvL07}
{McDonald} I., {van Loon} J.~T., 2007, A\&A, 476, 1261

\bibitem[{{McDonald} {et~al.}(2009){McDonald}, {van Loon}, {Decin}, {Boyer},
  {Dupree}, {Evans}, {Gehrz}, \& {Woodward}}]{MvLD+09}
{McDonald} I., {van Loon} J.~T., {Decin} L., {Boyer} M.~L., {Dupree} A.~K.,
  {Evans} A., {Gehrz} R.~D., {Woodward} C.~E., 2009, MNRAS, 394, 831

\bibitem[{{McDonald} {et~al.}(2010){McDonald}, {van Loon}, {Dupree}, \&
  {Boyer}}]{MvLDB10}
{McDonald} I., {van Loon} J.~T., {Dupree} A.~K., {Boyer} M.~L., 2010, MNRAS,
  405, 1711

\bibitem[{{McDonald} {et~al.}(2011{\natexlab{d}}){McDonald}, {van Loon},
  {Sloan}, {Dupree}, {Zijlstra}, {Boyer}, {Gehrz}, {Evans}, {Woodward}, \&
  {Johnson}}]{MvLS+11}
{McDonald} I., {van Loon} J.~T., {Sloan} G.~C., {Dupree} A.~K., {Zijlstra}
  A.~A., {Boyer} M.~L., {Gehrz} R.~D., {Evans} A., {Woodward} C.~E., {Johnson}
  C.~I., 2011{\natexlab{d}}, MNRAS, 417, 20

\bibitem[{{McDonald} \& {Zijlstra}(2014)}]{MZ14}
{McDonald} I., {Zijlstra} A., 2014, ArXiv e-prints

\bibitem[{{McDonald} {et~al.}(2012){McDonald}, {Zijlstra}, \& {Boyer}}]{MZB12}
{McDonald} I., {Zijlstra} A.~A., {Boyer} M.~L., 2012, MNRAS, 427, 343

\bibitem[{{M{\'e}sz{\'a}ros} {et~al.}(2009){M{\'e}sz{\'a}ros}, {Avrett}, \&
  {Dupree}}]{MAD09}
{M{\'e}sz{\'a}ros} S., {Avrett} E.~H., {Dupree} A.~K., 2009, AJ, 138, 615

\bibitem[{{Moehler} {et~al.}(2004){Moehler}, {Koester}, {Zoccali}, {Ferraro},
  {Heber}, {Napiwotzki}, \& {Renzini}}]{MKZ+04}
{Moehler} S., {Koester} D., {Zoccali} M., {Ferraro} F.~R., {Heber} U.,
  {Napiwotzki} R., {Renzini} A., 2004, A\&A, 420, 515

\bibitem[{{Momany} {et~al.}(2012){Momany}, {Saviane}, {Smette}, {Bayo},
  {Girardi}, {Marconi}, {Milone}, \& {Bressan}}]{MSS+12}
{Momany} Y., {Saviane} I., {Smette} A., {Bayo} A., {Girardi} L., {Marconi} G.,
  {Milone} A.~P., {Bressan} A., 2012, A\&A, 537, A2

\bibitem[{{Monaco} {et~al.}(2005){Monaco}, {Bellazzini}, {Bonifacio},
  {Ferraro}, {Marconi}, {Pancino}, {Sbordone}, \& {Zaggia}}]{MBB+05}
{Monaco} L., {Bellazzini} M., {Bonifacio} P., {Ferraro} F.~R., {Marconi} G.,
  {Pancino} E., {Sbordone} L., {Zaggia} S., 2005, A\&A, 441, 141

\bibitem[{{Paxton} {et~al.}(2011){Paxton}, {Bildsten}, {Dotter}, {Herwig},
  {Lesaffre}, \& {Timmes}}]{PBD+11}
{Paxton} B., {Bildsten} L., {Dotter} A., {Herwig} F., {Lesaffre} P., {Timmes}
  F., 2011, ApJS, 192, 3

\bibitem[{{Paxton} {et~al.}(2013){Paxton}, {Cantiello}, {Arras}, {Bildsten},
  {Brown}, {Dotter}, {Mankovich}, {Montgomery}, {Stello}, {Timmes}, \&
  {Townsend}}]{PCA+13}
{Paxton} B., {Cantiello} M., {Arras} P., {Bildsten} L., {Brown} E.~F., {Dotter}
  A., {Mankovich} C., {Montgomery} M.~H., {Stello} D., {Timmes} F.~X.,
  {Townsend} R., 2013, ApJS, 208, 4

\bibitem[{{Pietrinferni} {et~al.}(2004){Pietrinferni}, {Cassisi}, {Salaris}, \&
  {Castelli}}]{PCSC04}
{Pietrinferni} A., {Cassisi} S., {Salaris} M., {Castelli} F., 2004, ApJ, 612,
  168

\bibitem[{{Piotto} {et~al.}(2002){Piotto}, {King}, {Djorgovski}, {Sosin},
  {Zoccali}, {Saviane}, {De Angeli}, {Riello}, {Recio-Blanco}, {Rich},
  {Meylan}, \& {Renzini}}]{PKD+02}
{Piotto} G., {King} I.~R., {Djorgovski} S.~G., {Sosin} C., {Zoccali} M.,
  {Saviane} I., {De Angeli} F., {Riello} M., {Recio-Blanco} A., {Rich} R.~M.,
  {Meylan} G., {Renzini} A., 2002, A\&A, 391, 945

\bibitem[{{Piotto} {et~al.}(2005){Piotto}, {Villanova}, {Bedin}, {Gratton},
  {Cassisi}, {Momany}, {Recio-Blanco}, {Lucatello}, {Anderson}, {King},
  {Pietrinferni}, \& {Carraro}}]{PVB+05}
{Piotto} G., {Villanova} S., {Bedin} L.~R., {Gratton} R., {Cassisi} S.,
  {Momany} Y., {Recio-Blanco} A., {Lucatello} S., {Anderson} J., {King} I.~R.,
  {Pietrinferni} A., {Carraro} G., 2005, ApJ, 621, 777

\bibitem[{{Planck Collaboration} {et~al.}(2013){Planck Collaboration}, {Ade},
  {Aghanim}, {Armitage-Caplan}, {Arnaud}, {Ashdown}, {Atrio-Barandela},
  {Aumont}, {Baccigalupi}, {Banday}, \& et~al.}]{Planck-2013-XVI}
{Planck Collaboration}, {Ade} P.~A.~R., {Aghanim} N., {Armitage-Caplan} C.,
  {Arnaud} M., {Ashdown} M., {Atrio-Barandela} F., {Aumont} J., {Baccigalupi}
  C., {Banday} A.~J., et~al., 2013, ArXiv e-prints

\bibitem[{{Reimers}(1975)}]{Reimers75}
{Reimers} D., 1975, Memoires of the Soci{\'e}t{\'e} Royale des Sciences de
  Li{\`e}ge, 8, 369

\bibitem[{{Richer} {et~al.}(1997){Richer}, {Fahlman}, {Ibata}, {Pryor}, {Bell},
  {Bolte}, {Bond}, {Harris}, {Hesser}, {Holland}, {Ivanans}, {Mandushev},
  {Stetson}, \& {Wood}}]{RFI+97}
{Richer} H.~B., {Fahlman} G.~G., {Ibata} R.~A., {Pryor} C., {Bell} R.~A.,
  {Bolte} M., {Bond} H.~E., {Harris} W.~E., {Hesser} J.~E., {Holland} S.,
  {Ivanans} N., {Mandushev} G., {Stetson} P.~B., {Wood} M.~A., 1997, ApJ, 484,
  741

\bibitem[{{Roediger} {et~al.}(2013){Roediger}, {Courteau}, {Graves}, \&
  {Schiavon}}]{RCGS13}
{Roediger} J.~C., {Courteau} S., {Graves} G., {Schiavon} R., 2013, ArXiv
  e-prints

\bibitem[{{Roediger} {et~al.}(2014){Roediger}, {Courteau}, {Graves}, \&
  {Schiavon}}]{RCGS14}
{Roediger} J.~C., {Courteau} S., {Graves} G., {Schiavon} R.~P., 2014, ApJS,
  210, 10

\bibitem[{{Rosenberg} {et~al.}(2000{\natexlab{a}}){Rosenberg}, {Aparicio},
  {Saviane}, \& {Piotto}}]{RASP00}
{Rosenberg} A., {Aparicio} A., {Saviane} I., {Piotto} G., 2000{\natexlab{a}},
  A\&AS, 145, 451

\bibitem[{{Rosenberg} {et~al.}(2000{\natexlab{b}}){Rosenberg}, {Piotto},
  {Saviane}, \& {Aparicio}}]{RPSA00}
{Rosenberg} A., {Piotto} G., {Saviane} I., {Aparicio} A., 2000{\natexlab{b}},
  A\&AS, 144, 5

\bibitem[{{Rosenberg} {et~al.}(1999){Rosenberg}, {Saviane}, {Piotto}, \&
  {Aparicio}}]{RSPA99}
{Rosenberg} A., {Saviane} I., {Piotto} G., {Aparicio} A., 1999, AJ, 118, 2306

\bibitem[{{Rutledge} {et~al.}(1997){Rutledge}, {Hesser}, {Stetson}, {Mateo},
  {Simard}, {Bolte}, {Friel}, \& {Copin}}]{RHS+97}
{Rutledge} G.~A., {Hesser} J.~E., {Stetson} P.~B., {Mateo} M., {Simard} L.,
  {Bolte} M., {Friel} E.~D., {Copin} Y., 1997, PASP, 109, 883

\bibitem[{{Salaris} \& {Weiss}(1997)}]{SW97}
{Salaris} M., {Weiss} A., 1997, A\&A, 327, 107

\bibitem[{{Salaris} \& {Weiss}(1998)}]{SW98}
---, 1998, A\&A, 335, 943

\bibitem[{{Salaris} \& {Weiss}(2002)}]{SW02}
---, 2002, A\&A, 388, 492

\bibitem[{{Sarajedini} {et~al.}(2007){Sarajedini}, {Bedin}, {Chaboyer},
  {Dotter}, {Siegel}, {Anderson}, {Aparicio}, {King}, {Majewski},
  {Mar{\'{\i}}n-Franch}, {Piotto}, {Reid}, \& {Rosenberg}}]{SBC+07}
{Sarajedini} A., {Bedin} L.~R., {Chaboyer} B., {Dotter} A., {Siegel} M.,
  {Anderson} J., {Aparicio} A., {King} I., {Majewski} S., {Mar{\'{\i}}n-Franch}
  A., {Piotto} G., {Reid} I.~N., {Rosenberg} A., 2007, AJ, 133, 1658

\bibitem[{{Schr{\"o}der} \& {Cuntz}(2005)}]{SC05}
{Schr{\"o}der} K.-P., {Cuntz} M., 2005, ApJ, 630, L73

\bibitem[{{Schr{\"o}der} \& {Cuntz}(2007)}]{SC07}
---, 2007, A\&A, 465, 593

\bibitem[{{Serenelli} \& {Basu}(2010)}]{SB10}
{Serenelli} A.~M., {Basu} S., 2010, ApJ, 719, 865

\bibitem[{{Smith} {et~al.}(2014){Smith}, {Dupree}, \& {Strader}}]{SDS14}
{Smith} G., {Dupree} A., {Strader} J., 2014, ArXiv e-prints

\bibitem[{{Sneden}(2004)}]{Sneden04}
{Sneden} C., 2004, MmSAI, 75, 267

\bibitem[{{Thygesen} {et~al.}(2014){Thygesen}, {Sbordone}, {Andrievsky},
  {Korotin}, {Yong}, {Zaggia}, {Ludwig}, {Collet}, {Asplund}, {D'Antona},
  {Mel{\'e}ndez}, \& {D'Ercole}}]{TSA+14}
{Thygesen} A.~O., {Sbordone} L., {Andrievsky} S., {Korotin} S., {Yong} D.,
  {Zaggia} S., {Ludwig} H.-G., {Collet} R., {Asplund} M., {D'Antona} F.,
  {Mel{\'e}ndez} J., {D'Ercole} A., 2014, ArXiv e-prints

\bibitem[{{{\v C}erniauskas} {et~al.}(2014){{\v C}erniauskas}, {Ku{\v
  c}inskas}, {Bonifacio}, {Andrievsky}, {Korotin}, \& {Dobrovolskas}}]{CKB+14}
{{\v C}erniauskas} A., {Ku{\v c}inskas} A., {Bonifacio} P., {Andrievsky} S.~M.,
  {Korotin} S.~A., {Dobrovolskas} V., 2014, ArXiv e-prints

\bibitem[{{Valcarce} {et~al.}(2013{\natexlab{a}}){Valcarce}, {Catelan},
  {Alonso-Garc{\'{\i}}a}, {Cort{\'e}s}, \& {De Medeiros}}]{VCAG+13}
{Valcarce} A.~A.~R., {Catelan} M., {Alonso-Garc{\'{\i}}a} J., {Cort{\'e}s} C.,
  {De Medeiros} J.~R., 2013{\natexlab{a}}, ArXiv e-prints

\bibitem[{{Valcarce} {et~al.}(2013{\natexlab{b}}){Valcarce}, {Catelan}, \& {De
  Medeiros}}]{VCDM13}
{Valcarce} A.~A.~R., {Catelan} M., {De Medeiros} J.~R., 2013{\natexlab{b}},
  A\&A, 553, A62

\bibitem[{{Valcarce} {et~al.}(2012){Valcarce}, {Catelan}, \&
  {Sweigart}}]{VCS12}
{Valcarce} A.~A.~R., {Catelan} M., {Sweigart} A.~V., 2012, A\&A, 547, A5

\bibitem[{{VandenBerg} {et~al.}(2012){VandenBerg}, {Bergbusch}, {Dotter},
  {Ferguson}, {Michaud}, {Richer}, \& {Proffitt}}]{VBD+12}
{VandenBerg} D.~A., {Bergbusch} P.~A., {Dotter} A., {Ferguson} J.~W., {Michaud}
  G., {Richer} J., {Proffitt} C.~R., 2012, ApJ, 755, 15

\bibitem[{{VandenBerg} {et~al.}(2006){VandenBerg}, {Bergbusch}, \&
  {Dowler}}]{VBD06}
{VandenBerg} D.~A., {Bergbusch} P.~A., {Dowler} P.~D., 2006, ApJS, 162, 375

\bibitem[{{VandenBerg} {et~al.}(2013){VandenBerg}, {Brogaard}, {Leaman}, \&
  {Casagrande}}]{VBLC13}
{VandenBerg} D.~A., {Brogaard} K., {Leaman} R., {Casagrande} L., 2013, ApJ,
  775, 134

\bibitem[{{Villanova} {et~al.}(2012){Villanova}, {Geisler}, {Piotto}, \&
  {Gratton}}]{VGPG12}
{Villanova} S., {Geisler} D., {Piotto} G., {Gratton} R.~G., 2012, ApJ, 748, 62

\bibitem[{{Villanova} {et~al.}(2009){Villanova}, {Piotto}, \&
  {Gratton}}]{VPG09}
{Villanova} S., {Piotto} G., {Gratton} R.~G., 2009, A\&A, 499, 755

\bibitem[{{Woitke}(2006)}]{Woitke06b}
{Woitke} P., 2006, A\&A, 460, L9

\bibitem[{{Zinn} \& {West}(1984)}]{ZW84}
{Zinn} R., {West} M.~J., 1984, ApJS, 55, 45

\end{thebibliography}

\label{lastpage}

\end{document}